\documentclass[12pt,preprint]{aastex}
\usepackage{epsfig}
\usepackage{natbib}                
\shorttitle{Eclipsing Binaries with Companions}
\shortauthors{Gies et al.}

\newcommand{\noprint}[1]{}
\newcommand{\figsetstart}{{\bf Fig. Set} }
\newcommand{\figsetend}{}
\newcommand{\figsetgrpstart}{}
\newcommand{\figsetgrpend}{}
\newcommand{\figsetnum}[1]{{\bf #1.}}
\newcommand{\figsettitle}[1]{ {\bf #1} }
\newcommand{\figsetgrpnum}[1]{\noprint{#1}}
\newcommand{\figsetgrptitle}[1]{\noprint{#1}}
\newcommand{\figsetplot}[1]{\noprint{#1}}
\newcommand{\figsetgrpnote}[1]{\noprint{#1}}

\begin{document}

\received{2015 August 9}

\title{Kepler Eclipsing Binaries with Stellar Companions}

\author{D. R. Gies, R. A. Matson, Z. Guo, K. V. Lester}
\affil{Center for High Angular Resolution Astronomy and
 Department of Physics and Astronomy,
 Georgia State University, P. O. Box 5060, Atlanta, GA 30302-5060, USA} 
\email{gies@chara.gsu.edu, rmatson@chara.gsu.edu, 
        guo@chara.gsu.edu,  lester@chara.gsu.edu}

\author{J. A. Orosz}
\affil{Department of Astronomy, San Diego State University, 
 San Diego, CA 92182-1221, USA} 
\email{jorosz@mail.sdsu.edu}

\and 

\author{G. J. Peters}
\affil{Space Sciences Center and Department of Physics and Astronomy, 
 University of Southern California, Los Angeles, CA 90089-1341, USA}
\email{gjpeters@mucen.usc.edu}

\slugcomment{Submitted to AJ; R1 10/05/2015}

\paperid{AJ-12791}


\begin{abstract}
Many short-period binary stars have distant orbiting companions 
that have played a role in driving the binary components into
close separation.  Indirect detection of a tertiary star is 
possible by measuring apparent changes in eclipse times of
eclipsing binaries as the binary orbits the common center of 
mass.  Here we present an analysis of the eclipse timings of 
41 eclipsing binaries observed throughout the NASA {\it Kepler} 
mission of long duration and precise photometry.   This subset 
of binaries is characterized by relatively deep and frequent 
eclipses of both stellar components.  We present preliminary 
orbital elements for seven probable triple stars among this 
sample, and we discuss apparent period changes in seven additional
eclipsing binaries that may be related to motion about a 
tertiary in a long period orbit.  The results will be used in 
ongoing investigations of the spectra and light curves of these
binaries for further evidence of the presence of third stars. 
\end{abstract}




\section{Introduction}                              

The process of star formation involves a large scale transfer 
of angular momentum from inner cloud to outer disk that often
results in the formation of binary stars \citep{2013ARA+A..51..269D}. 
The reduction in separation to produce close pairs may involve
the action of a third star through the eccentric Kozai-Lidov 
mechanism \citep{2014ApJ...793..137N}, and indeed 
\citet{2006A+A...450..681T} showed that most short period binaries
are orbited by a distant third star ($96\%$ for $P< 3$ days).  
Such tertiary companions can be directly observed in nearby cases through 
high angular resolution methods, but for most binaries we need other 
means to detect them.  A particularly attractive method is to 
search for light travel time effects in the times of minima 
of eclipsing binaries that are caused by the displacement of 
the binary around the center of mass of the binary - tertiary
system \citep{1990BAICz..41..231M,2005ASPC..335..103P}. 

The NASA {\it Kepler} mission has provided us with a singular 
opportunity to search for eclipse timing variations due to 
orbital motion about a third star.   The {\it Kepler} and 
follow-on {\it K2} missions have created long duration light 
curves\footnote{http://keplerebs.villanova.edu/}
for some 3000 eclipsing binaries 
\citep{2011AJ....141...83P,2011AJ....142..160S,2015A+A...579A..19A,2015MNRAS.452.3561L}. 
The search for eclipse time variations among the {\it Kepler} binaries 
has been led by \citet{2014AJ....147...45C} (short period systems through Q16)
and \citet{2015arXiv150307295O} (long period systems).  
These surveys indicate that $\approx 18\%$ of the observed systems
display variations consistent with orbital motion about a third star. 
Detailed case studies of specific triples have been presented by 
\citet{2013ApJ...768...33R}, \citet{2015MNRAS.448..946B}, and 
\citet{2015AJ....149..197Z}, among others.   
Furthermore, {\it Kepler} light curves have uncovered a number of remarkable cases in which 
the orientation of the outer system is such that eclipses by the tertiary are also observed 
\citep{2011Sci...331..562C,2011Sci...332..216D,2012A+A...545L...4A,2013MNRAS.428.1656B}.

We embarked on a search for eclipse timing variations among 
a subset of 41 eclipsing binaries that were identified prior to the 
start of {\it Kepler} observations \citep{2012AJ....143..137G}. 
We subsequently completed a large set of spectroscopic observations 
of this sample in order to derive spectroscopic orbital elements, 
estimate the stellar properties, compare with evolutionary codes, 
and explore the pulsational properties of the component stars. 
Our first paper documented the eclipse times in observations made over
quarters Q0 -- Q9 (2009.3 -- 2011.5).  Now with the {\it Kepler} 
mission complete with observations through Q17 (ending 2013.4), we present 
here the eclipse timings for our sample of 41 binaries over the entire 
duration of the mission.  These results help document the frequency 
of tertiary companions in this well-observed group of stars and 
provide period, epoch, and third-light constraints for our ongoing 
spectroscopic and light curve analyses.  We present the full set 
of timing measurements in Section 2, and we offer preliminary 
orbital elements for seven probable triple systems in Section 3. 
We summarize the results and discuss their implications in Section 4. 


\section{Eclipse Timing Measurements}               

The {\it Kepler} light curves of the target eclipsing binaries 
represent the latest reprocessing of all the observations 
made between quarters Q0 and Q17.  These are long cadence 
measurements that are obtained in net exposure times of 
29.4244 minutes.  The associated times given in our first paper
\citep{2012AJ....143..137G} were based upon UTC (Coordinated Universal Time)
while the current set uses TDB (Barycentric Dynamical 
Time\footnote{https://archive.stsci.edu/kepler/release\_notes/release\_notes19/DataRelease\_19\_20130204.pdf}), 
and here we report the times in reduced Barycentric Julian Date 
(BJD -- 2,400,000 days).   We used the Simple Aperture Photometry
(SAP) flux except in the case of KIC~04678873.  This star has 
a nearby companion that creates large variations in the summed 
flux throughout and between observing quarters.  We used
PyKE\footnote{http://keplerscience.arc.nasa.gov/DataAnalysisTools.shtml}
software to extract the flux of this target using a larger aperture
that included both stars.  The final light curve has a significant 
``third light'' flux contribution from the companion but is free 
from large fluctuations between quarters.  These light curves 
were detrended as described in our first paper, and the subsequent
analysis is based on flux versions normalized to unity at the 
maximum between eclipses.  The list of targets appears in Table~1. 

\placetable{tab1}      

The eclipse timing measurements were made in almost the same 
way as described in our first paper.  We first formed an
eclipse template as a function of orbital phase by assuming a 
preliminary eclipse epoch $T$ and binary orbital period $P$
and then forming a phase rebinned version of the specific 
primary or secondary eclipse from all the available fluxes.
The template was then smoothed by convolution with a Gaussian function
to avoid the introduction of spurious patterns in systems where 
the orbital period is a near integer multiple of the observational 
cadence \citep{2014AJ....147...45C}.
The eclipse template was recentered on the minimum by 
fitting a parabola to the lowest $20\%$ of the eclipse flux, 
and the epoch of minimum adjusted accordingly. 
We then fit the template to each eclipse (after local flux normalization)
using the code MPFIT \citep{2009ASPC..411..251M}
to find best match with parameters for the phase offset and 
an eclipse depth scaling factor $D_E$.  The latter parameter 
allows the eclipse template to be transformed by 
\begin{equation}
F({\rm revised}) = 1-D_E+D_E~F({\rm template})
\end{equation}
so that temporal changes in the eclipse depth may be tracked. 
The eclipse depth changes are interesting in the context of 
long term changes (due to apsidal motion for example), 
starspot activity, and quarter-to-quarter blending of the 
target flux with that from stars with small angular separations. 
Finally the phase offset parameter is multiplied by $P$ to find
the time of eclipse.  Our measurements appear in Table~2 
(given in full as a machine readable table in the electronic version)
that lists the target KIC number, a 1 or 2 for primary (deeper)
or secondary eclipse, the eclipse time $T_E$ and its uncertainty, 
and the depth scaling parameter $D_E$ and its uncertainty.
A comparison of our eclipse timing measurements with those from 
\citet{2014AJ....147...45C} and \citet{2015AJ....149..197Z} shows 
excellent agreement, although there may be a small constant offset 
between the results due to differences in measurement techniques. 

\placetable{tab2}      

The eclipse times were then fit using a quadratic function 
for both the primary and secondary eclipses independently
\begin{equation}
T_E = T + P E + Q E^2
\end{equation}
where $E$ is the integer eclipse cycle number in the time series.
The final parameter is related to the period derivative by
\begin{equation}
\dot{P} / P = 2 Q / P^2.
\end{equation}
We included the quadratic term $Q E^2$ to search for long term 
trends related to mass transfer in interacting binaries and 
to orbital motion about a tertiary in a very long period orbit (Section 4).
However, in cases where the time span of observations is similar
to a tertiary orbital period (see KIC~04848423), the curve is more
complicated than a simple parabola and the value of $Q$ has no 
physical meaning.  We treated the primary and secondary eclipse 
times independently in order to search for cases of non-zero 
orbital eccentricity (in which the intervals between primary and secondary
eclipse may differ from half the period) and possible apsidal motion 
(causing differences in measured period; \citealt{2003IBVS.5482....1O}).  
Table~1 lists the fitting results in two rows per target
for each of the primary and secondary eclipses (KIC number appended
with a P or S, respectively).   The columns of Table~1 give the KIC number, 
the average uncertainty in eclipse time measurement $I$ (internal scatter),
the standard deviation of the full set of measurements $E$ (external scatter), 
the epoch of minimum $T$ near the middle of the time series, the 
orbital period $P$, the period derivative $\dot{P} / P$, the standard 
deviation of the eclipse depth factor $\sigma (D_E)$, and a code to 
references on eclipse timings for the target.  The numbers in parentheses
give the uncertainties in terms of the last digit quoted.  
The final column includes the code ECC to identify those systems with a 
non-zero eccentricity as indicated by significantly differing epochs 
of eclipse, $(|T({\rm sec})-T({\rm pri})-P/2| ~{\rm mod} ~P) > 0$.

A graphical representation of the results is shown in Figure~Set~1
(given in full in the electronic version).  Each figure shows the net 
variation of eclipse timing ($O$ = observed) after subtraction of 
the linear trend ($C = T + P E$ = calculated). In several cases the 
{\it Kepler} measurements are augmented with timings from before 
and after {\it Kepler} from \citet{2015AJ....149..197Z} (see \S3). 
The plots appear often as scatter diagrams with no trends visible 
in either the primary or secondary eclipse measurements (for example, 
KIC~5738698; Fig.\ 1.16).  However, trends do appear in many cases 
that are related to starspots, apsidal motion (KIC~04544587; Fig. 1.6;
\citealt{2013MNRAS.434..925H}), long term period changes, or the 
light time changes associated with orbital motion about a third star.  
We focus on the latter group in the next section.  

\placefigure{fig1}     

The variations due to starspots are often found only in the timings for the 
eclipse of the cooler secondary component (for example, KIC~10581918 = WX~Dra, 
an Algol-type binary; Fig.\ 1.35).  These variations are caused by the change 
in the surface intensity and center of light over the visible hemisphere
\citep{2002A+A...387..969K,2013ApJ...774...81T}. 
For example, suppose a dark spot group rotates into view on the 
approaching portion of the eclipsed star.  This will cause the overall
flux to drop around the time of eclipse, and the eclipse minimum will 
occur slightly later as the transiting star passes over the center of 
light on the receding (brighter) portion of the visible hemisphere.  
\citet{2012Sci...337.1511O} and \citet{2015ApJ...807..170H} have
documented how the starspot illumination changes lead to a general 
anticorrelation between the local flux derivative and eclipse time
deviation, and this trend is found for example in the eclipse light curve 
of KIC~10581918 (see Fig.\ 1.35 of \citealt{2012AJ....143..137G}). 
We suspect that most of the cases that show ``random-walk'' 
kinds of variations on relatively short time scales are due to 
starspot activity on one or both components.  

Finally we note that the $O-C$ diagram for the eccentric binary 
KIC~04544587 (Fig.\ 1.6) looks quite different from what appeared in our 
earlier work (Fig.\ 2.6 of \citealt{2012AJ....143..137G}) because we now 
determine separate periods for the primary and secondary eclipse times 
so that the differing slopes in the earlier figure are absent here. 


\section{Preliminary Orbital Solutions}             

The light time effect variations due to a third star may be difficult
to detect if the third star's orbital period is longer than the 
duration of the {\it Kepler} time series (4 years).  In such cases, 
we may only sample a section of the linear or curved parts of the 
$O-C$ variation, so a full orbital fit must await further eclipse 
observations.  On the other hand, the $O-C$ variations from 
shorter period companions may be sufficient for a preliminary 
orbital solution.  We selected seven potential targets 
for orbital fits (listed in Table~3) 
from those that showed two inflection points in their $O-C$ 
changes in measurements for both the primary and secondary 
eclipses.  There are supplementary timing measurements from 
\citet{2015AJ....149..197Z} for five of these seven candidate 
systems, and we included these other timings in our orbital 
fits.  However, because our eclipse fitting method differs from 
that adopted by Zasche et al., we first compared 
our eclipse times with theirs for the same {\it Kepler} observations
to find a small mean difference $\Delta T$ (mean of $T_E$ from this work 
minus that from Zasche et al.) that was added to the supplementary data.  
These small offsets are listed in column 11 of Table~3.  
The $O-C$ diagrams in Figure~Set~1 include these supplementary timings 
for the five targets discussed by \citet{2015AJ....149..197Z}. 

We fit the eclipse timings using a linear plus Keplerian model 
\citep{1952ApJ...116..211I,1959AJ.....64..149I,1990BAICz..41..231M,2005ASPC..335..103P}:
\begin{equation}
T_E = T_1 + P E + {{A}\over{\sqrt{1 - e^2 \cos^2 \omega}}} 
 \left[{{1-e^2}\over{1+e\cos \nu}}\sin(\nu+\omega)+e\sin\omega \right]
\end{equation}
where the semiamplitude is $A=(a_{12} \sin i) \sqrt{1 - e^2 \cos^2 \omega} /c$, 
$a_{12}$ is the semimajor axis of the binary in its orbit around the common
center of mass of the triple system, $i$ is the inclination, 
$e$ is the eccentricity, $\omega$ is the longitude of periastron,
and $\nu$ is the true anomaly of the binary in its orbit about the 
center of mass of the triple system. 
The parameter $\nu$ is found from the outer orbit epoch of periastron $T_3$, period $P_3$, 
$e$, and time of eclipse.  Note that we are ignoring physical perturbations
of the inner orbit, because their associated amplitudes tend to be much lower than 
those of the light travel time effect for longer period orbits of the third star 
like those we are considering here \citep{2012AJ....143..137G,2013ApJ...768...33R}.   
We made fits of the primary eclipse timings using the model formulation above
and the MPFIT solver for parameters $T_1$, $P$, $A$, $e$, $\omega$, $P_3$, and $T_3$. 
For completeness, we then fit the secondary eclipse times with all these 
parameters fixed except the epoch $T_2$.    Our final results are listed in 
Table~3 together with a mass function for the third star 
\citep{1973A&AS...12....1F,1990BAICz..41..231M},
\begin{equation}
f(m_3) = {{(m_3\sin i)^3}\over{(m_1+m_2+m_3)^2}} = 
 {{1}\over{P_3^2}} \left[{{173.15~A}\over{\sqrt{1-e^2\cos^2\omega}}}\right]^3 
\end{equation}
where $A$ is given in units of days and $P_3$ in units of years.
Again numbers in parentheses give the uncertainty in the last quoted digit that
were taken as the maximum of the estimated errors from MPFIT and the absolute
differences of results from trials made with some parameters fixed.  
We caution that these formal uncertainties are probably underestimates 
because the observations only record about one orbital cycle.  

\placetable{tab3}      

The derived orbital fits are shown as solid lines in Figure~Set~1 for 
these seven candidate triples.  The fits are satisfactory in every case, 
and the solutions are often driven by measurements made beyond the 
time span of the {\it Kepler} observations.  \citet{2015AJ....149..197Z}
discuss the solutions for five of these triples, and in general our 
results agree within the uncertainties.   Orbits for the other two, 
KIC~04574310 and KIC~04848423, are shown here for the first time. 


\section{Discussion}                                

Our analysis of the eclipse timing variations has led to the identification 
of seven probable triples among our sample of 41 targets.  Although
this is a small sample, the detection rate of triples by the light travel 
time effect is $7/41=17\%$, which is consistent with estimates from 
other larger samples of {\it Kepler} eclipsing binaries 
\citep{2013ApJ...768...33R,2014AJ....147...45C,2015arXiv150307295O}.
The observed semiamplitudes range from 29 seconds for KIC~4574310
to 26 minutes for KIC~02305372.  The associated mass functions generally 
indicate that the companions are probably lower mass, K-type stars
(with the exception of KIC~04574310, which has a low mass function 
more consistent with a brown dwarf companion), but we 
caution that such mass estimates are lower limits because 
of the $\sin i$ factor in the mass function (equation 5).   
We might have detected even smaller semiamplitudes 
in principle, because there are a number of cases where the 
standard deviation of the timings is less than 10~seconds (see 
the external scatter $E$ in column 3 of Table~1).  Note that 
the inner binary periods are generally 5 days or less (with the 
exception of KIC~12071006 with $P=6.096$ days), and planets orbiting eclipsing binaries
have only been detected in systems with periods in excess of 7 days
\citep{2014IAUS..293..125W}.  This may indicate that Kozai modulations with a third 
star have tended to clear away lower mass components in binaries 
with short orbital periods \citep{2015MNRAS.453.3554M,2015PNAS..112.9264M}.

It is certainly possible that other systems in our sample have distant third
stars that have orbits longer than 4 years, in which we record only a 
small piece of the total $O-C$ variation.   For example, there are seven 
systems with very large and similar parabolic variations observed in both
the primary and secondary eclipses (KIC~03241619, KIC~03440230, KIC~04851217,
KIC~05621294, KIC~06206751, KIC~09602595, KIC~10736223; see Table~1). 
These systems may represent those that were recorded during wide orbit 
phases of rapid change.  If we imagine that the orbital $O-C$ variations 
are approximately sinusoidal, then during the {\it Kepler} window of 
observations we might expect about half to be found in the nearly linear
part (undetected in the $O-C$ variations) and the other half to be 
found in curved portions (yielding a significant $\dot{P}/P$ variation).
The latter group would have equal numbers of positive and negative values
of period derivative, and indeed we find nearly equal numbers 
(three positive and four negative) among those cases with large 
parabolic variations.  Similar numbers of positive and negative valued
quadratic changes in $O-C$ were also found among eclipsing binaries 
in the SuperWASP survey \citep{2015A+A...578A.136L}, consistent with the idea
that many of these are triples with incompletely covered, long orbits. 
We caution that some of the weaker parabolic trends may result from 
long term variations in starspot activity and not orbital motion
(for example, the case of KIC~02708156 in Fig.\ 1.2). 
These cases may often be identified by correlated variations between 
$O-C$ and the eclipse depth scaling parameter $D_E$.  

Our analysis of the eclipse timing measurements has provided us 
with a reliable set of orbital ephemerides and a list of probable 
triples among this sample of eclipsing binaries.  In future papers, 
we will use these results in combined spectroscopic and photometric
analyses to determine the stellar properties, evolutionary state, 
and pulsational characteristics.  We plan to investigate the 
influence of ``third light'' components on the interpretation of the 
orbital light curve and to search for direct evidence of the 
spectral lines of tertiaries among the triple star candidates. 


\acknowledgments

{\it Kepler} was competitively selected as the tenth Discovery mission. 
Funding for this mission is provided by NASA's Science Mission Directorate. 
This study was supported by NASA grants NNX12AC81G, NNX13AC21G, and NNX13AC20G.
This material is based upon work supported by the National Science Foundation 
under Grant No.~AST-1411654. 
Institutional support has been provided from the GSU College
of Arts and Sciences and from the Research Program Enhancement
fund of the Board of Regents of the University System of Georgia,
administered through the GSU Office of the Vice President for Research
and Economic Development.
All of the data presented in this paper were obtained from the 
Mikulski Archive for Space Telescopes (MAST). STScI is operated by the 
Association of Universities for Research in Astronomy, Inc., under 
NASA contract NAS5-26555. Support for MAST for non-HST data is provided 
by the NASA Office of Space Science via grant NNX09AF08G and by other 
grants and contracts.


Facilities: \facility{Kepler}



\bibliography{apj-jour.bib,ms}

\begin{thebibliography}{34}
\expandafter\ifx\csname natexlab\endcsname\relax\def\natexlab#1{#1}\fi

\bibitem[{{Armstrong} {et~al.}(2012){Armstrong}, {Pollacco}, {Watson}, {Faedi},
  {G{\'o}mez Maqueo Chew}, {Cegla}, {McDaid}, {Burton}, {McCormac}, \&
  {Skillen}}]{2012A+A...545L...4A}
{Armstrong}, D., {Pollacco}, D., {Watson}, C.~A., {et~al.} 2012, \aap, 545, L4

\bibitem[{{Armstrong} {et~al.}(2015){Armstrong}, {Kirk}, {Lam}, {McCormac},
  {Walker}, {Brown}, {Osborn}, {Pollacco}, \& {Spake}}]{2015A+A...579A..19A}
{Armstrong}, D.~J., {Kirk}, J., {Lam}, K.~W.~F., {et~al.} 2015, \aap, 579, A19

\bibitem[{{Borkovits} {et~al.}(2015){Borkovits}, {Rappaport}, {Hajdu}, \&
  {Sztakovics}}]{2015MNRAS.448..946B}
{Borkovits}, T., {Rappaport}, S., {Hajdu}, T., \& {Sztakovics}, J. 2015,
  \mnras, 448, 946

\bibitem[{{Borkovits} {et~al.}(2013){Borkovits}, {Derekas}, {Kiss},
  {Kir{\'a}ly}, {Forg{\'a}cs-Dajka}, {B{\'{\i}}r{\'o}}, {Bedding}, {Bryson},
  {Huber}, \& {Szab{\'o}}}]{2013MNRAS.428.1656B}
{Borkovits}, T., {Derekas}, A., {Kiss}, L.~L., {et~al.} 2013, \mnras, 428, 1656

\bibitem[{{Carter} {et~al.}(2011){Carter}, {Fabrycky}, {Ragozzine}, {Holman},
  {Quinn}, {Latham}, {Buchhave}, {Van Cleve}, {Cochran}, {Cote}, {Endl},
  {Ford}, {Haas}, {Jenkins}, {Koch}, {Li}, {Lissauer}, {MacQueen}, {Middour},
  {Orosz}, {Rowe}, {Steffen}, \& {Welsh}}]{2011Sci...331..562C}
{Carter}, J.~A., {Fabrycky}, D.~C., {Ragozzine}, D., {et~al.} 2011, Science,
  331, 562

\bibitem[{{Conroy} {et~al.}(2014){Conroy}, {Pr{\v s}a}, {Stassun}, {Orosz},
  {Fabrycky}, \& {Welsh}}]{2014AJ....147...45C}
{Conroy}, K.~E., {Pr{\v s}a}, A., {Stassun}, K.~G., {et~al.} 2014, \aj, 147, 45

\bibitem[{{Derekas} {et~al.}(2011){Derekas}, {Kiss}, {Borkovits}, {Huber},
  {Lehmann}, {Southworth}, {Bedding}, {Balam}, {Hartmann}, {Hrudkova},
  {Ireland}, {Kov{\'a}cs}, {Mez{\H o}}, {Mo{\'o}r}, {Niemczura}, {Sarty},
  {Szab{\'o}}, {Szab{\'o}}, {Telting}, {Tkachenko}, {Uytterhoeven}, {Benk{\H
  o}}, {Bryson}, {Maestro}, {Simon}, {Stello}, {Schaefer}, {Aerts}, {ten
  Brummelaar}, {De Cat}, {McAlister}, {Maceroni}, {M{\'e}rand}, {Still},
  {Sturmann}, {Sturmann}, {Turner}, {Tuthill}, {Christensen-Dalsgaard},
  {Gilliland}, {Kjeldsen}, {Quintana}, {Tenenbaum}, \&
  {Twicken}}]{2011Sci...332..216D}
{Derekas}, A., {Kiss}, L.~L., {Borkovits}, T., {et~al.} 2011, Science, 332, 216

\bibitem[{{Duch{\^e}ne} \& {Kraus}(2013)}]{2013ARA+A..51..269D}
{Duch{\^e}ne}, G., \& {Kraus}, A. 2013, \araa, 51, 269

\bibitem[{{Frieboes-Conde} \& {Herczeg}(1973)}]{1973A&AS...12....1F}
{Frieboes-Conde}, H., \& {Herczeg}, T. 1973, \aaps, 12, 1

\bibitem[{{Gies} {et~al.}(2012){Gies}, {Williams}, {Matson}, {Guo}, {Thomas},
  {Orosz}, \& {Peters}}]{2012AJ....143..137G}
{Gies}, D.~R., {Williams}, S.~J., {Matson}, R.~A., {et~al.} 2012, \aj, 143, 137

\bibitem[{{Hambleton} {et~al.}(2013){Hambleton}, {Kurtz}, {Pr{\v s}a}, {Guzik},
  {Pavlovski}, {Bloemen}, {Southworth}, {Conroy}, {Littlefair}, \&
  {Fuller}}]{2013MNRAS.434..925H}
{Hambleton}, K.~M., {Kurtz}, D.~W., {Pr{\v s}a}, A., {et~al.} 2013, \mnras,
  434, 925

\bibitem[{{Holczer} {et~al.}(2015){Holczer}, {Shporer}, {Mazeh}, {Fabrycky},
  {Nachmani}, {McQuillan}, {Sanchis-Ojeda}, {Orosz}, {Welsh}, {Ford}, \&
  {Jontof-Hutter}}]{2015ApJ...807..170H}
{Holczer}, T., {Shporer}, A., {Mazeh}, T., {et~al.} 2015, \apj, 807, 170

\bibitem[{{Irwin}(1952)}]{1952ApJ...116..211I}
{Irwin}, J.~B. 1952, \apj, 116, 211

\bibitem[{{Irwin}(1959)}]{1959AJ.....64..149I}
---. 1959, \aj, 64, 149

\bibitem[{{Kalimeris} {et~al.}(2002){Kalimeris}, {Rovithis-Livaniou}, \&
  {Rovithis}}]{2002A+A...387..969K}
{Kalimeris}, A., {Rovithis-Livaniou}, H., \& {Rovithis}, P. 2002, \aap, 387,
  969

\bibitem[{{LaCourse} {et~al.}(2015){LaCourse}, {Jek}, {Jacobs}, {Winarski},
  {Boyajian}, {Rappaport}, {Sanchis-Ojeda}, {Conroy}, {Nelson}, {Barclay},
  {Fischer}, {Schmitt}, {Wang}, {Stassun}, {Pepper}, {Coughlin}, {Shporer}, \&
  {Pr{\v s}a}}]{2015MNRAS.452.3561L}
{LaCourse}, D.~M., {Jek}, K.~J., {Jacobs}, T.~L., {et~al.} 2015, \mnras, 452,
  3561

\bibitem[{{Lee} {et~al.}(2015){Lee}, {Hong}, \& {Hinse}}]{2015AJ....149...93L}
{Lee}, J.~W., {Hong}, K., \& {Hinse}, T.~C. 2015, \aj, 149, 93

\bibitem[{{Lohr} {et~al.}(2015){Lohr}, {Norton}, {Payne}, {West}, \&
  {Wheatley}}]{2015A+A...578A.136L}
{Lohr}, M.~E., {Norton}, A.~J., {Payne}, S.~G., {West}, R.~G., \& {Wheatley},
  P.~J. 2015, \aap, 578, A136

\bibitem[{{Markwardt}(2009)}]{2009ASPC..411..251M}
{Markwardt}, C.~B. 2009, in Astronomical Society of the Pacific Conference
  Series, Vol. 411, Astronomical Data Analysis Software and Systems XVIII, ed.
  D.~A. {Bohlender}, D.~{Durand}, \& P.~{Dowler}, 251

\bibitem[{{Martin} {et~al.}(2015){Martin}, {Mazeh}, \&
  {Fabrycky}}]{2015MNRAS.453.3554M}
{Martin}, D.~V., {Mazeh}, T., \& {Fabrycky}, D.~C. 2015, \mnras, 453, 3554

\bibitem[{{Mayer}(1990)}]{1990BAICz..41..231M}
{Mayer}, P. 1990, Bulletin of the Astronomical Institutes of Czechoslovakia,
  41, 231

\bibitem[{{Mu{\~n}oz} \& {Lai}(2015)}]{2015PNAS..112.9264M}
{Mu{\~n}oz}, D.~J., \& {Lai}, D. 2015, Proceedings of the National Academy of
  Science, 112, 9264

\bibitem[{{Naoz} \& {Fabrycky}(2014)}]{2014ApJ...793..137N}
{Naoz}, S., \& {Fabrycky}, D.~C. 2014, \apj, 793, 137

\bibitem[{{Orosz}(2015)}]{2015arXiv150307295O}
{Orosz}, J.~A. 2015, ArXiv e-prints, 1503.07295

\bibitem[{{Orosz} {et~al.}(2012){Orosz}, {Welsh}, {Carter}, {Fabrycky},
  {Cochran}, {Endl}, {Ford}, {Haghighipour}, {MacQueen}, {Mazeh},
  {Sanchis-Ojeda}, {Short}, {Torres}, {Agol}, {Buchhave}, {Doyle}, {Isaacson},
  {Lissauer}, {Marcy}, {Shporer}, {Windmiller}, {Barclay}, {Boss}, {Clarke},
  {Fortney}, {Geary}, {Holman}, {Huber}, {Jenkins}, {Kinemuchi}, {Kruse},
  {Ragozzine}, {Sasselov}, {Still}, {Tenenbaum}, {Uddin}, {Winn}, {Koch}, \&
  {Borucki}}]{2012Sci...337.1511O}
{Orosz}, J.~A., {Welsh}, W.~F., {Carter}, J.~A., {et~al.} 2012, Science, 337,
  1511

\bibitem[{{Otero}(2003)}]{2003IBVS.5482....1O}
{Otero}, S.~A. 2003, Information Bulletin on Variable Stars, 5482, 1

\bibitem[{{Pribulla} {et~al.}(2005){Pribulla}, {Chochol}, {Tremko}, \&
  {Kreiner}}]{2005ASPC..335..103P}
{Pribulla}, T., {Chochol}, D., {Tremko}, J., \& {Kreiner}, J.~M. 2005, in
  Astronomical Society of the Pacific Conference Series, Vol. 335, The
  Light-Time Effect in Astrophysics: Causes and cures of the O-C diagram, ed.
  {C.~Sterken}, 103

\bibitem[{{Pr{\v s}a} {et~al.}(2011){Pr{\v s}a}, {Batalha}, {Slawson}, {Doyle},
  {Welsh}, {Orosz}, {Seager}, {Rucker}, {Mjaseth}, {Engle}, {Conroy},
  {Jenkins}, {Caldwell}, {Koch}, \& {Borucki}}]{2011AJ....141...83P}
{Pr{\v s}a}, A., {Batalha}, N., {Slawson}, R.~W., {et~al.} 2011, \aj, 141, 83

\bibitem[{{Rappaport} {et~al.}(2013){Rappaport}, {Deck}, {Levine}, {Borkovits},
  {Carter}, {El Mellah}, {Sanchis-Ojeda}, \& {Kalomeni}}]{2013ApJ...768...33R}
{Rappaport}, S., {Deck}, K., {Levine}, A., {et~al.} 2013, \apj, 768, 33

\bibitem[{{Slawson} {et~al.}(2011){Slawson}, {Pr{\v s}a}, {Welsh}, {Orosz},
  {Rucker}, {Batalha}, {Doyle}, {Engle}, {Conroy}, {Coughlin}, {Gregg},
  {Fetherolf}, {Short}, {Windmiller}, {Fabrycky}, {Howell}, {Jenkins}, {Uddin},
  {Mullally}, {Seader}, {Thompson}, {Sanderfer}, {Borucki}, \&
  {Koch}}]{2011AJ....142..160S}
{Slawson}, R.~W., {Pr{\v s}a}, A., {Welsh}, W.~F., {et~al.} 2011, \aj, 142, 160

\bibitem[{{Tokovinin} {et~al.}(2006){Tokovinin}, {Thomas}, {Sterzik}, \&
  {Udry}}]{2006A+A...450..681T}
{Tokovinin}, A., {Thomas}, S., {Sterzik}, M., \& {Udry}, S. 2006, \aap, 450,
  681

\bibitem[{{Tran} {et~al.}(2013){Tran}, {Levine}, {Rappaport}, {Borkovits},
  {Csizmadia}, \& {Kalomeni}}]{2013ApJ...774...81T}
{Tran}, K., {Levine}, A., {Rappaport}, S., {et~al.} 2013, \apj, 774, 81

\bibitem[{{Welsh} {et~al.}(2014){Welsh}, {Orosz}, {Carter}, \&
  {Fabrycky}}]{2014IAUS..293..125W}
{Welsh}, W.~F., {Orosz}, J.~A., {Carter}, J.~A., \& {Fabrycky}, D.~C. 2014, in
  IAU Symposium, Vol. 293, Formation, Detection, and Characterization of
  Extrasolar Habitable Planets, ed. N.~{Haghighipour}, 125

\bibitem[{{Zasche} {et~al.}(2015){Zasche}, {Wolf}, {Ku{\v c}{\'a}kov{\'a}},
  {Vra{\v s}til}, {Jury{\v s}ek}, {Ma{\v s}ek}, \&
  {Jel{\'{\i}}nek}}]{2015AJ....149..197Z}
{Zasche}, P., {Wolf}, M., {Ku{\v c}{\'a}kov{\'a}}, H., {et~al.} 2015, \aj, 149,
  197

\end{thebibliography}


\begin{deluxetable}{lrrllccc}
\tabletypesize{\scriptsize}
\tablewidth{0pt}
\tablenum{1}
\tablecaption{Eclipsing Binary Properties\label{tab1}}
\tablehead{
\colhead{}  &
\colhead{$I$}      &
\colhead{$E$}  &
\colhead{$T$}  &
\colhead{$P$}  &
\colhead{$\dot{P}/P$}  &
\colhead{}  &
\colhead{}  \\
\colhead{KIC}  &
\colhead{(sec)}  &
\colhead{(sec)}  &
\colhead{(BJD-2,400,000)}  &
\colhead{(day)}  &
\colhead{($10^{-6}$ year$^{-1}$)}  &
\colhead{$\sigma (D_E)$}  &
\colhead{References\tablenotemark{a}} 
}
\startdata
02305372P &  25.3 & 178.1 & 55693.58496 (2)  & 1.40469191 (3)   & 9.63 (4)         &   0.018 & C14, Z15 \\
02305372S &  35.0 & 222.3 & 55701.31436 (5)  & 1.4046910 (2)    & 11.4 (2)         &   0.038 & \nodata \\
02708156P &  21.2 &  51.8 & 55690.03950 (2)  & 1.89126793 (6)   & $-$1.91 (6)\phs  &   0.008 & C14 \\
02708156S &  33.0 & 128.7 & 55692.87715 (5)  & 1.8912707 (2)    & $-$0.07 (4)\phs  &   0.033 & \nodata \\
03241619P &  26.6 &  58.7 & 55694.50030 (2)  & 1.70334737 (8)   & 2.18 (9)         &   0.015 & \nodata \\
03241619S &  25.5 &  50.5 & 55691.94541 (3)  & 1.70334676 (5)   & 2.09 (6)         &   0.041 & \nodata \\
03327980P &  29.4 &   4.5 & 55699.074202 (3) & 4.23102186 (3)   & $-$0.021 (2)\phs &   0.002 & \nodata \\
03327980S &  38.8 &   4.4 & 55688.494827 (3) & 4.23102201 (3)   & $-$0.013 (2)\phs &   0.002 & \nodata \\
03440230P &  24.4 & 112.8 & 55690.39643 (2)  & 2.88110022 (9)   & $-$5.87 (6)\phs  &   0.004 & C14, Z15 \\
03440230S &  33.8 & 135.8 & 55700.48186 (7)  & 2.8811004 (3)    & $-$4.8 (2)\phs   &   0.018 & \nodata \\
04544587P &  24.9 &  13.9 & 55689.673044 (7) & 2.18909811 (3)   & 0.45 (2)         &   0.006 & H13, ECC \\
04544587S &  27.3 &  13.7 & 55688.910266 (9) & 2.18913048 (4)   & $-$0.17 (1)\phs  &   0.006 & \nodata \\
04574310P &  18.4 &  20.1 & 55644.346499 (6) & 1.30622004 (1)   & 0.80 (2)         &   0.003 & C14 \\
04574310S &  12.0 &  22.6 & 55646.30602 (1)  & 1.30622004 (2)   & 0.45 (2)         &   0.007 & \nodata \\
04660997P &  15.4 &  37.5 & 55654.43913 (1)  & 0.56256064 (2)   & $-$0.50 (4)\phs  &   0.024 & C14 \\
04660997S &  25.7 &  53.2 & 55647.40777 (1)  & 0.56256075 (3)   & $-$1.37 (5)\phs  &   0.075 & \nodata \\
04665989P &  22.4 &   1.8 & 55646.917282 (1) & 2.248067587 (7)  & 0.0125 (7)       &   0.001 & C14 \\
04665989S &  31.1 &   3.3 & 55648.041437 (2) & 2.24806759 (1)   & 0.0064 (9)       &   0.002 & \nodata \\
04678873P &  22.7 &  42.0 & 55364.39581 (3)  & 1.8788767 (2)    & 0.7 (1)          &   0.015 & \nodata \\
04678873S & 130.1 & 357.6 & 55365.3373 (3)   & 1.878880 (2)     & $-$16. (4)\phs   &   0.171 & \nodata \\
04848423P &  58.5 & 213.1 & 55928.7090 (1)   & 3.0036456 (8)    & 22.2 (8)         &   0.004 & \nodata \\
04848423S &  59.0 & 215.4 & 55933.2150 (1)   & 3.0036465 (8)    & 22.0 (8)         &   0.004 & \nodata \\
04851217P &  67.9 &  38.1 & 55643.10900 (1)  & 2.47028357 (5)   & 1.71 (4)         &   0.004 & C14, ECC \\
04851217S &  49.4 &  39.5 & 55649.23501 (1)  & 2.47028209 (5)   & 1.93 (3)         &   0.003 & \nodata \\
05444392P &  16.8 &  24.1 & 55688.84830 (1)  & 1.51952885 (3)   & 0.054 (8)        &   0.018 & C14, O15 \\
05444392S &  25.2 &  47.9 & 55689.60795 (2)  & 1.51952893 (7)   & $-$0.06 (2)\phs  &   0.019 & \nodata \\
05513861P &  23.6 & 350.6 & 55625.52615 (3)  & 1.51020817 (6)   & 17.70 (7)        &   0.001 & C14, O15 \\
05513861S &  25.6 & 349.8 & 55627.79154 (3)  & 1.51020833 (6)   & 17.74 (7)        &   0.001 & \nodata \\
05621294P &  15.2 &  46.7 & 55693.432093 (6) & 0.938904885 (9)  & $-$2.40 (2)\phs  &   0.004 & C14, L15, Z15, ECC \\
05621294S &  52.8 & 105.5 & 55690.14750 (3)  & 0.93890558 (5)   & $-$3.80 (9)\phs  &   0.021 & \nodata \\
05738698P &  31.2 &   3.6 & 55692.334870 (2) & 4.80877394 (3)   & 0.0062 (7)       &   0.002 & \nodata \\
05738698S &  32.1 &   4.4 & 55680.314028 (3) & 4.80877398 (4)   & $-$0.040 (3)\phs &   0.002 & \nodata \\
06206751P &  14.7 &  44.7 & 55683.783072 (8) & 1.24534282 (3)   & $-$1.36 (4)\phs  &   0.018 & C14 \\
06206751S &  23.6 &  73.9 & 55686.89665 (3)  & 1.24534280 (7)   & $-$1.01 (9)\phs  &   0.065 & \nodata \\
07368103P &  85.4 &  28.9 & 55698.67906 (1)  & 2.18251565 (7)   & $-$0.07 (2)\phs  &   0.011 & C14 \\
07368103S & 532.4 & 543.0 & 55697.5891 (3)   & 2.1825145 (9)    & $-$1.3 (8)\phs   &   0.075 & \nodata \\
08196180P &  28.0 &   6.8 & 55695.751468 (5) & 3.67166106 (4)   & $-$0.024 (4)\phs &   0.004 & ECC \\
08196180S &  30.1 &  27.2 & 55686.21595 (2)  & 3.6716582 (1)    & $-$0.07 (2)\phs  &   0.017 & \nodata \\
08262223P &  16.4 &   7.6 & 55692.217903 (3) & 1.61301476 (1)   & 0.018 (3)        &   0.003 & C14 \\
08262223S &  21.8 &  18.1 & 55689.798699 (7) & 1.61301470 (3)   & 0.055 (8)        &   0.005 & \nodata \\
08552540P &  16.8 &  33.0 & 55692.15027 (1)  & 1.06193424 (3)   & $-$0.11 (1)\phs  &   0.014 & C14 \\
08552540S &  18.1 &  28.2 & 55690.55782 (1)  & 1.06193427 (2)   & 0.064 (8)        &   0.033 & \nodata \\
08553788P &  14.8 & 147.2 & 55690.62207 (2)  & 1.60616355 (5)   & $-$7.56 (6)\phs  &   0.004 & C14, O15, Z15 \\
08553788S &  26.9 & 144.9 & 55691.42530 (3)  & 1.60616375 (6)   & $-$7.15 (8)\phs  &   0.019 & \nodata \\
08823397P &  19.3 &   2.2 & 55686.100164 (1) & 1.506503703 (4)  & $-$0.0052 (5)\phs &  0.001 & C14 \\
08823397S &  15.8 &   4.3 & 55694.386282 (2) & 1.506503696 (6)  & $-$0.0094 (6)\phs &  0.006 & \nodata \\
09159301P &  26.5 &  18.1 & 55693.13726 (1)  & 3.04477179 (7)   & 0.22 (2)         &   0.003 & C14 \\
09159301S & 174.7 &  92.5 & 55685.52674 (5)  & 3.0447725 (4)    & 0.0 (1)          &   0.020 & \nodata \\
09357275P &  14.7 &   3.1 & 55643.386640 (1) & 1.588298217 (5)  & 0.0239 (8)       &   0.002 & C14 \\
09357275S &  16.7 &  14.5 & 55644.181081 (9) & 1.58829821 (3)   & $-$0.060 (7)\phs &   0.013 & \nodata \\
09402652P &  11.0 & 192.0 & 55689.37059 (1)  & 1.07310425 (3)   & $-$10.15 (4)\phs &   0.002 & C14, Z15 \\
09402652S &  11.7 & 192.3 & 55692.05303 (1)  & 1.07310419 (3)   & $-$10.12 (4)\phs &   0.002 & \nodata \\
09592855P &  16.7 &  19.2 & 55691.664073 (7) & 1.21932487 (2)   & 0.027 (4)        &   0.016 & C14 \\
09592855S &  19.3 &  22.4 & 55692.273798 (7) & 1.21932479 (2)   & 0.039 (6)        &   0.014 & \nodata \\
09602595P &  19.3 &  75.7 & 55642.26003 (4)  & 3.5565108 (2)    & $-$3.4 (1)\phs   &   0.004 & C14, ECC \\
09602595S &  18.2 & 118.8 & 55633.3721 (1)   & 3.5565149 (5)    & $-$5.1 (3)\phs   &   0.029 & \nodata \\
09851944P &  22.5 &  17.0 & 55646.150110 (9) & 2.16390177 (5)   & 0.030 (6)        &   0.002 & C14 \\
09851944S &  21.7 &  16.2 & 55634.248526 (8) & 2.16390178 (4)   & $-$0.06 (1)\phs  &   0.003 & \nodata \\
09899416P &  15.9 &   6.8 & 55689.965570 (2) & 1.332564486 (7)  & $-$0.048 (2)\phs &   0.005 & C14, ECC \\
09899416S &   9.8 &  25.8 & 55691.96413 (1)  & 1.33256397 (2)   & 0.99 (3)         &   0.006 & \nodata \\
10156064P &  33.0 &   6.6 & 55648.858147 (5) & 4.85593640 (6)   & 0.020 (2)        &   0.003 & \nodata \\
10156064S &  21.5 &   5.8 & 55660.998491 (6) & 4.85593649 (7)   & 0.025 (4)        &   0.003 & \nodata \\
10191056P &  15.6 &   4.9 & 55688.135138 (3) & 2.42749476 (2)   & $-$0.0081 (9)\phs &  0.003 & C14, ECC \\
10191056S &  14.6 &   4.6 & 55684.495255 (2) & 2.42749492 (1)   & $-$0.035 (1)\phs &   0.004 & \nodata \\
10206340P &  24.4 & 146.2 & 55710.0929 (1)   & 4.564401 (1)     & 3.1 (4)          &   0.011 & C14 \\
10206340S &  33.3 & 176.1 & 55684.9895 (1)   & 4.564409 (1)     & 1.0 (4)          &   0.026 & \nodata \\
10486425P &  34.9 &  93.4 & 55681.8090 (1)   & 5.2748191 (8)    & 1.8 (3)          &   0.053 & \nodata \\
10486425S &  94.3 & 372.7 & 55705.5464 (2)   & 5.274817 (3)     & 0.1 (2)          &   0.157 & \nodata \\
10581918P &  20.9 &  10.3 & 55744.790056 (4) & 1.80186368 (2)   & 0.146 (7)        &   0.010 & C14, Z15 \\
10581918S &  33.1 &  81.5 & 55754.70106 (4)  & 1.8018637 (2)    & $-$0.4 (1)\phs   &   0.039 & \nodata \\
10619109P &  17.9 &  57.1 & 55642.31502 (2)  & 2.04516550 (8)   & 2.26 (6)         &   0.009 & C14, ECC \\
10619109S &  49.0 & 187.5 & 55647.4299 (1)   & 2.0451674 (5)    & $-$1.2 (4)\phs   &   0.050 & \nodata \\
10661783P &  30.9 &   9.5 & 55692.541890 (4) & 1.23136327 (1)   & 0.048 (3)        &   0.004 & C14 \\
10661783S &  54.2 &  20.9 & 55691.926166 (7) & 1.23136329 (2)   & $-$0.038 (6)\phs &   0.006 & \nodata \\
10686876P &  25.3 & 127.1 & 55632.12313 (1)  & 2.61841633 (5)   & $-$6.50 (3)\phs  &   0.009 & Z15 \\
10686876S &  27.7 & 112.5 & 55646.52478 (2)  & 2.61841462 (8)   & $-$5.88 (6)\phs  &   0.027 & \nodata \\
10736223P &  17.3 &  38.3 & 55692.697641 (3) & 1.105094220 (6)  & 2.03 (1)         &   0.005 & C14 \\
10736223S &  28.5 &  42.4 & 55695.46087 (1)  & 1.10509441 (2)   & 2.01 (4)         &   0.013 & \nodata \\
10858720P &  11.4 &  12.3 & 55692.951773 (4) & 0.952377626 (8)  &  0.0 (1)         &   0.004 & C14 \\
10858720S &  12.4 &   6.2 & 55691.523165 (2) & 0.952377609 (4)  & $-$0.013 (1)\phs &   0.006 & \nodata \\
12071006P &  38.5 &  55.2 & 55693.45930 (4)  & 6.0960295 (3)    & $-$2.4 (1)\phs   &   0.006 & C14, ECC \\
12071006S & 154.0 & 199.3 & 55696.5059 (2)   & 6.096043 (2)     & 3.5 (7)          &   0.024 & \nodata \\
\enddata
\tablenotetext{a}{
C14 = \citet{2014AJ....147...45C};
H13 = \citet{2013MNRAS.434..925H};
L15 = \citet{2015AJ....149...93L};
O15 = \citet{2015arXiv150307295O};
Z15 = \citet{2015AJ....149..197Z};
ECC = non-zero eccentricity.
} 
\end{deluxetable}

\newpage

\begin{deluxetable}{cccccc}
\tablewidth{0pt}
\tablenum{2}
\tablecaption{Eclipse Timing Measurements\tablenotemark{a}\label{tab2}}
\tablehead{
\colhead{KIC}  &
\colhead{Eclipse}  &
\colhead{$T_E$}  &
\colhead{$\sigma (T_E)$} &
\colhead{}  &
\colhead{} \\
\colhead{Number}  &
\colhead{Type}  &
\colhead{(BJD-2,400,000)}  &
\colhead{(day)}  &
\colhead{$D_E$}  &
\colhead{$\sigma (D_E)$}  
}
\startdata
02305372 &  2 &  54965.26257 &  0.00052 &  0.993 &  0.007 \\
02305372 &  1 &  54965.96121 &  0.00031 &  1.010 &  0.005 \\
02305372 &  2 &  54966.66766 &  0.00054 &  0.993 &  0.007 \\
02305372 &  1 &  54967.36578 &  0.00030 &  1.009 &  0.005 \\
02305372 &  2 &  54968.07249 &  0.00054 &  0.996 &  0.008 \\
02305372 &  1 &  54968.77050 &  0.00029 &  1.012 &  0.005 \\
02305372 &  2 &  54969.47710 &  0.00054 &  0.992 &  0.008 \\
02305372 &  1 &  54970.17523 &  0.00029 &  1.013 &  0.005 \\
02305372 &  2 &  54970.88144 &  0.00053 &  0.996 &  0.007 \\
02305372 &  1 &  54971.57993 &  0.00029 &  1.012 &  0.005 \\
02305372 &  2 &  54972.28629 &  0.00050 &  0.994 &  0.007 \\
02305372 &  1 &  54972.98458 &  0.00029 &  1.012 &  0.005 \\
\enddata
\tablenotetext{a}{The full table is available in the electronic version.}
\end{deluxetable}

\begin{deluxetable}{clllllllcccc}
\tabletypesize{\scriptsize}
\tablewidth{0pt}
\setlength{\tabcolsep}{0.02in} 
\rotate
\tablenum{3}
\tablecaption{Preliminary Orbital Elements\label{tab3}}
\tablehead{
\colhead{KIC}  &
\colhead{$T_1$}  &
\colhead{$T_2$}  &
\colhead{$P$}  &
\colhead{$A$}  &
\colhead{ }  &
\colhead{$\omega$}  &
\colhead{$P_3$}  &
\colhead{$T_3$}  &
\colhead{$f(m_3)$}  &
\colhead{$\Delta T$}  &
\colhead{} \\
\colhead{Number}  &
\colhead{(BJD-2,450,000)}  &
\colhead{(BJD-2,450,000)}  &
\colhead{(day)}  &
\colhead{(day)}  &
\colhead{$e$}  &
\colhead{(deg)}  &
\colhead{(year)}  &
\colhead{(BJD-2,450,000)}  &
\colhead{($M_\odot$)}  &
\colhead{(day)}  &
\colhead{References\tablenotemark{a}} 
}
\startdata
02305372 &   5693.59615 (10) &   5699.92166 (5) &   1.4047174 (2) & 
 0.018 (3) &  0.59 (8) &  106 (6) &  11.338 (4) &  4718 (100) & 
 0.25 (12) & \phs 0.00186 &  Z15 \\
04574310 &   5644.346644 (2) &   5646.306053 (8) &   1.306220105 (7) & 
 0.000333 (10) &  0.670 (20) &  346 (5) &  \phn 3.717 (38) &  5937 (4) & 
 0.000032 (3) &  \nodata & new \\
04848423 &   5928.710960 (21) &   5933.217094 (27) &   3.0036281 (11) & 
 0.0047 (2) &  0.29 (4) &  210 (14) &  \phn 2.81 (10) &  6642 (19) & 
 0.076 (11) &  \nodata & new \\
05513861 &   5624.02384 (2) &   5629.30955 (9) &   1.5102097 (2) & 
 0.00826 (19) &  0.12 (2) &  \phn\phn  7.4 (12) & \phn  6.06 (12) & \phn 6239 (13) & 
 0.081 (6) & \phs 0.00070 & C14, Z15 \\
08553788 &   5690.61611 (4) &   5691.419455 (19) &   1.6061766 (8) & 
 0.0073 (4) &  0.720 (30) &  238.2 (11) &   8.6 (2) &  6428 (6) & 
 0.035 (5) & \phs 0.00083 & Z15 \\
09402652 &   5689.363683 (6) &   5692.046455 (34) &   1.07310717 (14) & 
 0.00442 (3) &  0.800 (3) &  267.4 (5)  & \phn  4.169 (21) &  6346 (2) & 
 0.0259 (6) & \phs 0.00015 &  Z15 \\
10686876 &   5632.11773 (8) &   5646.519555 (22) &   2.6184131 (2) & 
 0.0054 (5) &  0.47 (10) &  284 (2) &  \phn 6.63 (7) &  7016 (22) & 
 0.019 (5) & $-$0.00060 & Z15 \\
\enddata
\tablenotetext{a}{
C14 = \citet{2014AJ....147...45C};
Z15 = \citet{2015AJ....149..197Z}; 
new = this paper.}
\end{deluxetable}



\clearpage



\figsetstart
\figsetnum{1}
\figsettitle{Eclipse timing variations}

\figsetgrpstart
\figsetgrpnum{1.1}
\figsetgrptitle{r1}
\figsetplot{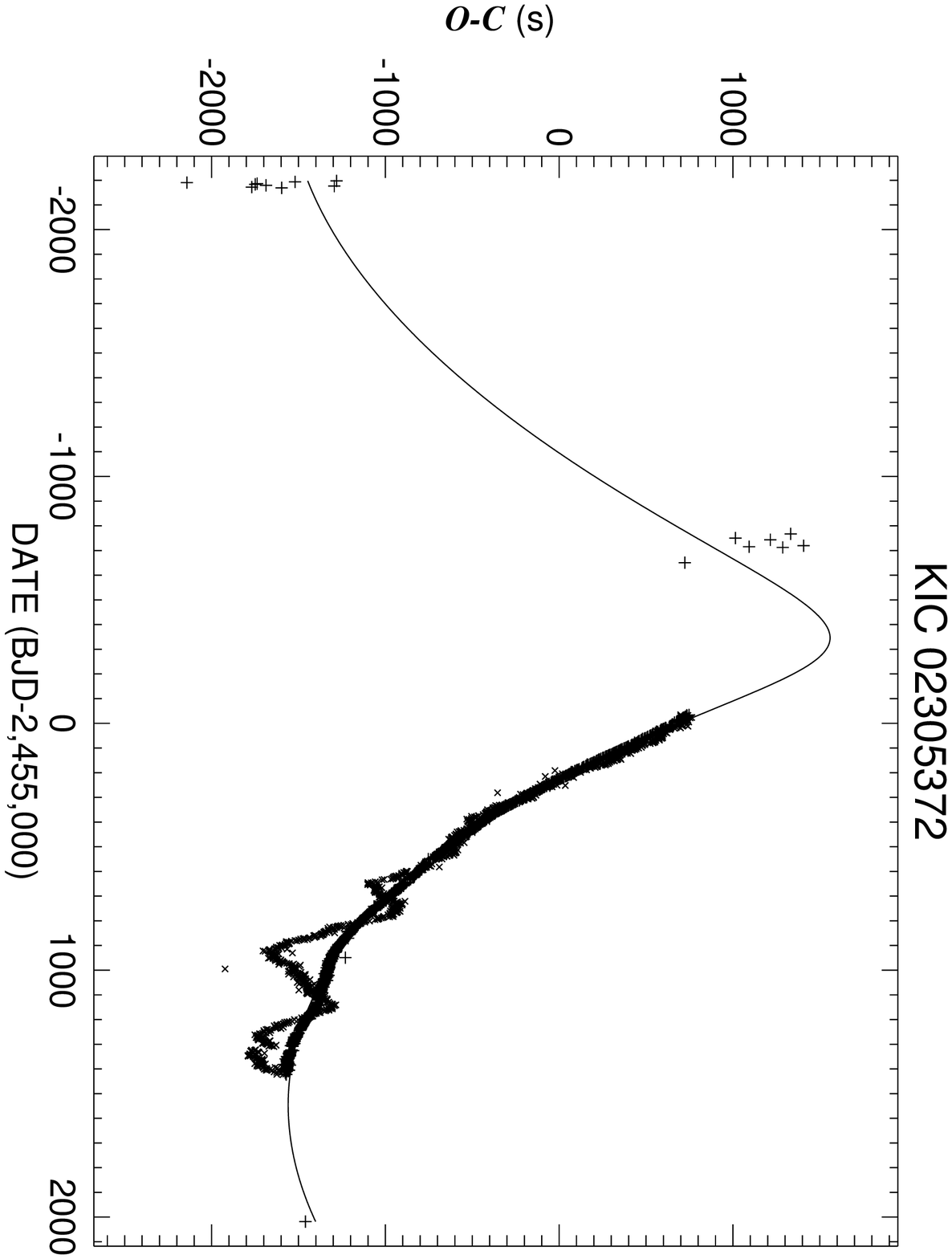}
\figsetgrpnote{The observed minus calculated eclipse times relative to
a linear ephemeris.  The primary and secondary eclipse
times are indicated by $+$ and $\times$ symbols, 
respectively. }
\figsetgrpend

\figsetgrpstart
\figsetgrpnum{1.2}
\figsetgrptitle{r2}
\figsetplot{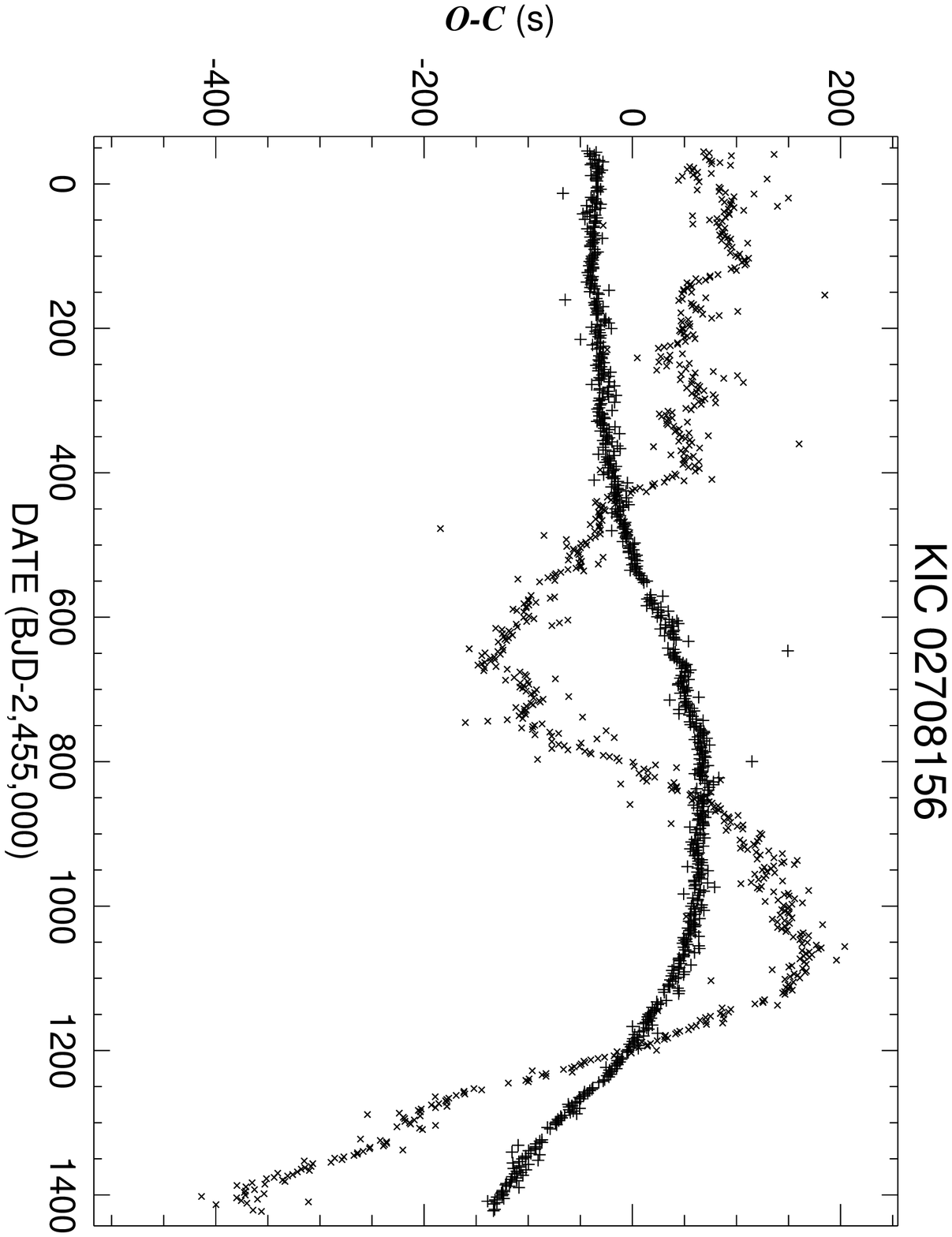}
\figsetgrpnote{The observed minus calculated eclipse times relative to
a linear ephemeris.  The primary and secondary eclipse
times are indicated by $+$ and $\times$ symbols, 
respectively. }
\figsetgrpend

\figsetgrpstart
\figsetgrpnum{1.3}
\figsetgrptitle{r3}
\figsetplot{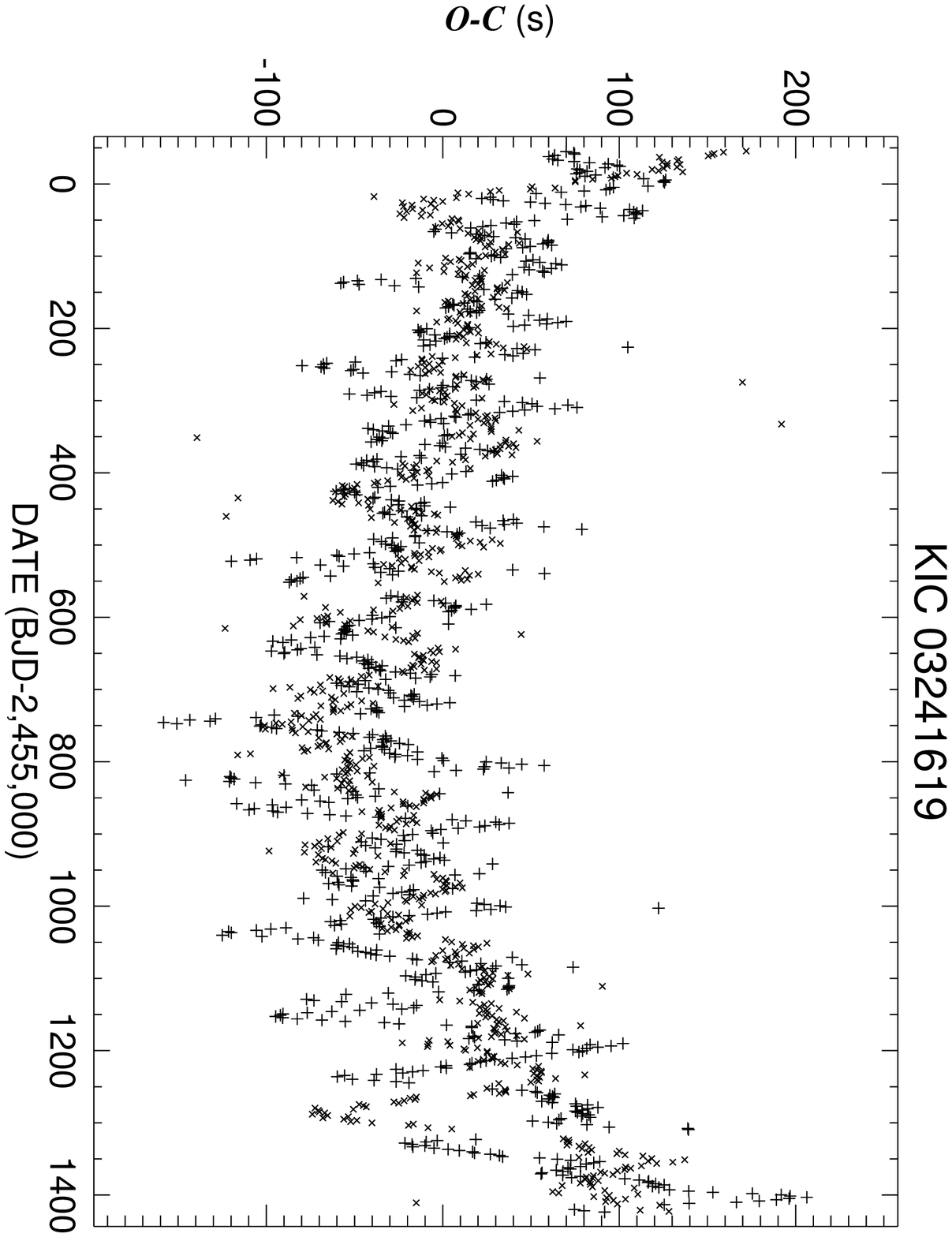}
\figsetgrpnote{The observed minus calculated eclipse times relative to
a linear ephemeris.  The primary and secondary eclipse
times are indicated by $+$ and $\times$ symbols, 
respectively. }
\figsetgrpend

\figsetgrpstart
\figsetgrpnum{1.4}
\figsetgrptitle{r4}
\figsetplot{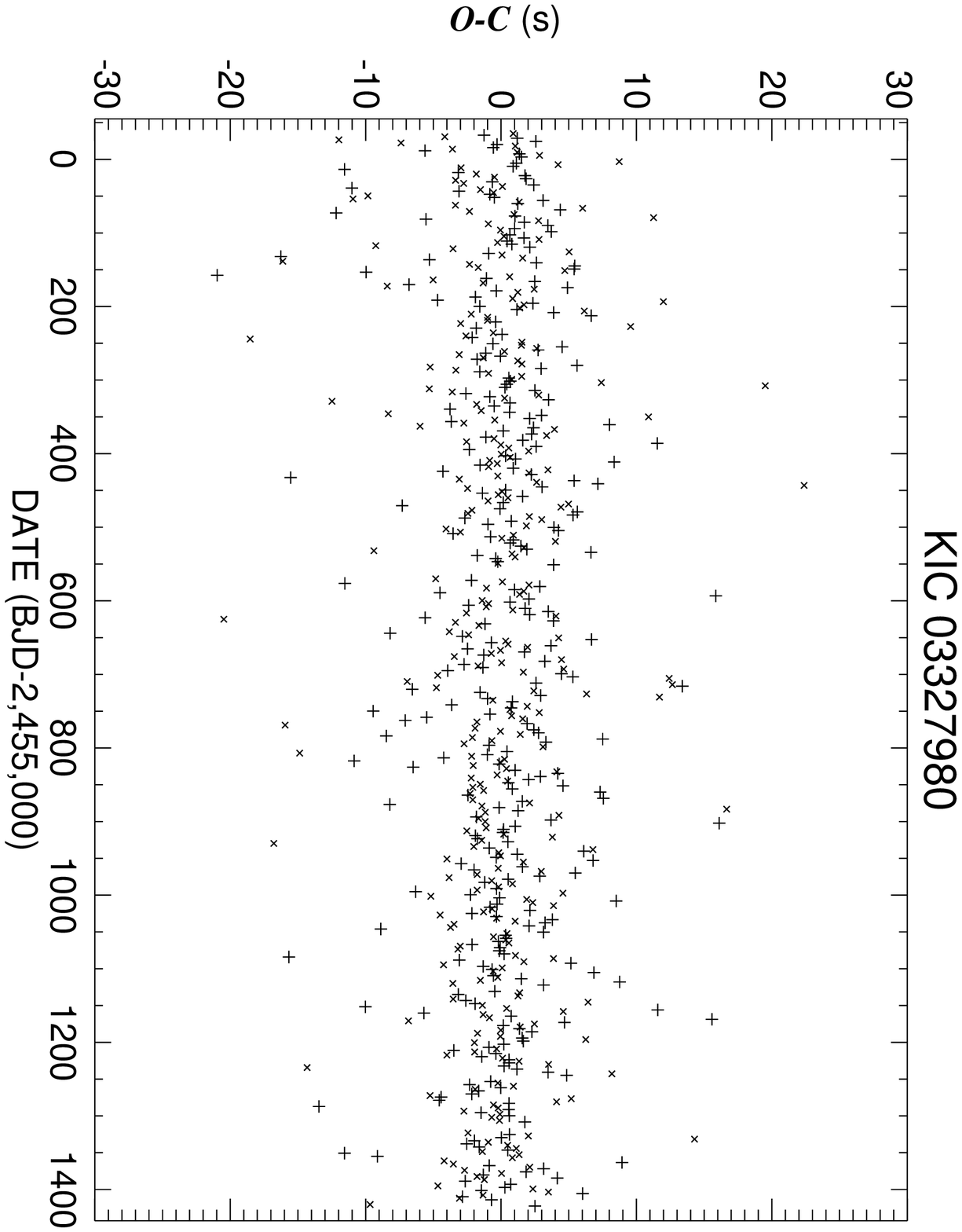}
\figsetgrpnote{The observed minus calculated eclipse times relative to
a linear ephemeris.  The primary and secondary eclipse
times are indicated by $+$ and $\times$ symbols, 
respectively. }
\figsetgrpend

\figsetgrpstart
\figsetgrpnum{1.5}
\figsetgrptitle{r5}
\figsetplot{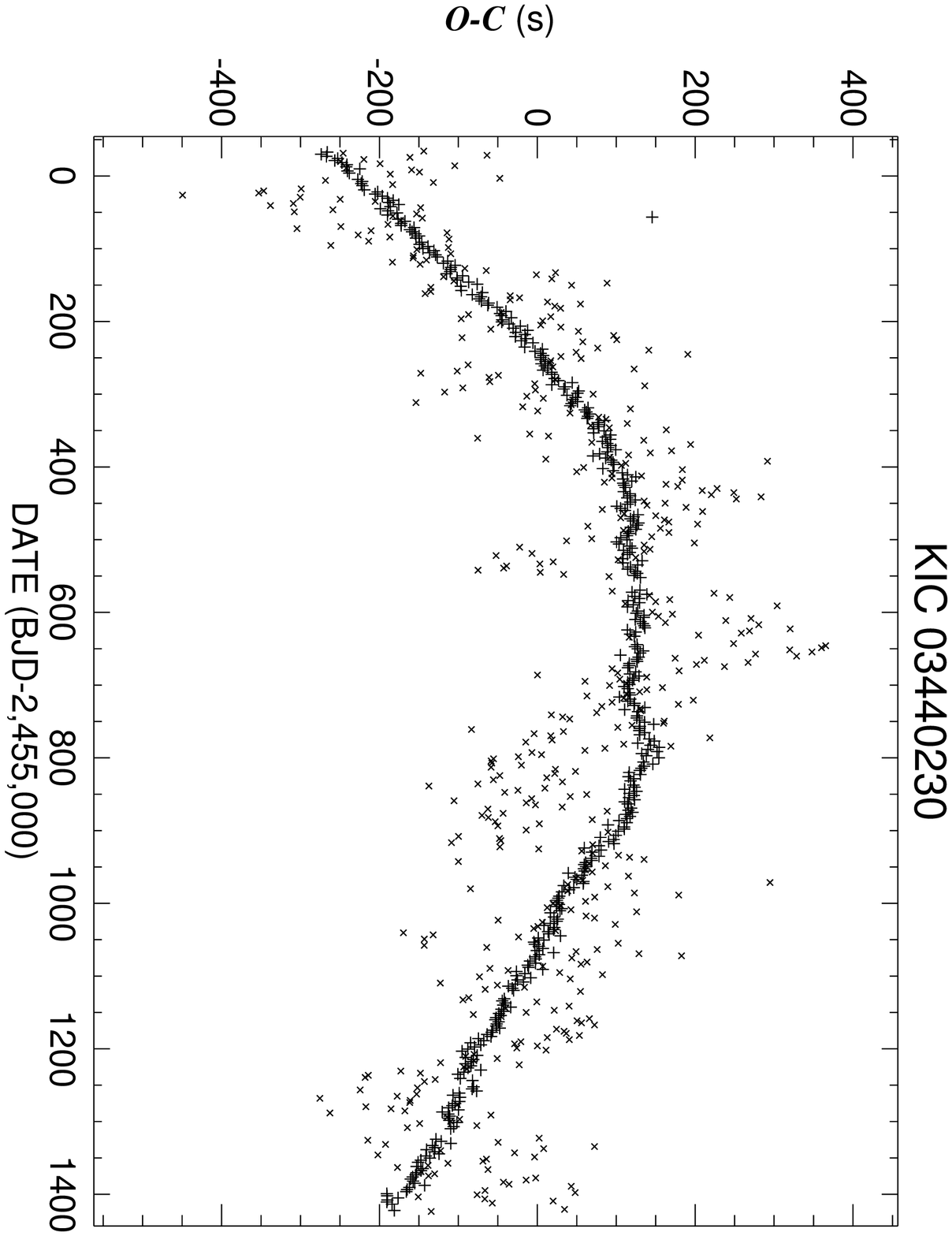}
\figsetgrpnote{The observed minus calculated eclipse times relative to
a linear ephemeris.  The primary and secondary eclipse
times are indicated by $+$ and $\times$ symbols, 
respectively. }
\figsetgrpend

\figsetgrpstart
\figsetgrpnum{1.6}
\figsetgrptitle{r6}
\figsetplot{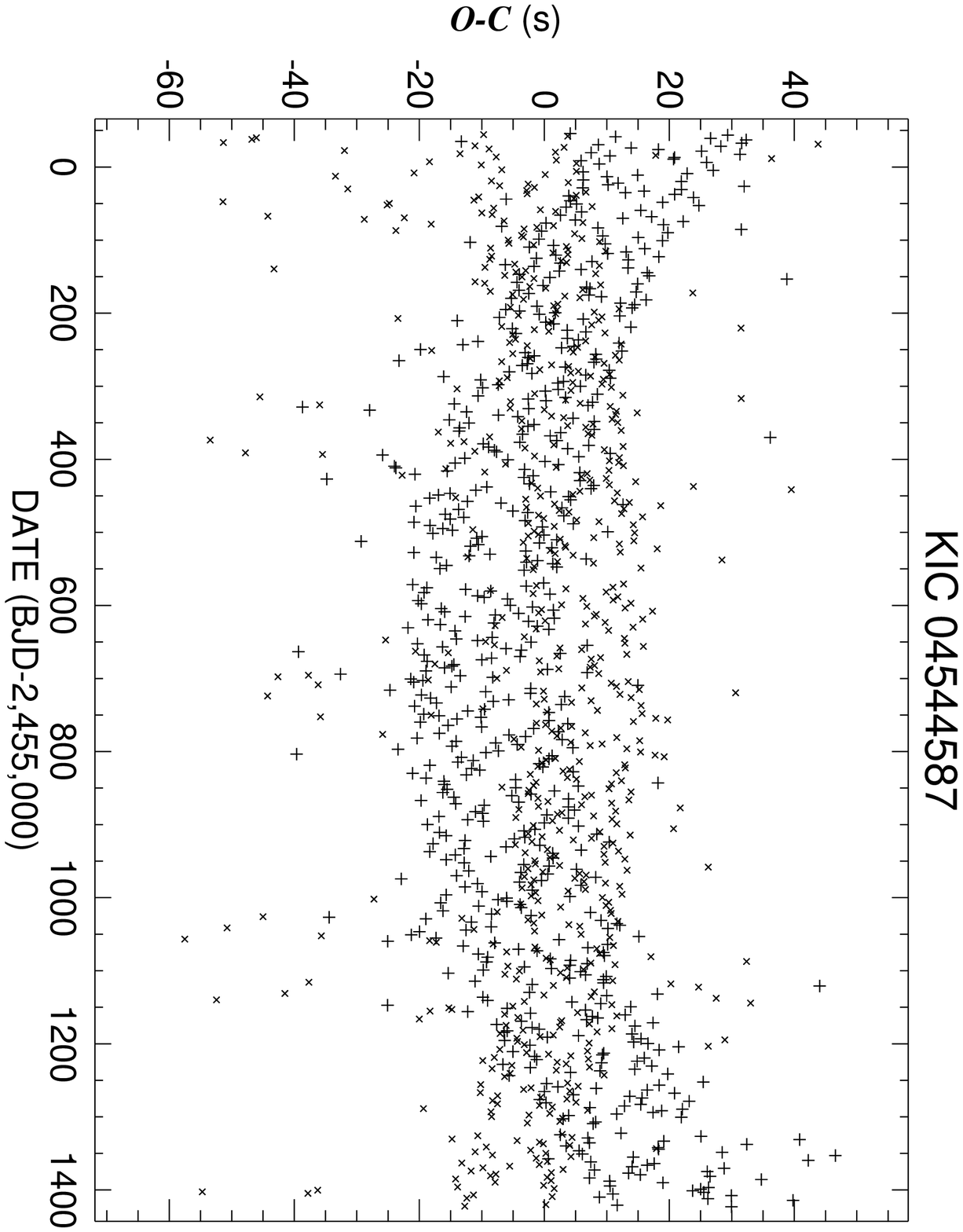}
\figsetgrpnote{The observed minus calculated eclipse times relative to
a linear ephemeris.  The primary and secondary eclipse
times are indicated by $+$ and $\times$ symbols, 
respectively. }
\figsetgrpend

\figsetgrpstart
\figsetgrpnum{1.7}
\figsetgrptitle{r7}
\figsetplot{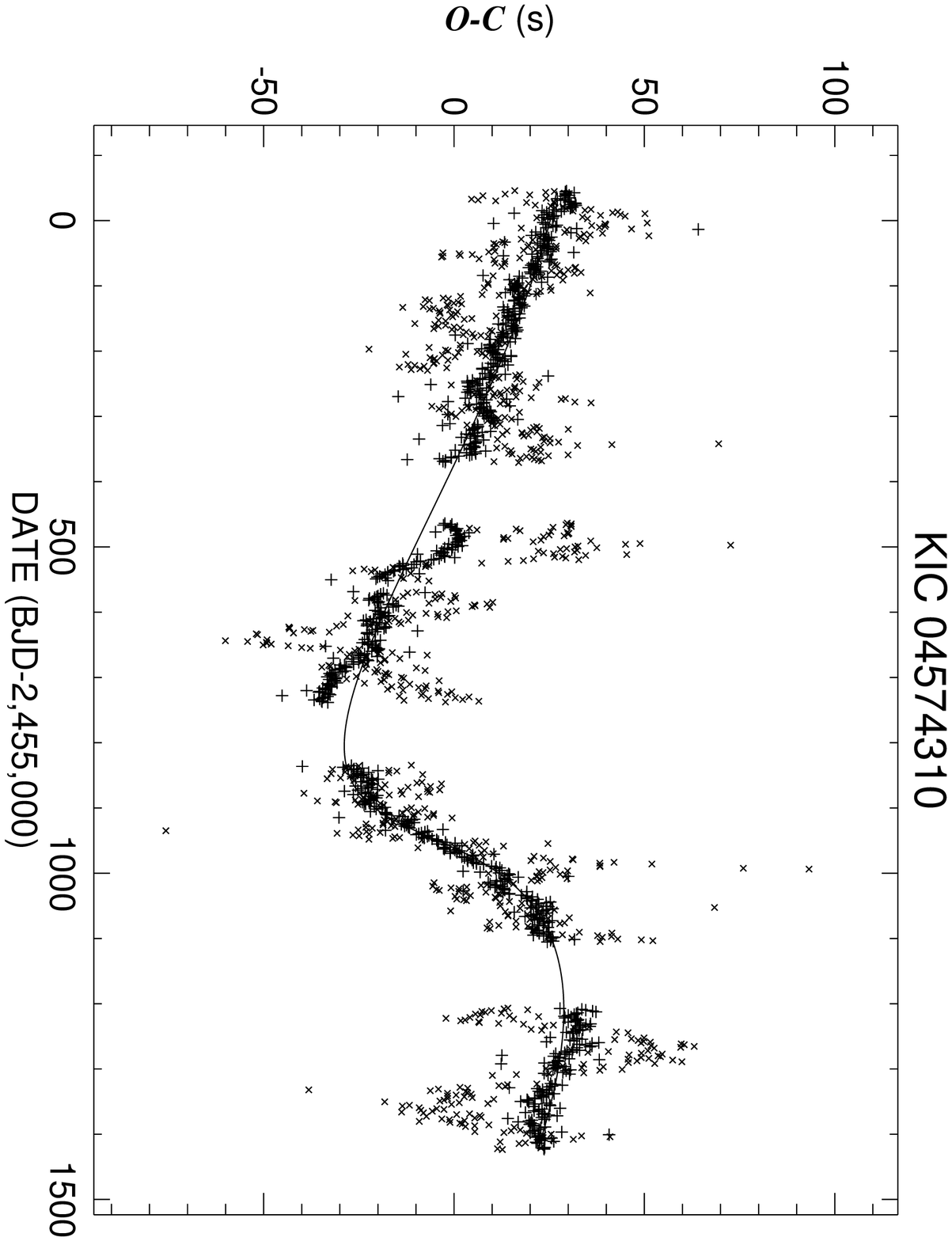}
\figsetgrpnote{The observed minus calculated eclipse times relative to
a linear ephemeris.  The primary and secondary eclipse
times are indicated by $+$ and $\times$ symbols, 
respectively. }
\figsetgrpend

\figsetgrpstart
\figsetgrpnum{1.8}
\figsetgrptitle{r8}
\figsetplot{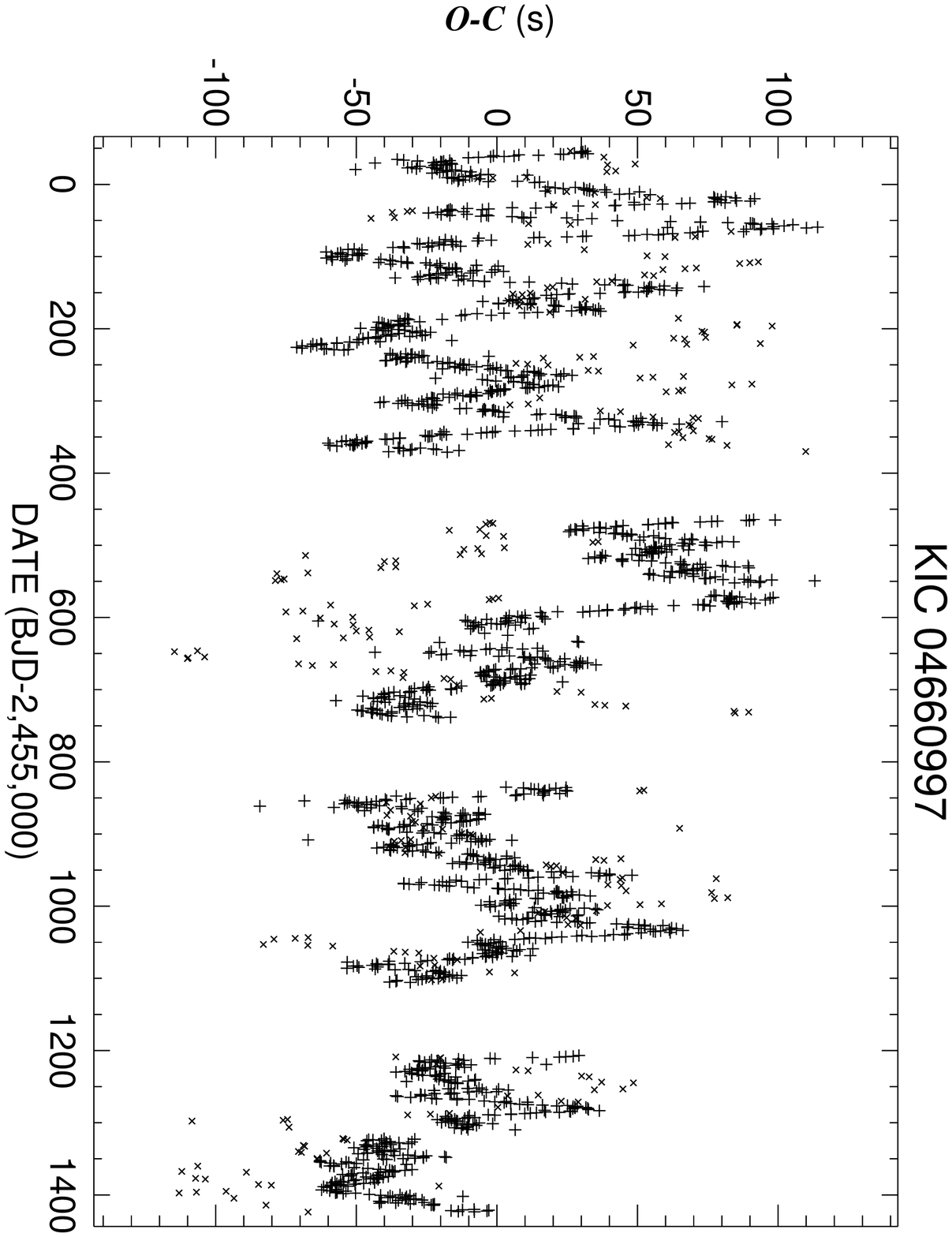}
\figsetgrpnote{The observed minus calculated eclipse times relative to
a linear ephemeris.  The primary and secondary eclipse
times are indicated by $+$ and $\times$ symbols, 
respectively. }
\figsetgrpend

\figsetgrpstart
\figsetgrpnum{1.9}
\figsetgrptitle{r9}
\figsetplot{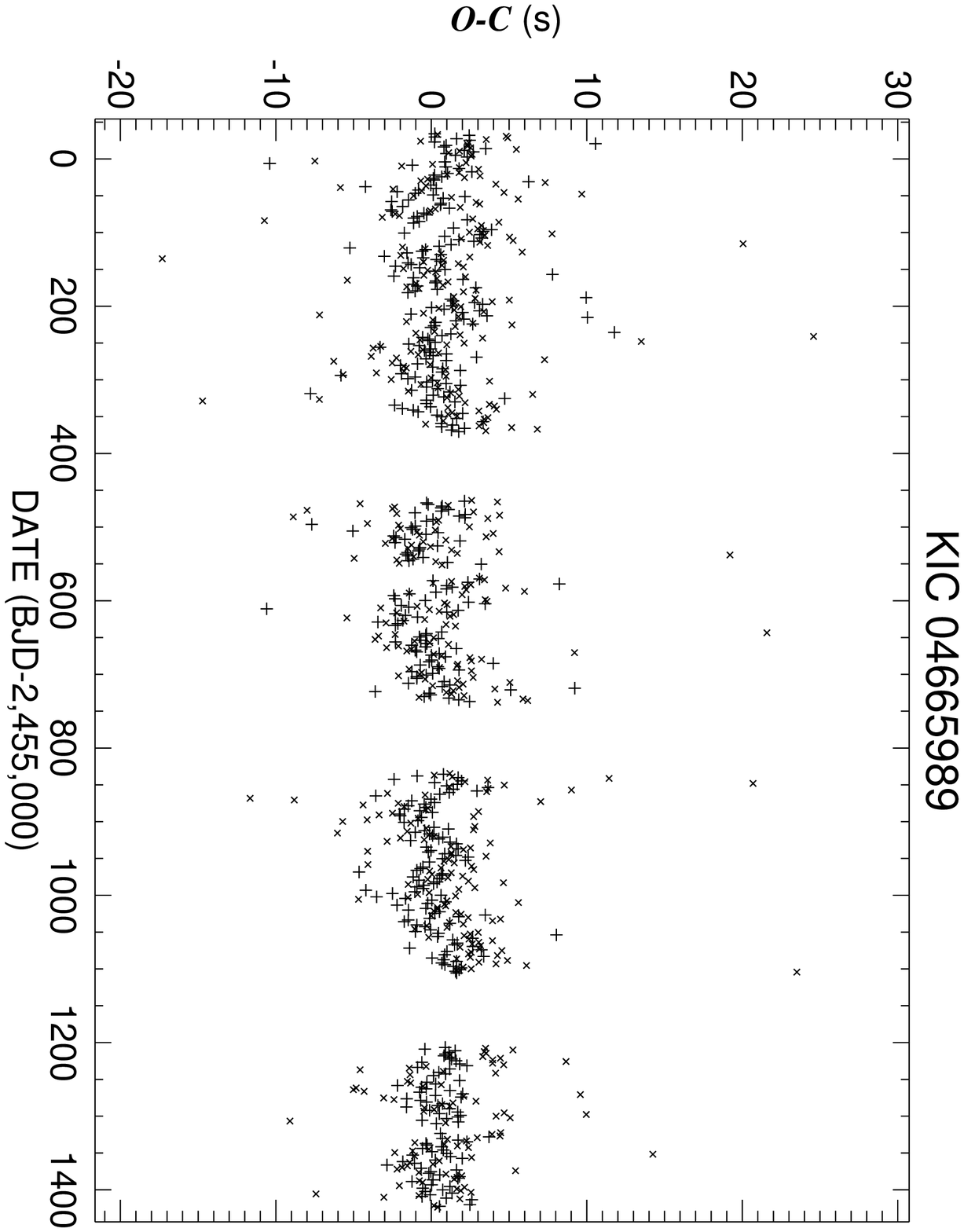}
\figsetgrpnote{The observed minus calculated eclipse times relative to
a linear ephemeris.  The primary and secondary eclipse
times are indicated by $+$ and $\times$ symbols, 
respectively. }
\figsetgrpend

\figsetgrpstart
\figsetgrpnum{1.10}
\figsetgrptitle{r10}
\figsetplot{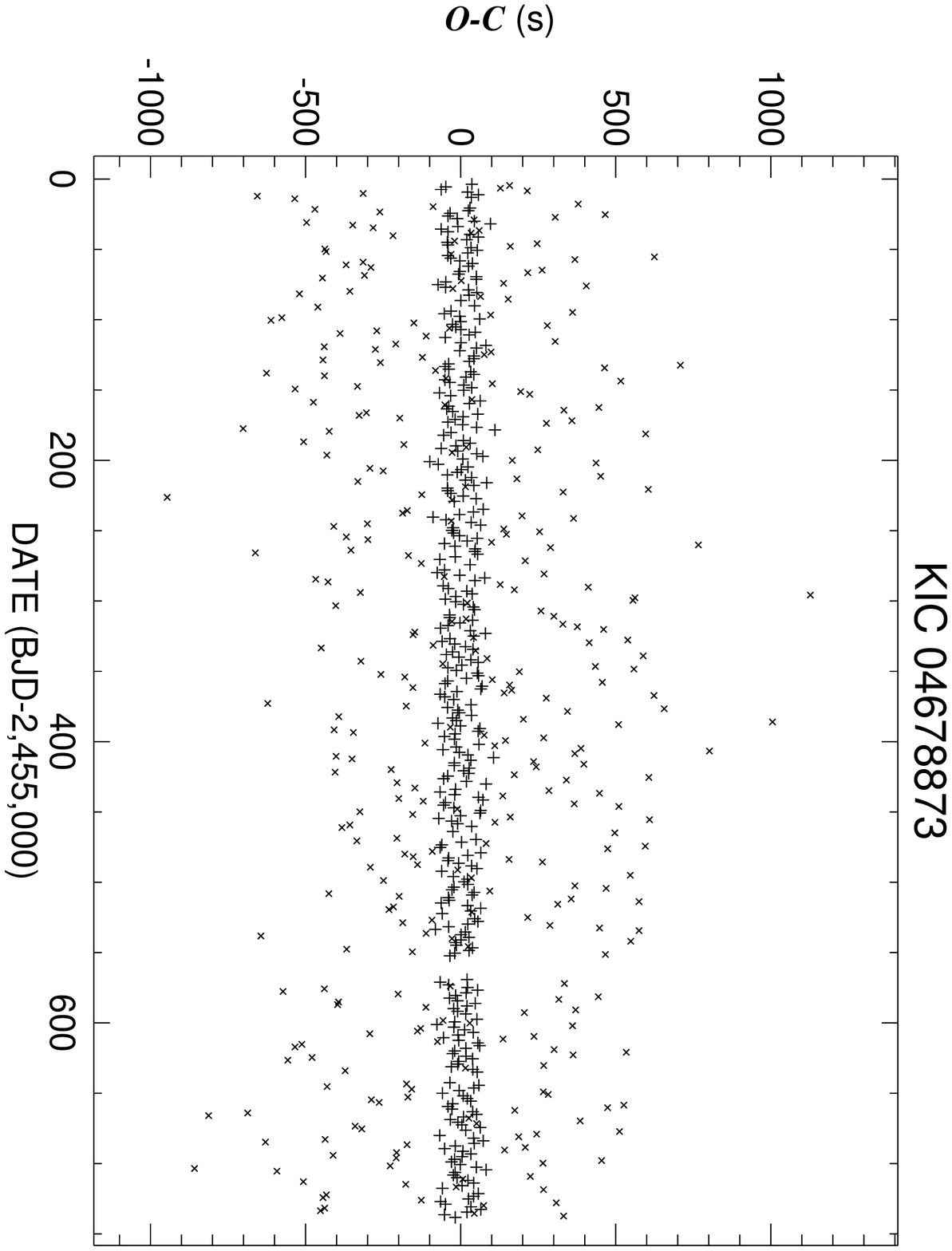}
\figsetgrpnote{The observed minus calculated eclipse times relative to
a linear ephemeris.  The primary and secondary eclipse
times are indicated by $+$ and $\times$ symbols, 
respectively. }
\figsetgrpend

\figsetgrpstart
\figsetgrpnum{1.11}
\figsetgrptitle{r11}
\figsetplot{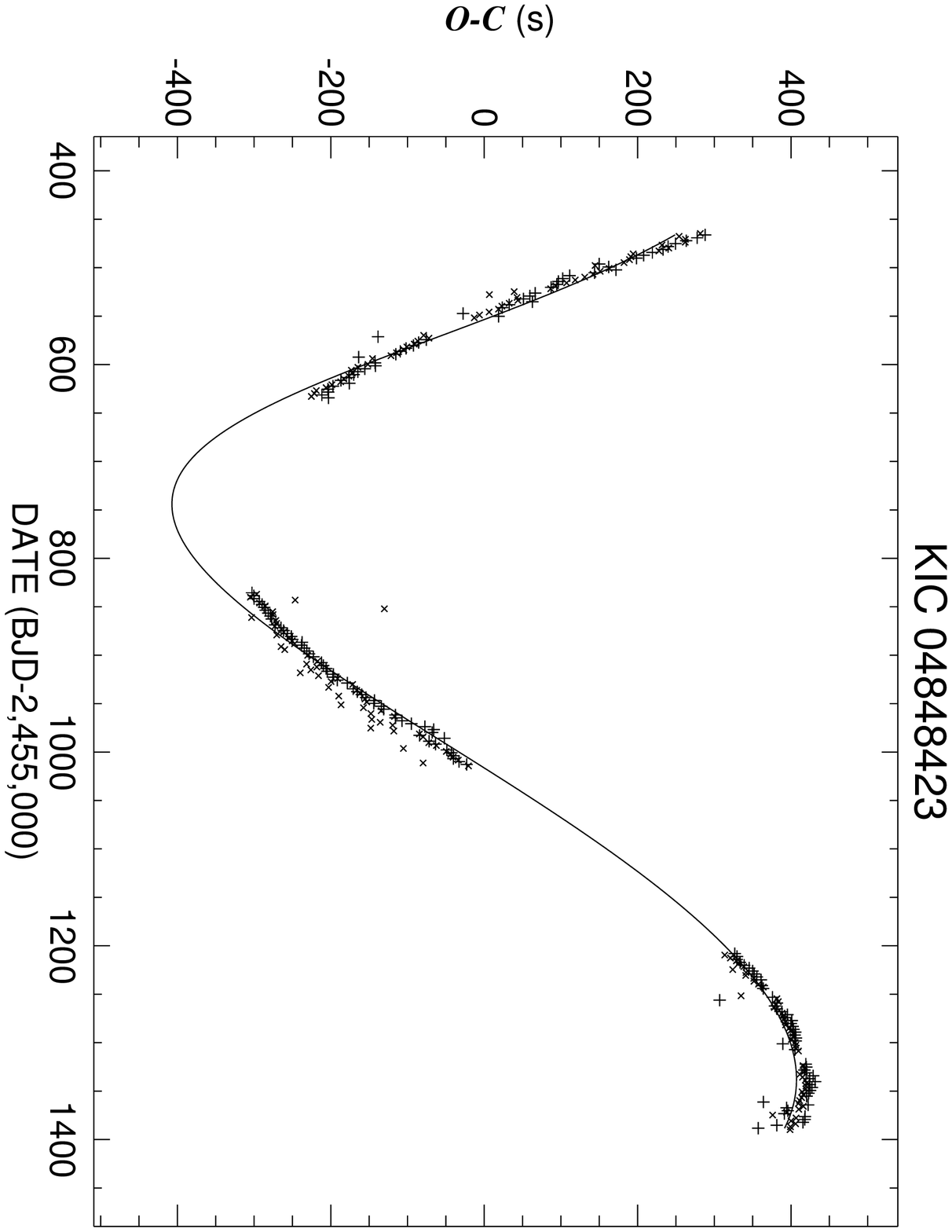}
\figsetgrpnote{The observed minus calculated eclipse times relative to
a linear ephemeris.  The primary and secondary eclipse
times are indicated by $+$ and $\times$ symbols, 
respectively. }
\figsetgrpend

\figsetgrpstart
\figsetgrpnum{1.12}
\figsetgrptitle{r12}
\figsetplot{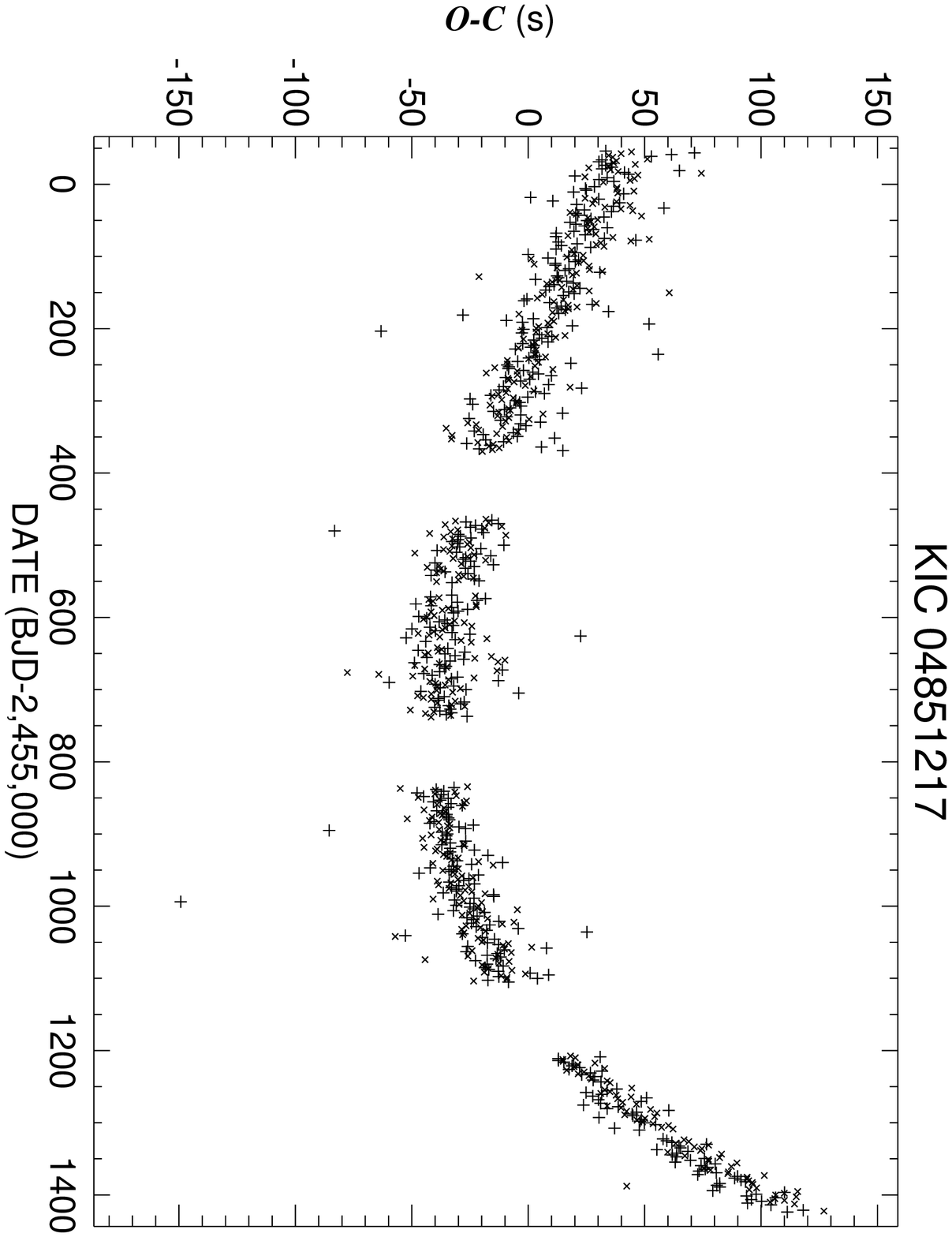}
\figsetgrpnote{The observed minus calculated eclipse times relative to
a linear ephemeris.  The primary and secondary eclipse
times are indicated by $+$ and $\times$ symbols, 
respectively. }
\figsetgrpend

\figsetgrpstart
\figsetgrpnum{1.13}
\figsetgrptitle{r13}
\figsetplot{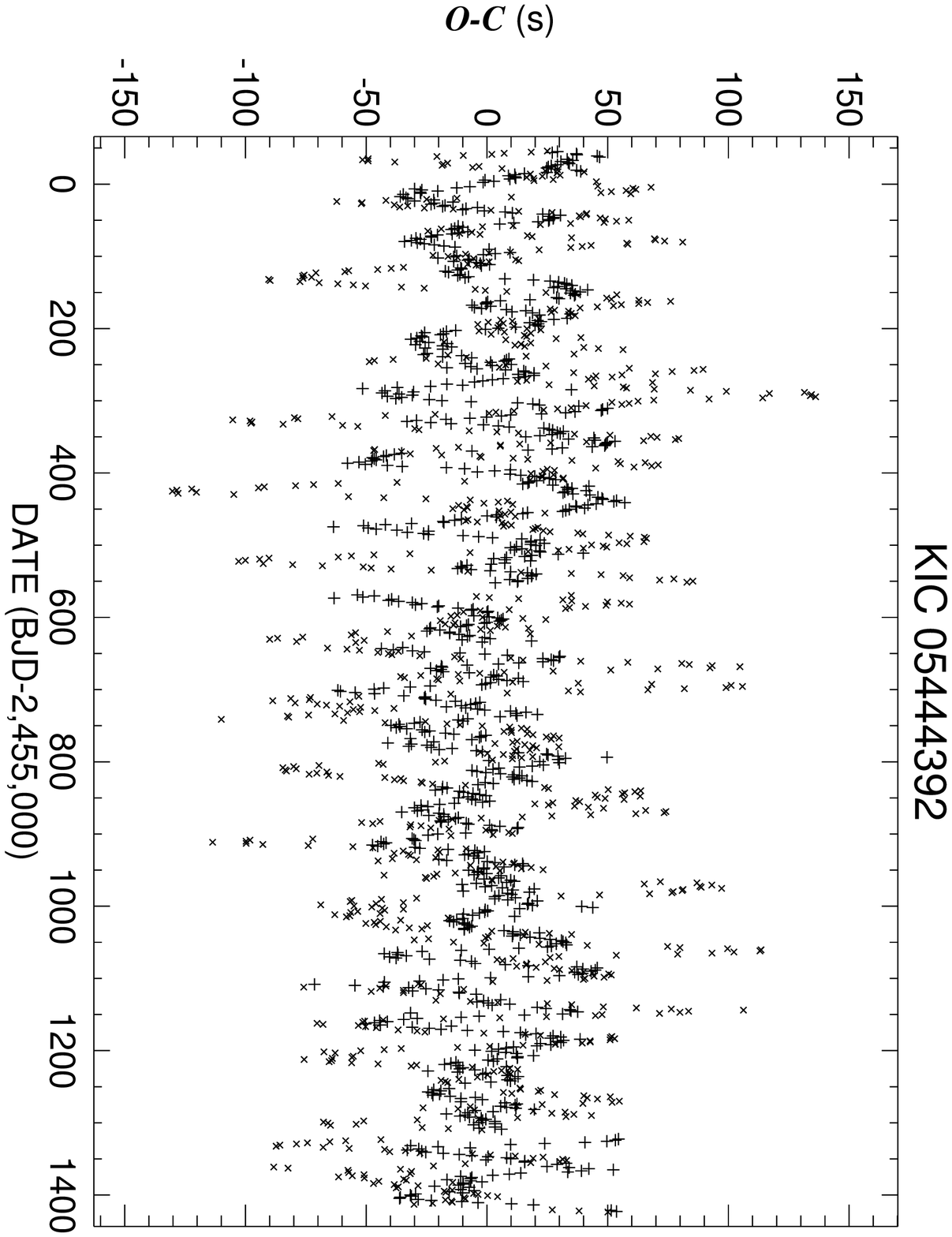}
\figsetgrpnote{The observed minus calculated eclipse times relative to
a linear ephemeris.  The primary and secondary eclipse
times are indicated by $+$ and $\times$ symbols, 
respectively. }
\figsetgrpend

\figsetgrpstart
\figsetgrpnum{1.14}
\figsetgrptitle{r14}
\figsetplot{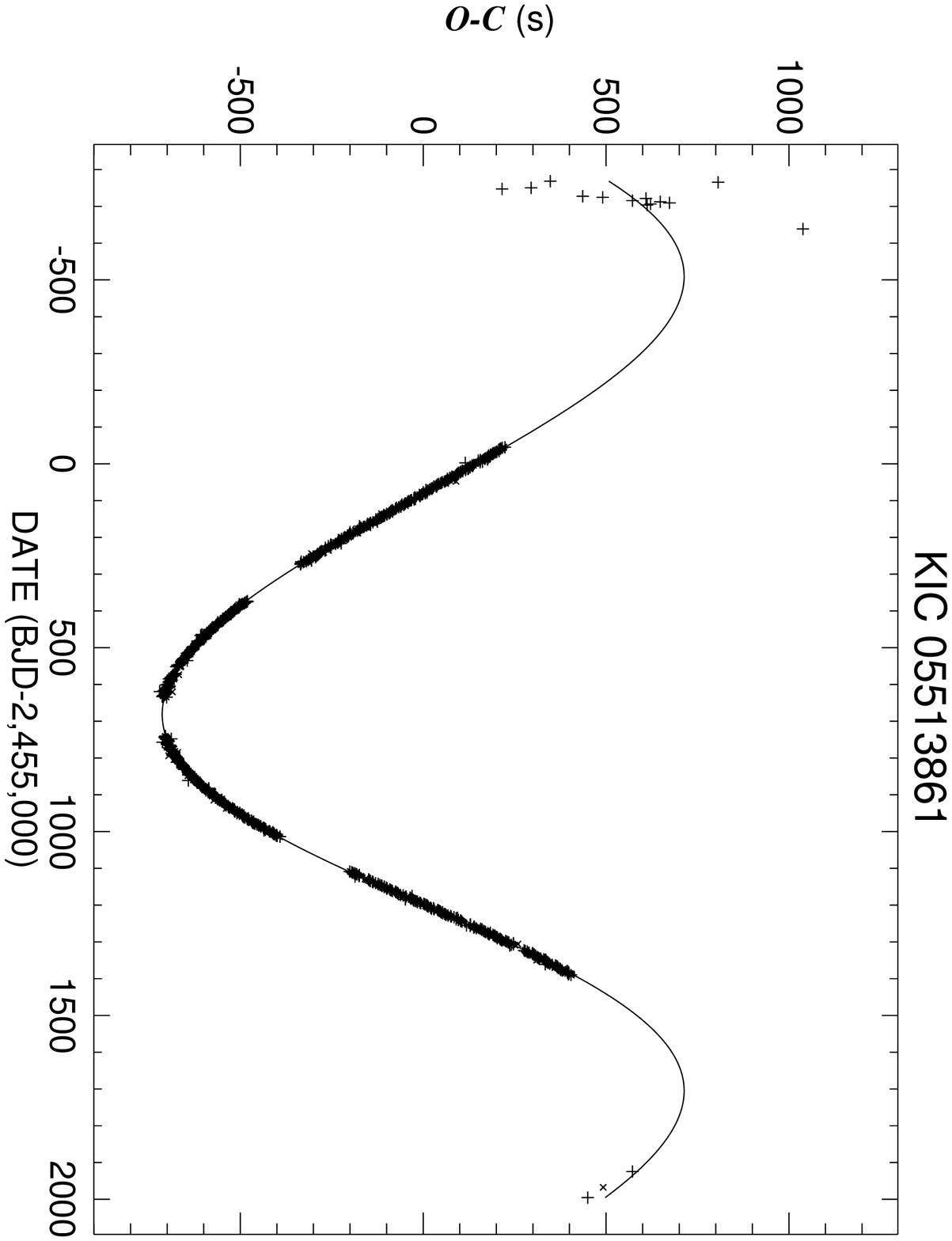}
\figsetgrpnote{The observed minus calculated eclipse times relative to
a linear ephemeris.  The primary and secondary eclipse
times are indicated by $+$ and $\times$ symbols, 
respectively. }
\figsetgrpend

\figsetgrpstart
\figsetgrpnum{1.15}
\figsetgrptitle{r15}
\figsetplot{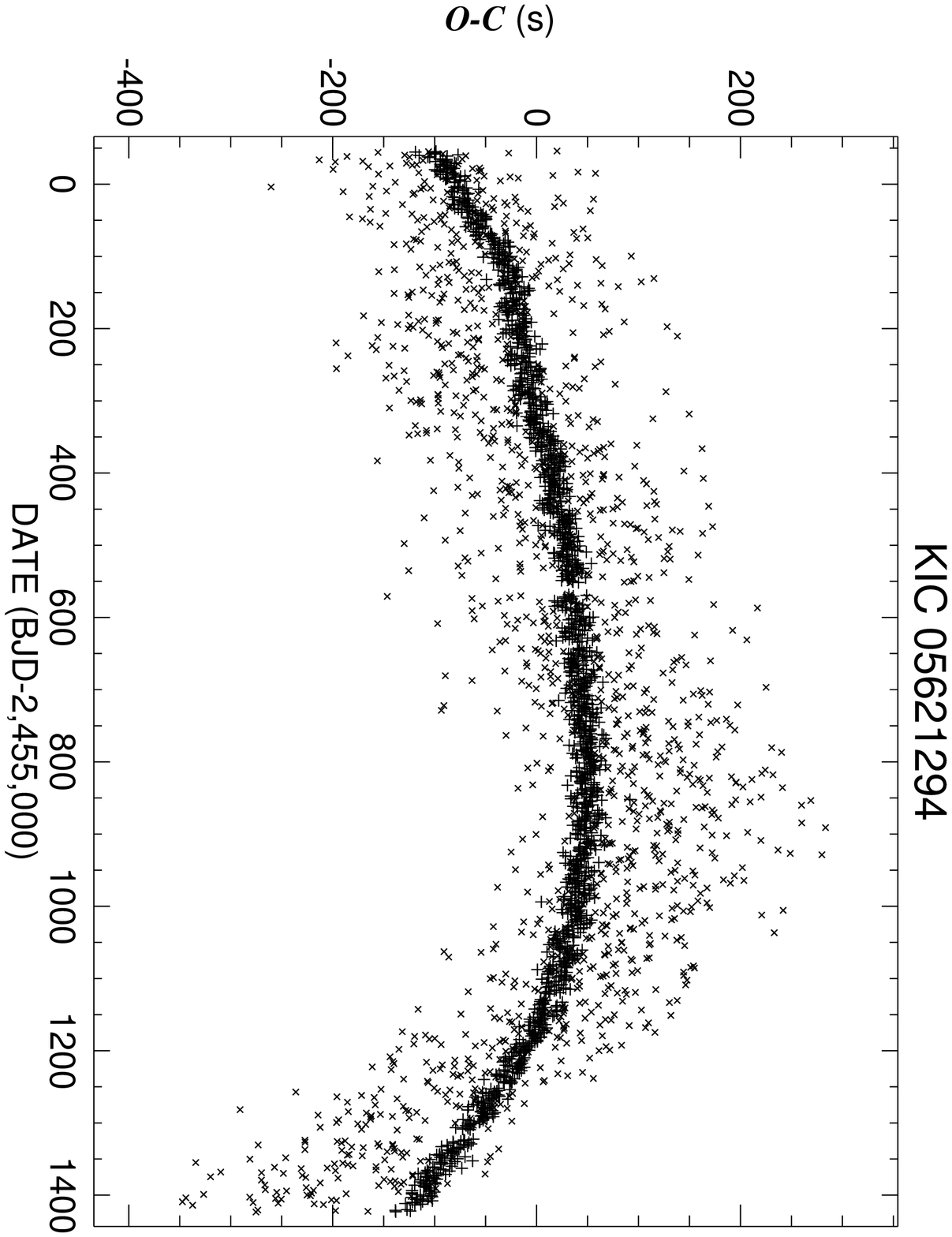}
\figsetgrpnote{The observed minus calculated eclipse times relative to
a linear ephemeris.  The primary and secondary eclipse
times are indicated by $+$ and $\times$ symbols, 
respectively. }
\figsetgrpend

\figsetgrpstart
\figsetgrpnum{1.16}
\figsetgrptitle{r16}
\figsetplot{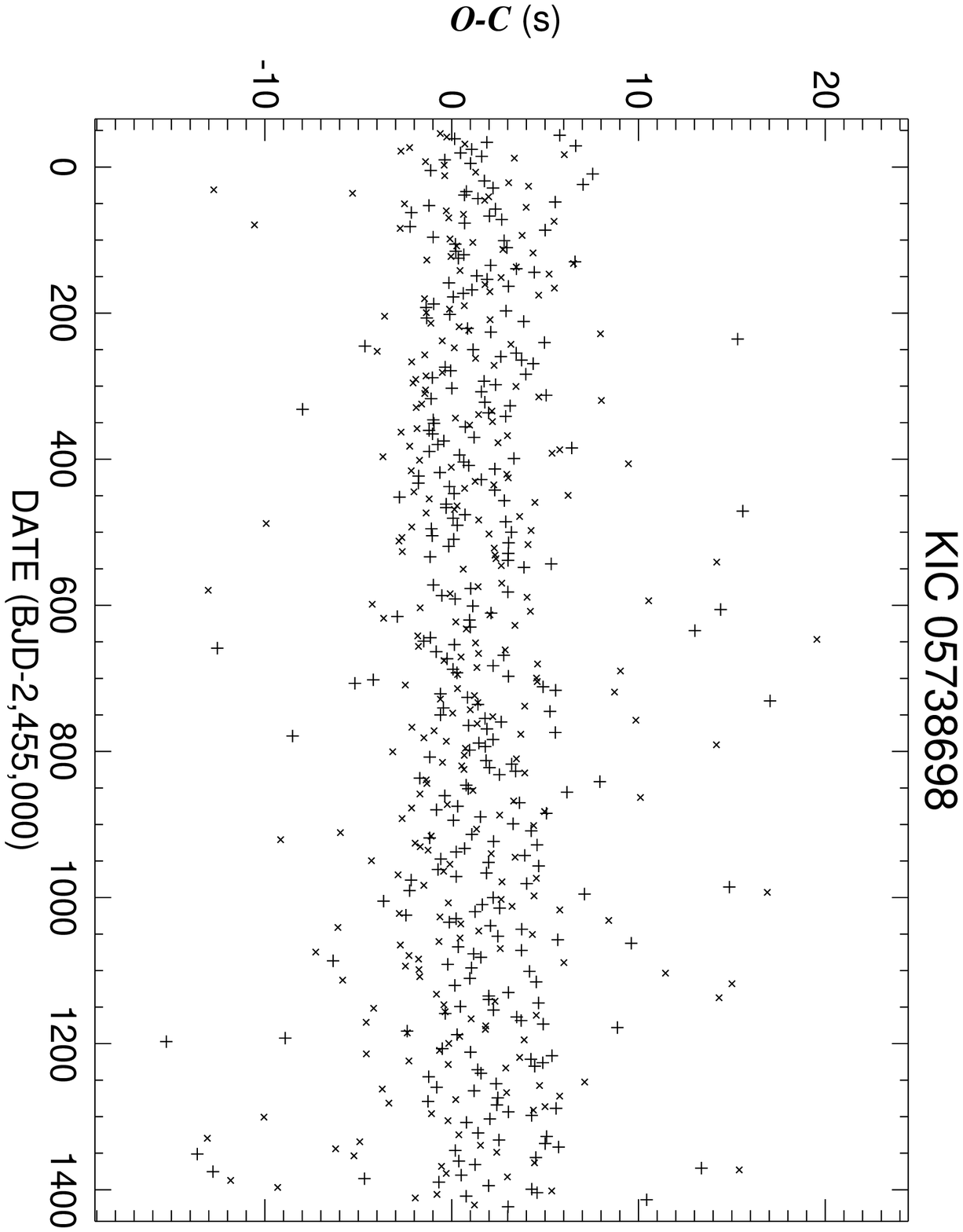}
\figsetgrpnote{The observed minus calculated eclipse times relative to
a linear ephemeris.  The primary and secondary eclipse
times are indicated by $+$ and $\times$ symbols, 
respectively. }
\figsetgrpend

\figsetgrpstart
\figsetgrpnum{1.17}
\figsetgrptitle{r17}
\figsetplot{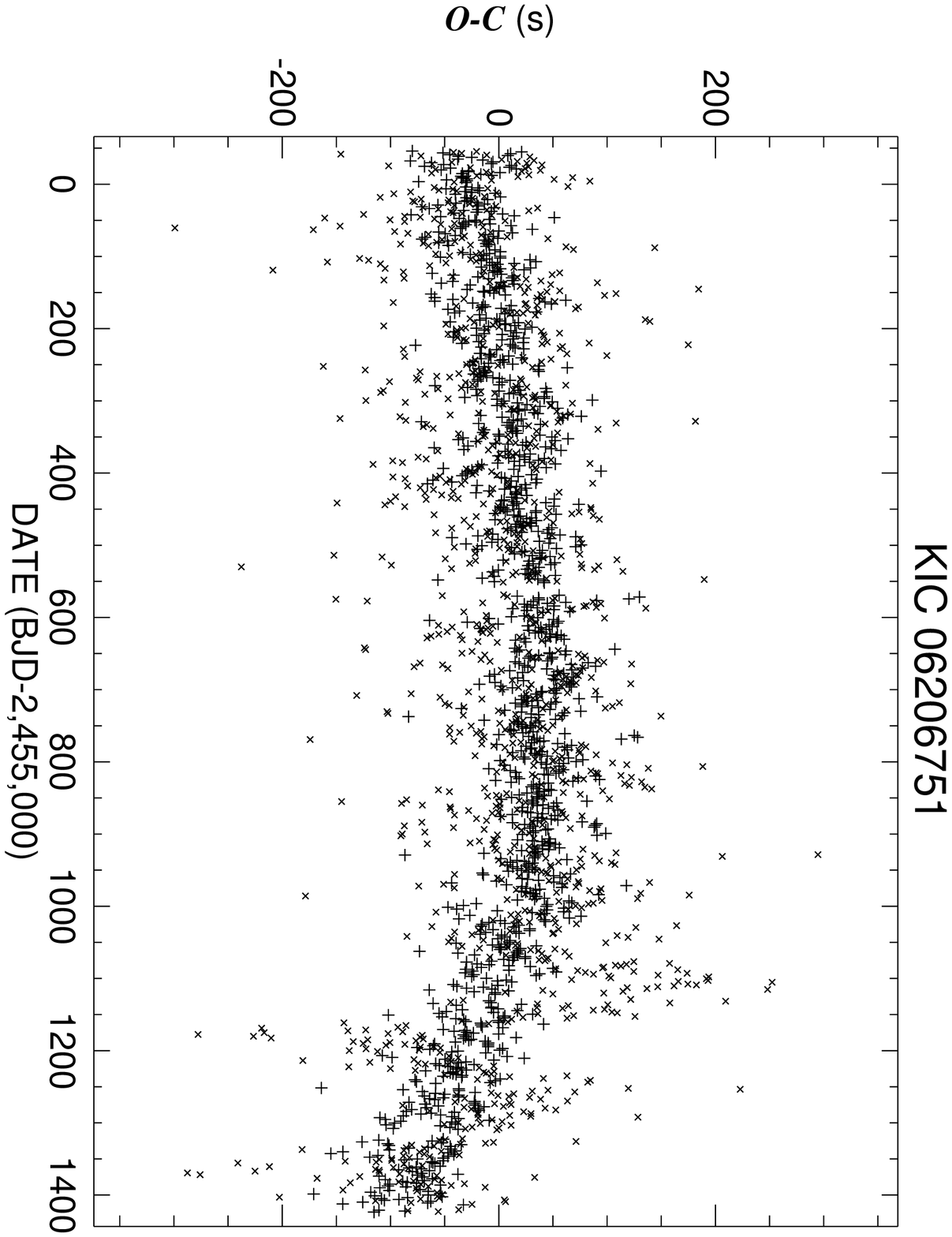}
\figsetgrpnote{The observed minus calculated eclipse times relative to
a linear ephemeris.  The primary and secondary eclipse
times are indicated by $+$ and $\times$ symbols, 
respectively. }
\figsetgrpend

\figsetgrpstart
\figsetgrpnum{1.18}
\figsetgrptitle{r18}
\figsetplot{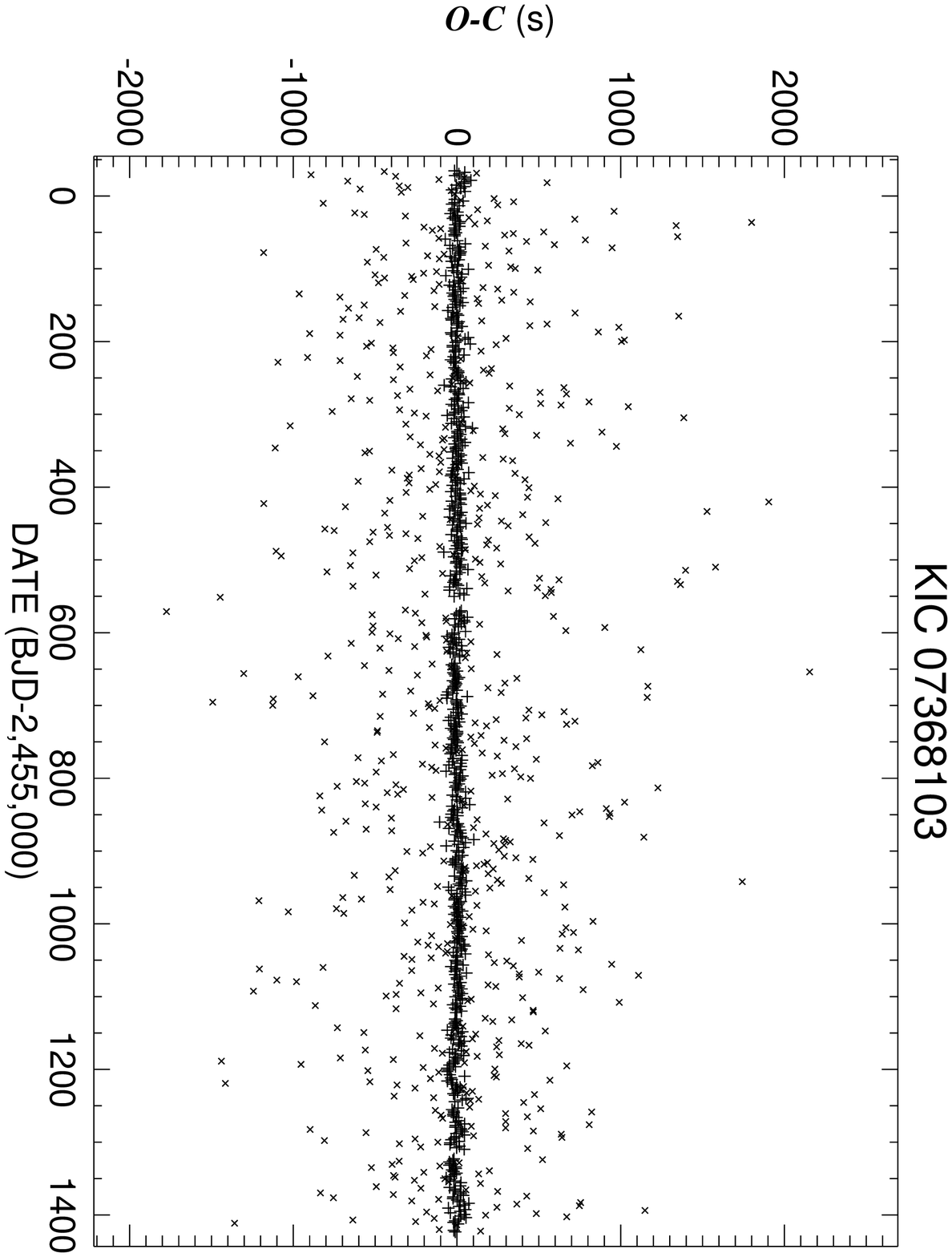}
\figsetgrpnote{The observed minus calculated eclipse times relative to
a linear ephemeris.  The primary and secondary eclipse
times are indicated by $+$ and $\times$ symbols, 
respectively. }
\figsetgrpend

\figsetgrpstart
\figsetgrpnum{1.19}
\figsetgrptitle{r19}
\figsetplot{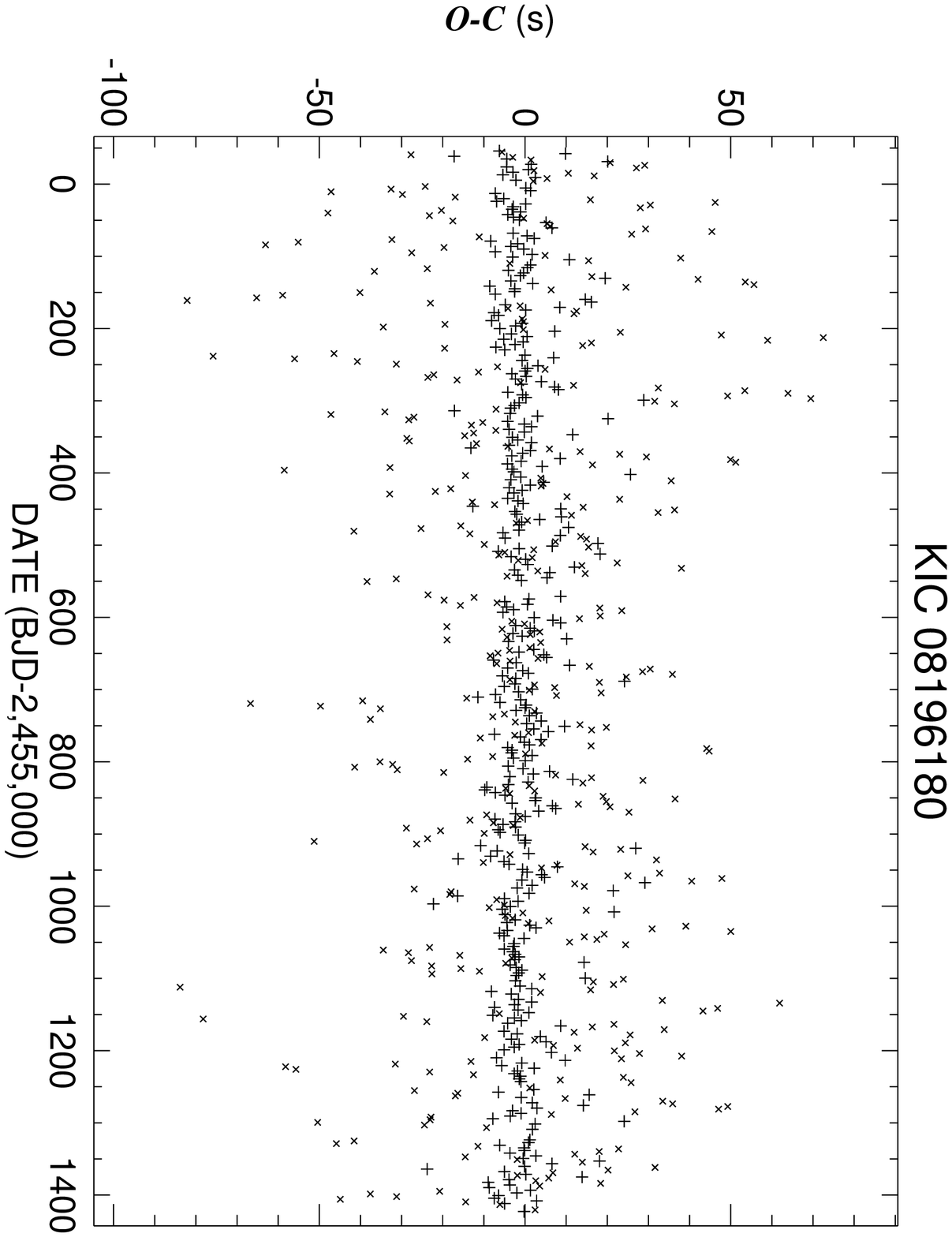}
\figsetgrpnote{The observed minus calculated eclipse times relative to
a linear ephemeris.  The primary and secondary eclipse
times are indicated by $+$ and $\times$ symbols, 
respectively. }
\figsetgrpend

\figsetgrpstart
\figsetgrpnum{1.20}
\figsetgrptitle{r20}
\figsetplot{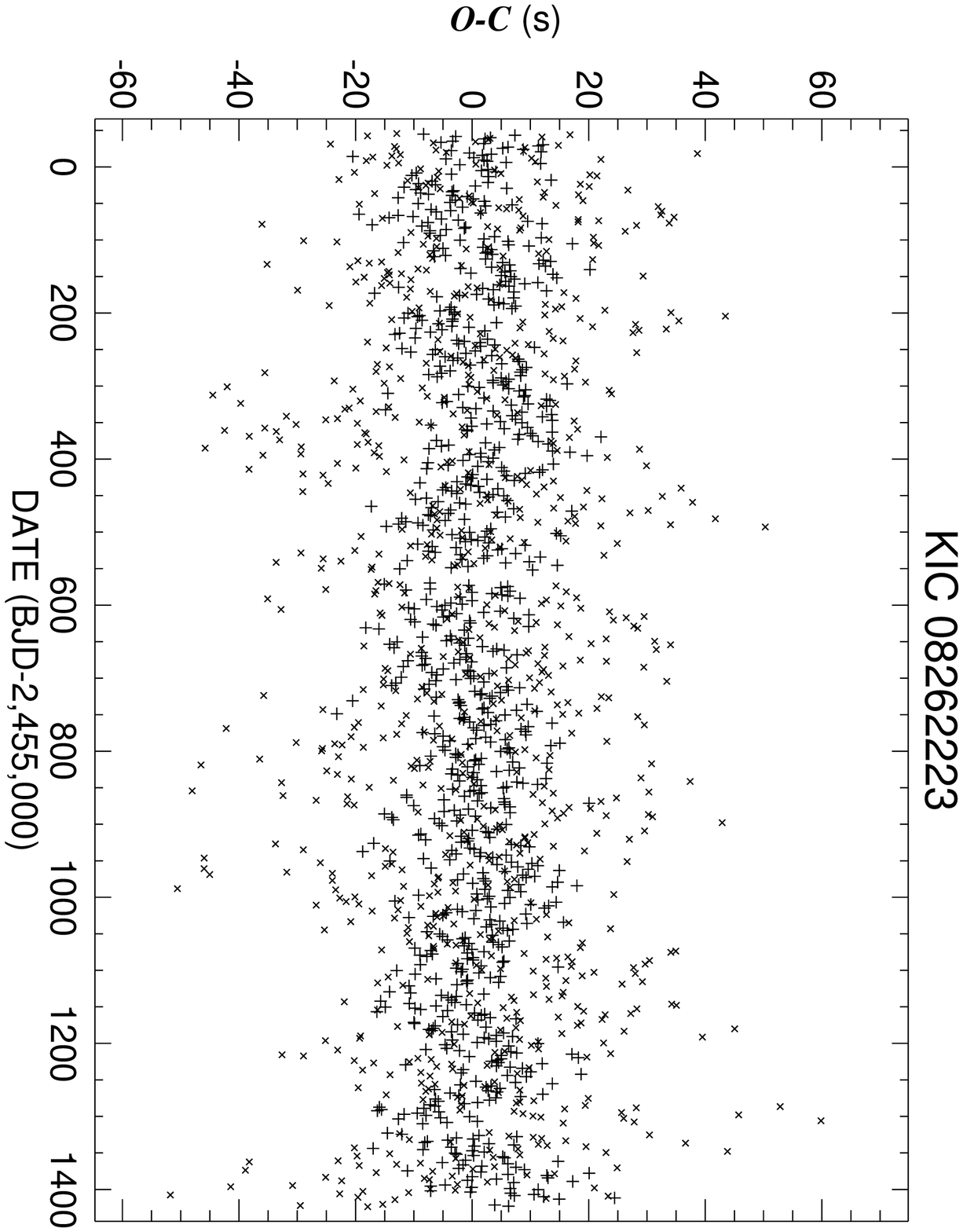}
\figsetgrpnote{The observed minus calculated eclipse times relative to
a linear ephemeris.  The primary and secondary eclipse
times are indicated by $+$ and $\times$ symbols, 
respectively. }
\figsetgrpend

\figsetgrpstart
\figsetgrpnum{1.21}
\figsetgrptitle{r21}
\figsetplot{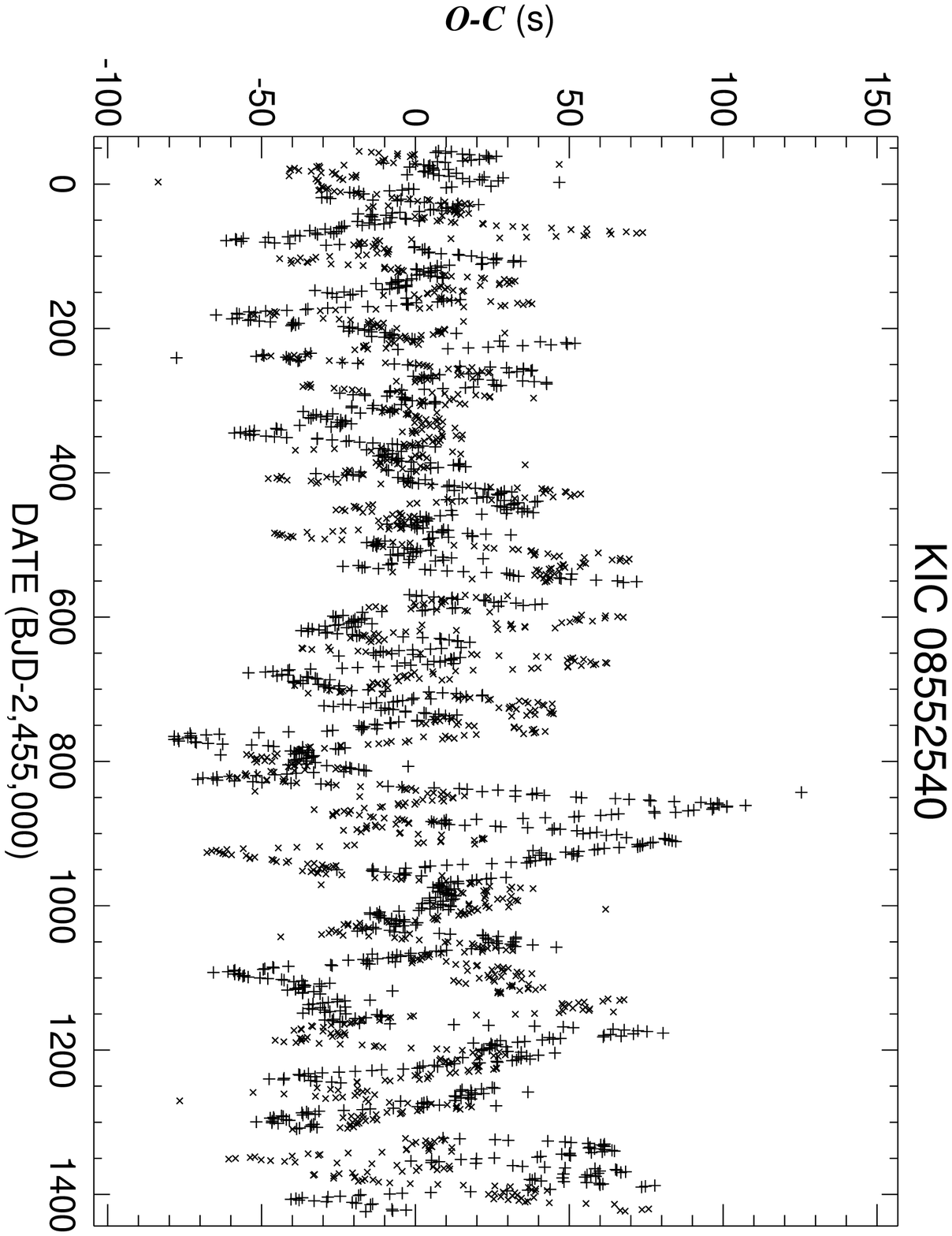}
\figsetgrpnote{The observed minus calculated eclipse times relative to
a linear ephemeris.  The primary and secondary eclipse
times are indicated by $+$ and $\times$ symbols, 
respectively. }
\figsetgrpend

\figsetgrpstart
\figsetgrpnum{1.22}
\figsetgrptitle{r22}
\figsetplot{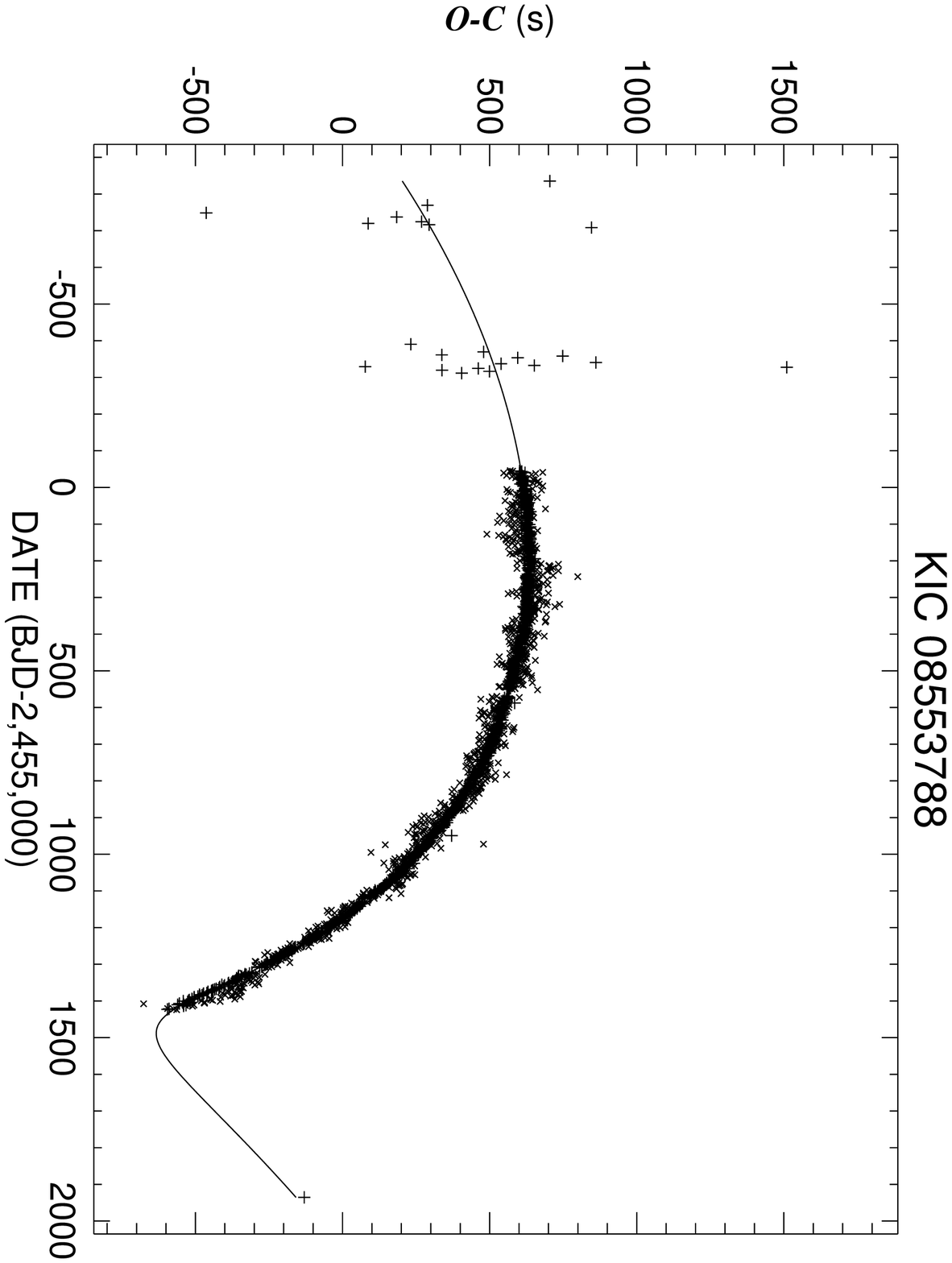}
\figsetgrpnote{The observed minus calculated eclipse times relative to
a linear ephemeris.  The primary and secondary eclipse
times are indicated by $+$ and $\times$ symbols, 
respectively. }
\figsetgrpend

\figsetgrpstart
\figsetgrpnum{1.23}
\figsetgrptitle{r23}
\figsetplot{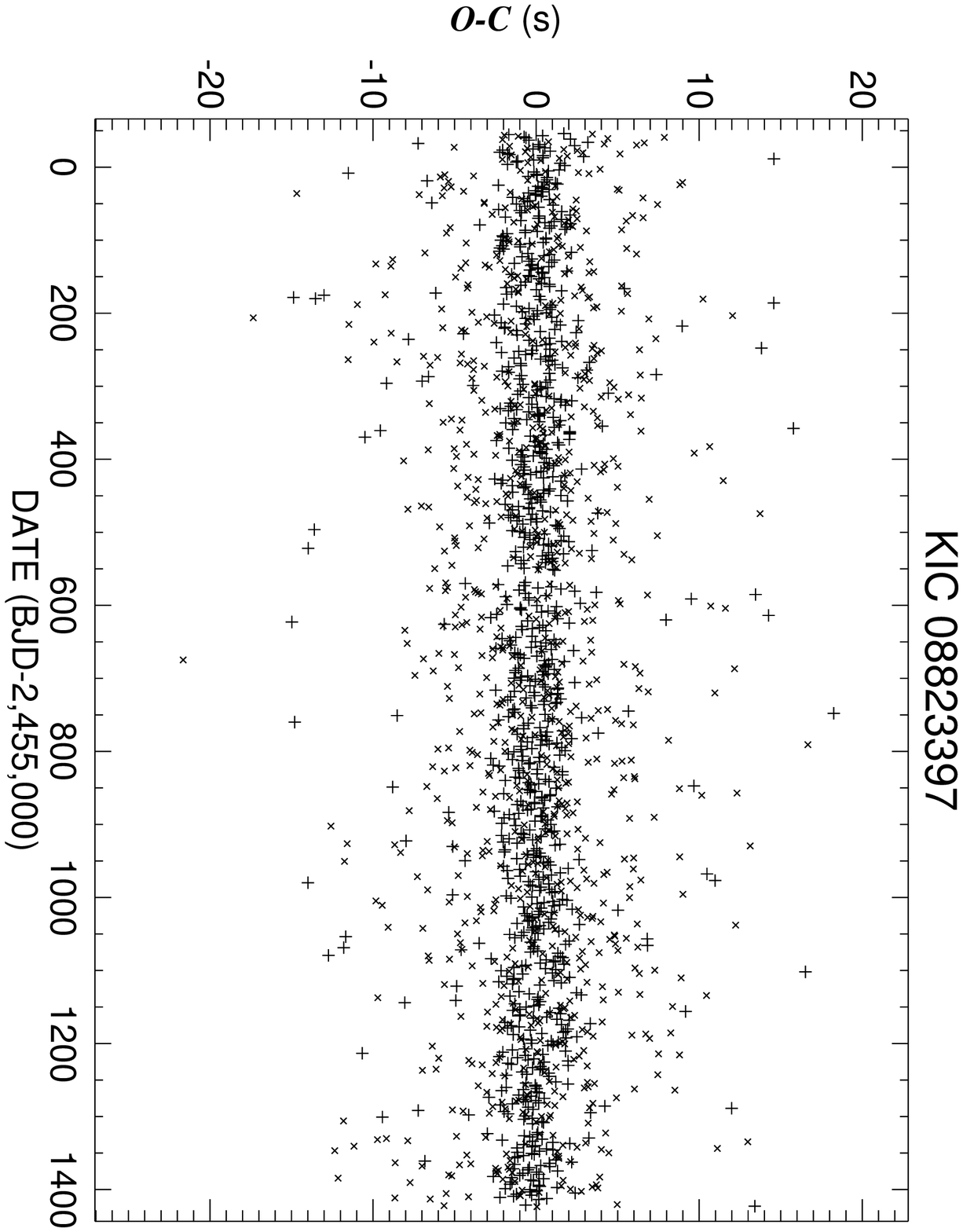}
\figsetgrpnote{The observed minus calculated eclipse times relative to
a linear ephemeris.  The primary and secondary eclipse
times are indicated by $+$ and $\times$ symbols, 
respectively. }
\figsetgrpend

\figsetgrpstart
\figsetgrpnum{1.24}
\figsetgrptitle{r24}
\figsetplot{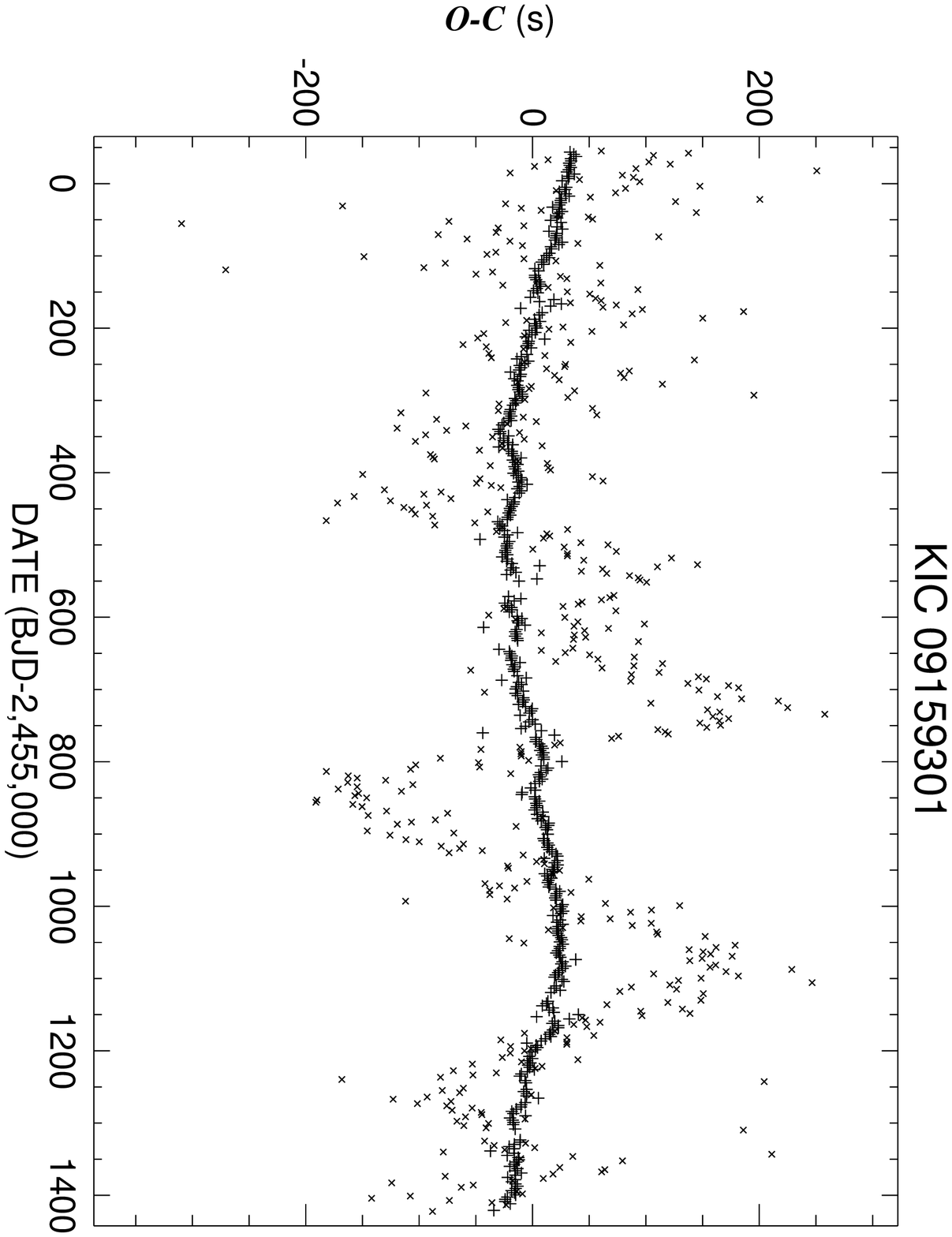}
\figsetgrpnote{The observed minus calculated eclipse times relative to
a linear ephemeris.  The primary and secondary eclipse
times are indicated by $+$ and $\times$ symbols, 
respectively. }
\figsetgrpend

\figsetgrpstart
\figsetgrpnum{1.25}
\figsetgrptitle{r25}
\figsetplot{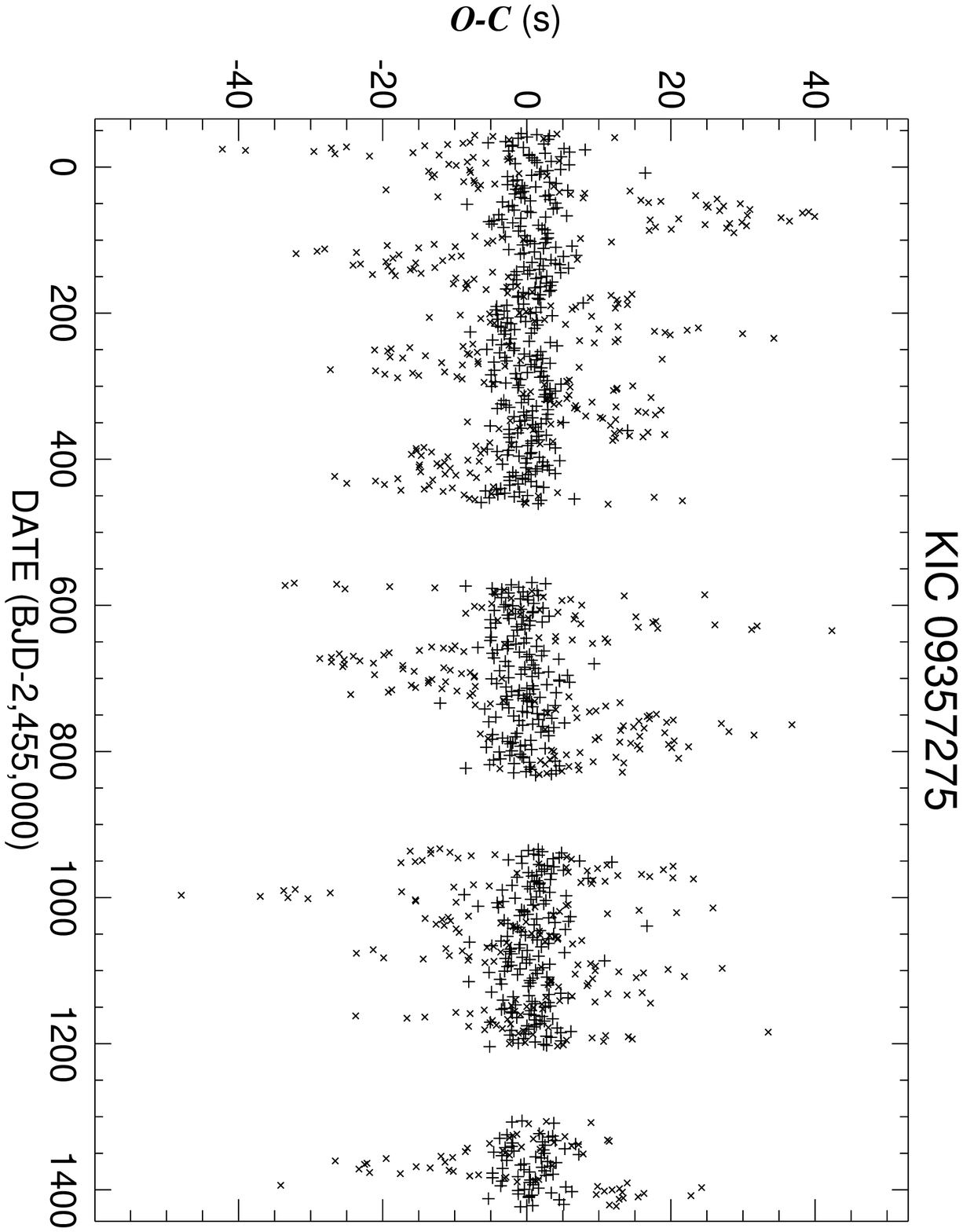}
\figsetgrpnote{The observed minus calculated eclipse times relative to
a linear ephemeris.  The primary and secondary eclipse
times are indicated by $+$ and $\times$ symbols, 
respectively. }
\figsetgrpend

\figsetgrpstart
\figsetgrpnum{1.26}
\figsetgrptitle{r26}
\figsetplot{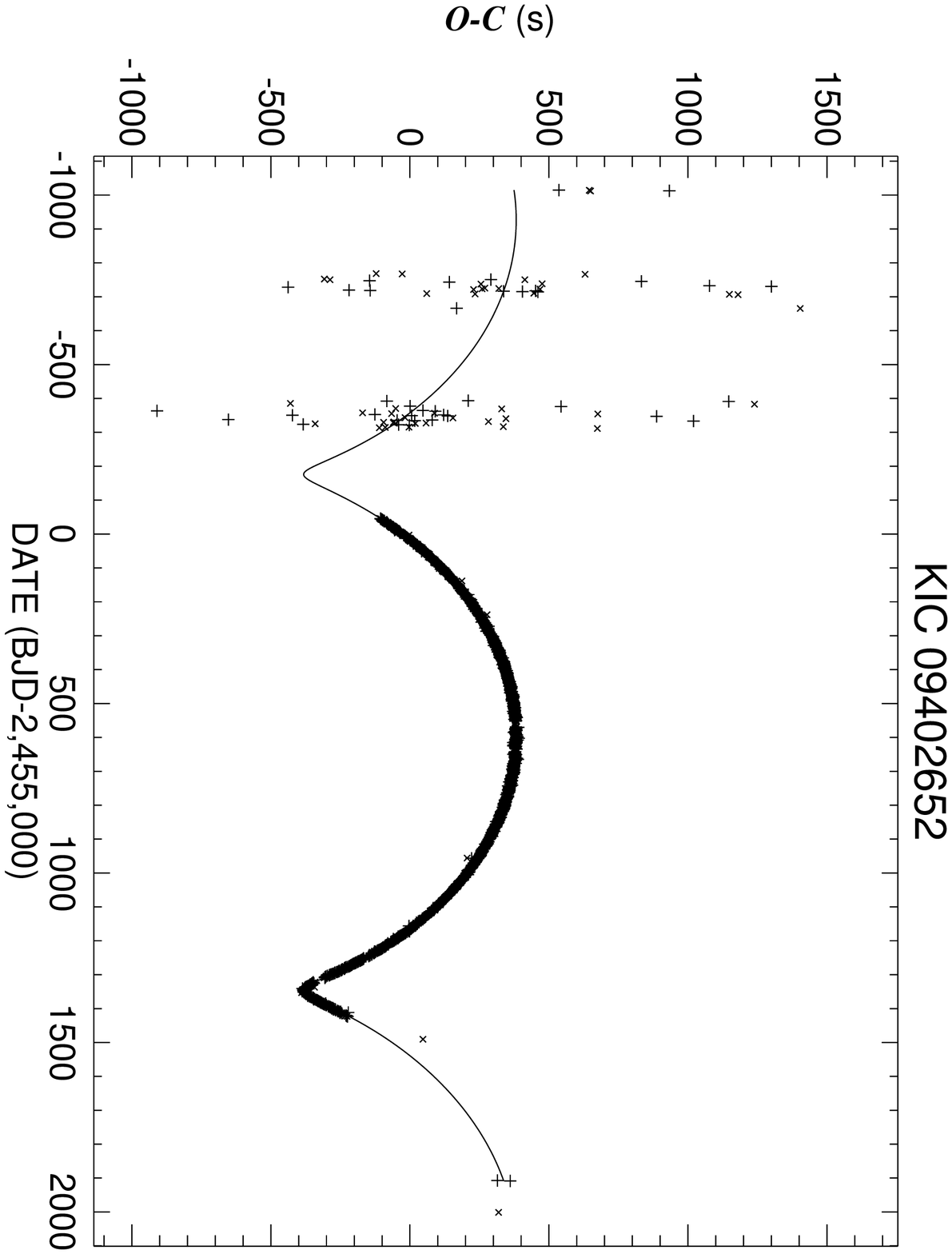}
\figsetgrpnote{The observed minus calculated eclipse times relative to
a linear ephemeris.  The primary and secondary eclipse
times are indicated by $+$ and $\times$ symbols, 
respectively. }
\figsetgrpend

\figsetgrpstart
\figsetgrpnum{1.27}
\figsetgrptitle{r27}
\figsetplot{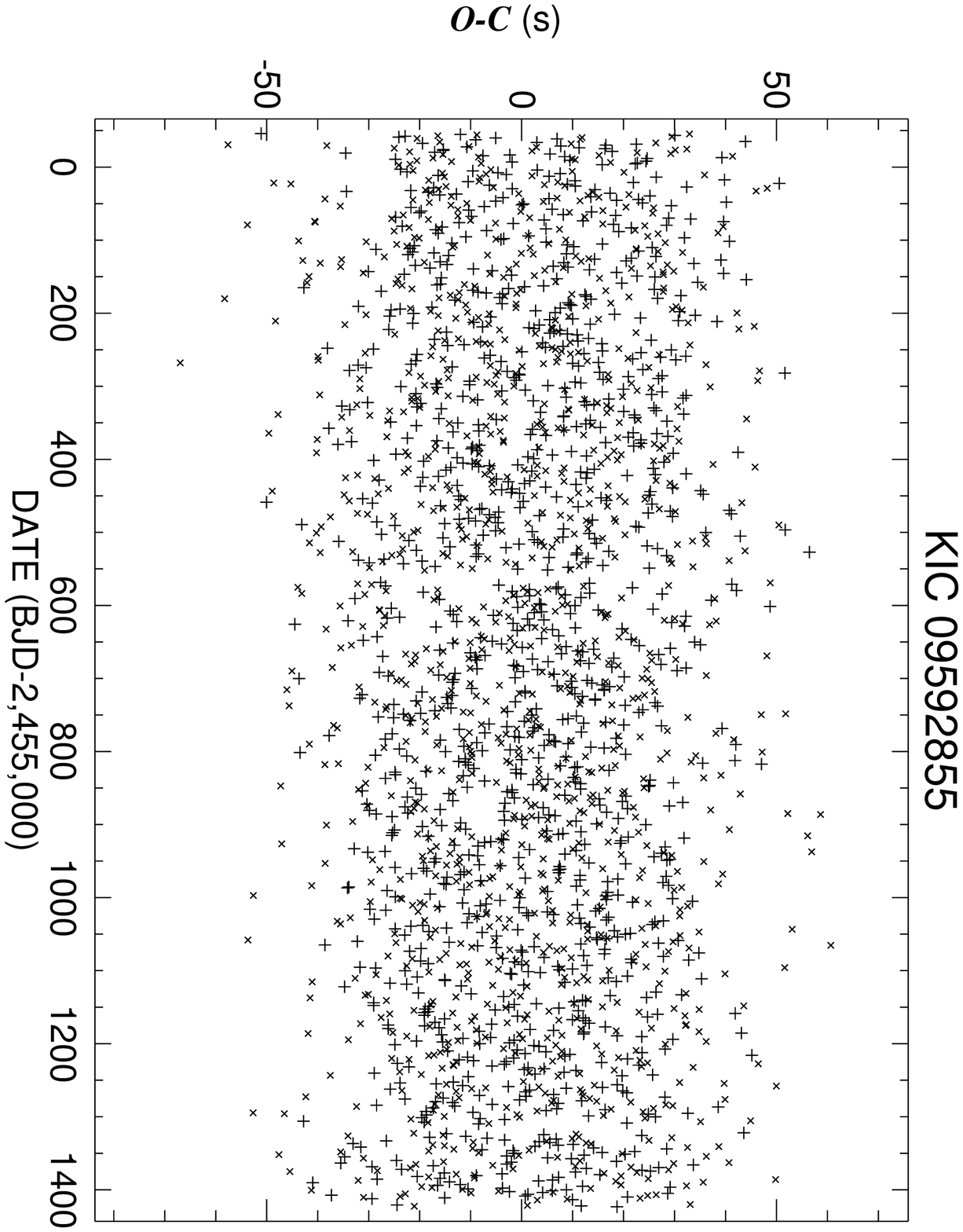}
\figsetgrpnote{The observed minus calculated eclipse times relative to
a linear ephemeris.  The primary and secondary eclipse
times are indicated by $+$ and $\times$ symbols, 
respectively. }
\figsetgrpend

\figsetgrpstart
\figsetgrpnum{1.28}
\figsetgrptitle{r28}
\figsetplot{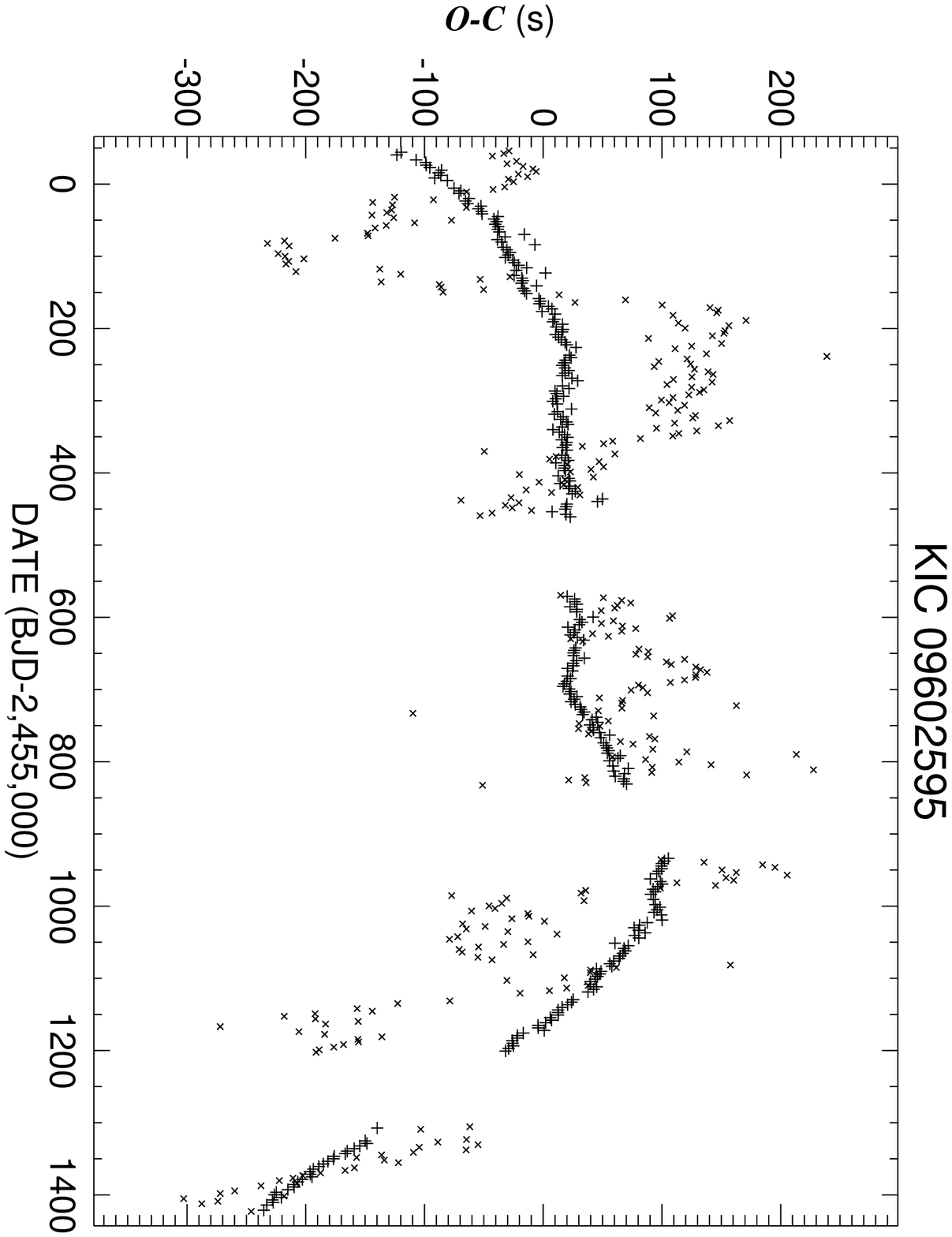}
\figsetgrpnote{The observed minus calculated eclipse times relative to
a linear ephemeris.  The primary and secondary eclipse
times are indicated by $+$ and $\times$ symbols, 
respectively. }
\figsetgrpend

\figsetgrpstart
\figsetgrpnum{1.29}
\figsetgrptitle{r29}
\figsetplot{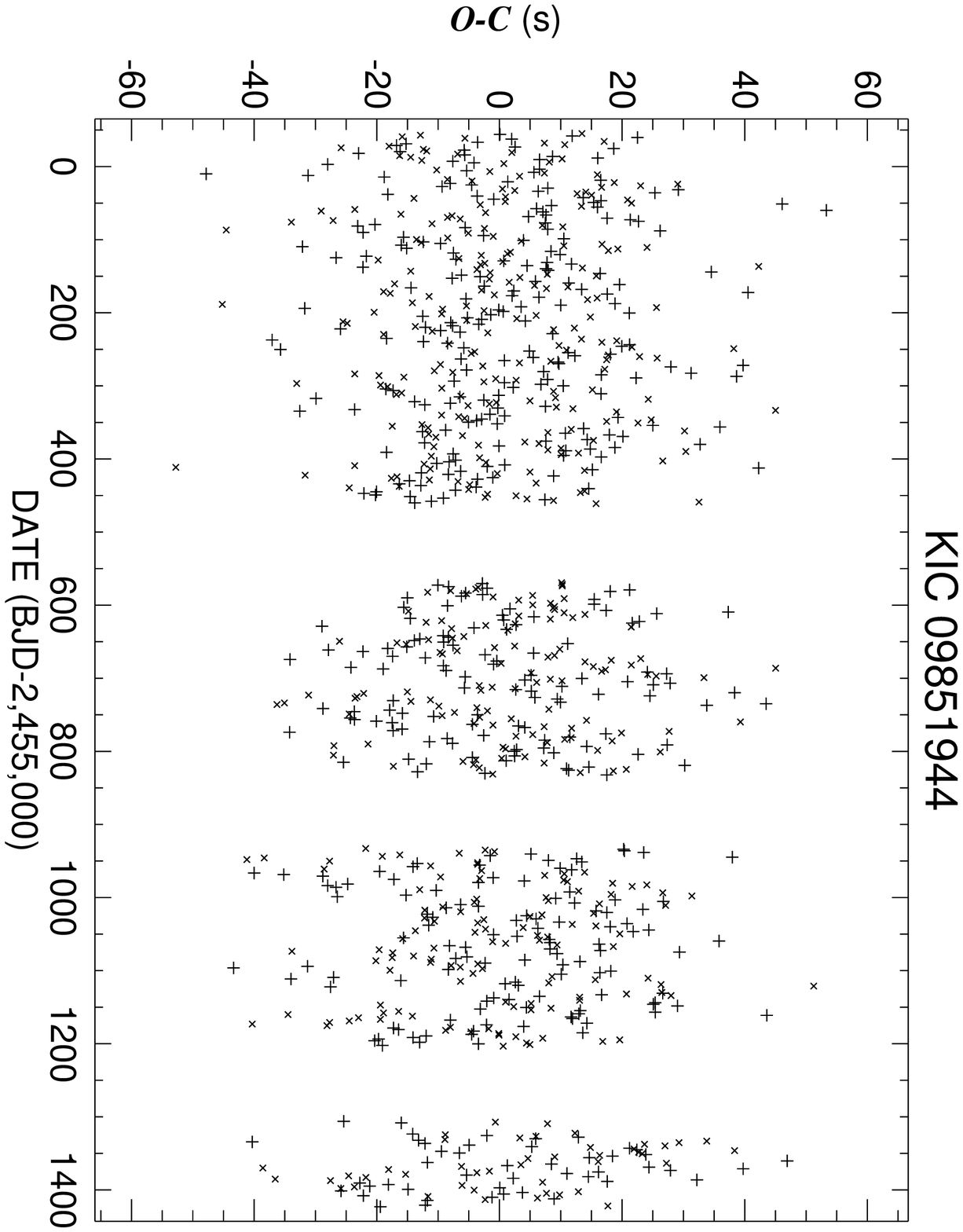}
\figsetgrpnote{The observed minus calculated eclipse times relative to
a linear ephemeris.  The primary and secondary eclipse
times are indicated by $+$ and $\times$ symbols, 
respectively. }
\figsetgrpend

\figsetgrpstart
\figsetgrpnum{1.30}
\figsetgrptitle{r30}
\figsetplot{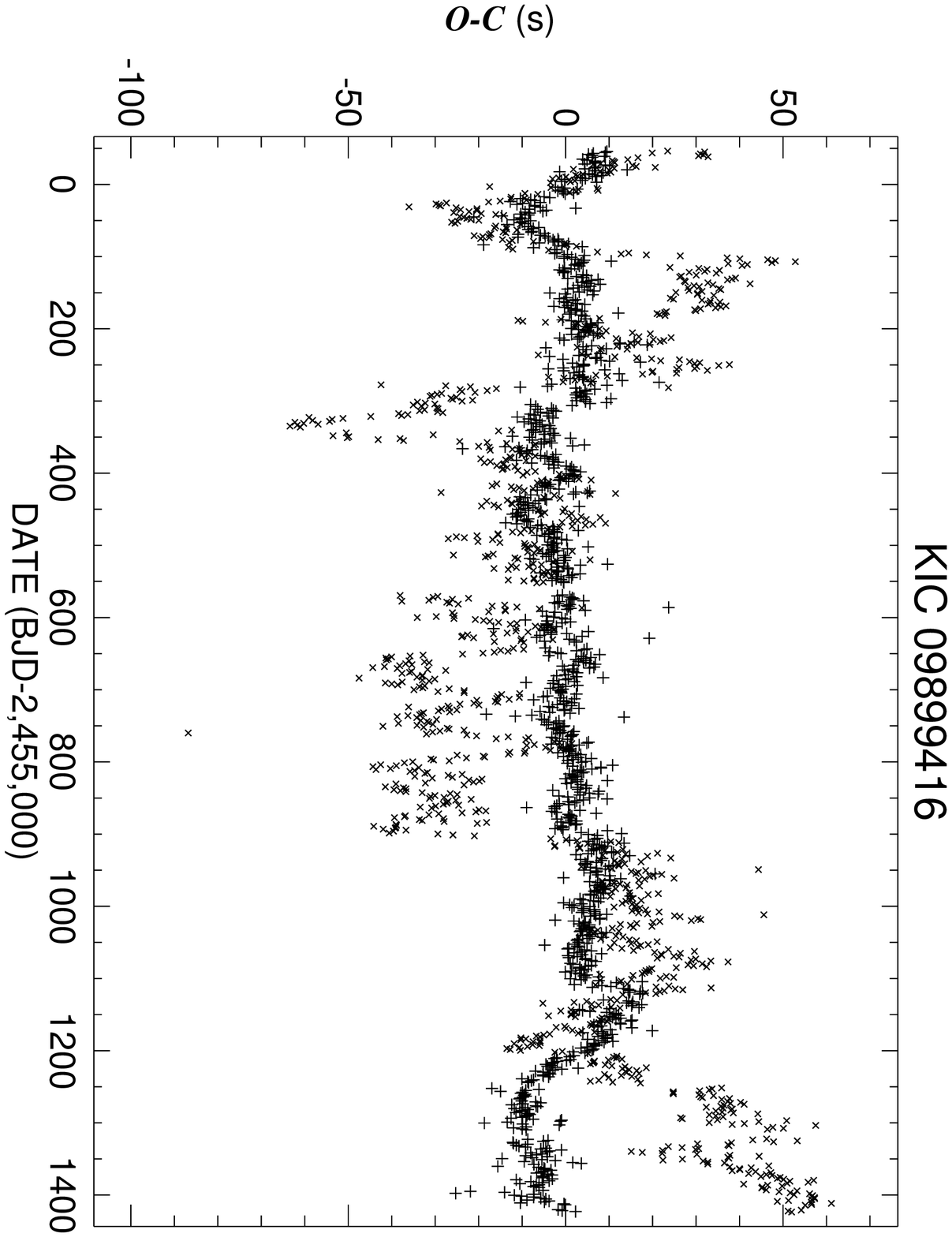}
\figsetgrpnote{The observed minus calculated eclipse times relative to
a linear ephemeris.  The primary and secondary eclipse
times are indicated by $+$ and $\times$ symbols, 
respectively. }
\figsetgrpend

\figsetgrpstart
\figsetgrpnum{1.31}
\figsetgrptitle{r31}
\figsetplot{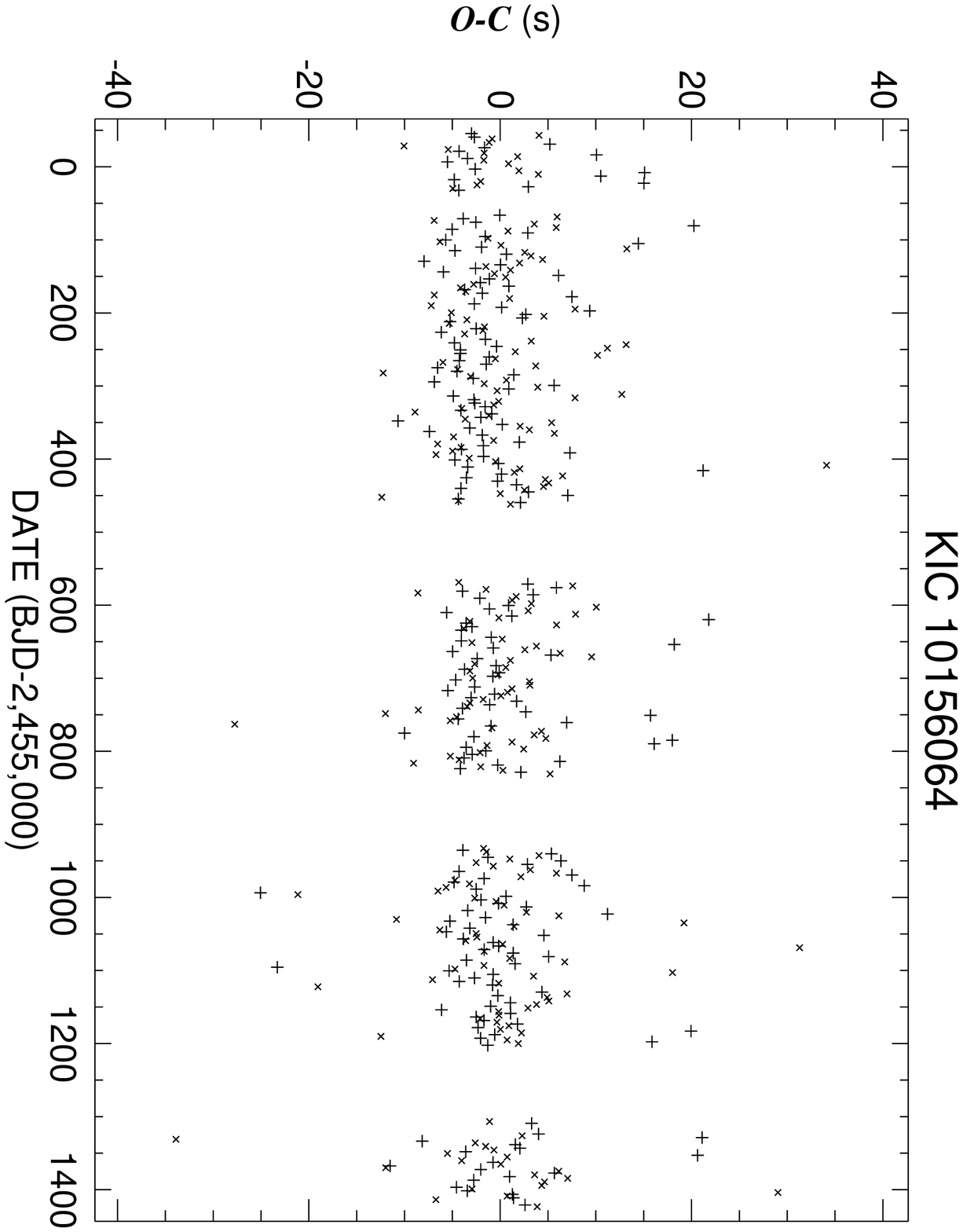}
\figsetgrpnote{The observed minus calculated eclipse times relative to
a linear ephemeris.  The primary and secondary eclipse
times are indicated by $+$ and $\times$ symbols, 
respectively. }
\figsetgrpend

\figsetgrpstart
\figsetgrpnum{1.32}
\figsetgrptitle{r32}
\figsetplot{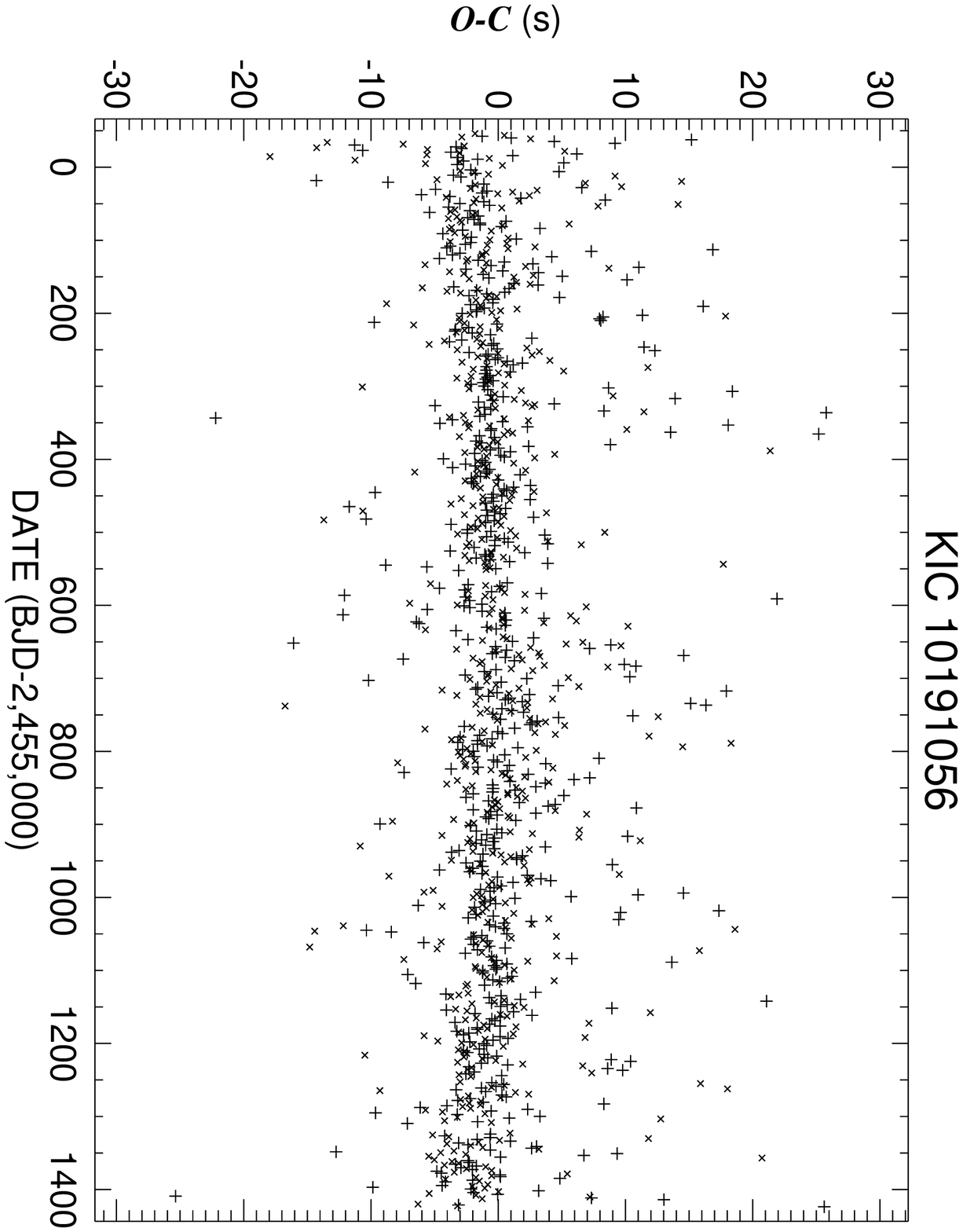}
\figsetgrpnote{The observed minus calculated eclipse times relative to
a linear ephemeris.  The primary and secondary eclipse
times are indicated by $+$ and $\times$ symbols, 
respectively. }
\figsetgrpend

\figsetgrpstart
\figsetgrpnum{1.33}
\figsetgrptitle{r33}
\figsetplot{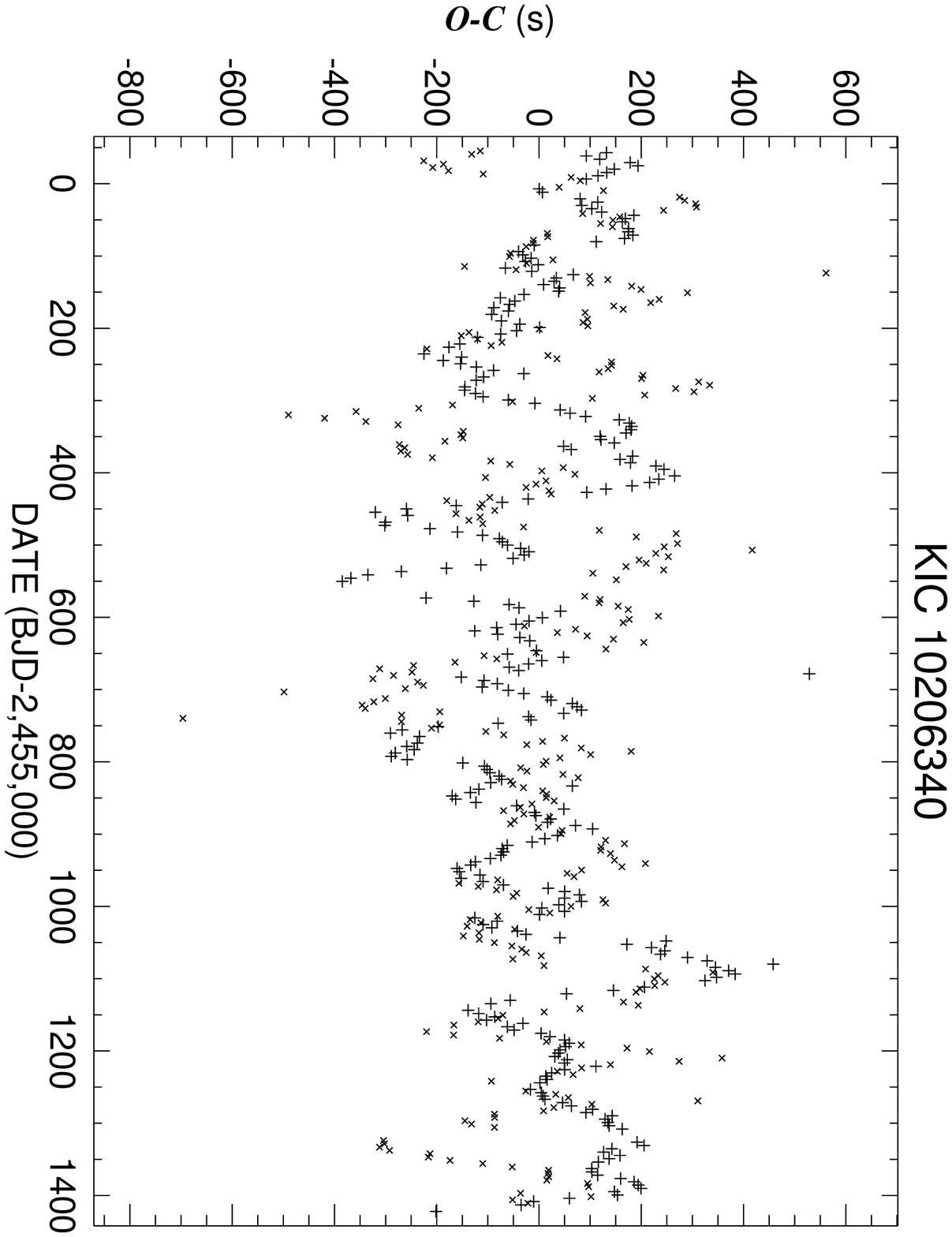}
\figsetgrpnote{The observed minus calculated eclipse times relative to
a linear ephemeris.  The primary and secondary eclipse
times are indicated by $+$ and $\times$ symbols, 
respectively. }
\figsetgrpend

\figsetgrpstart
\figsetgrpnum{1.34}
\figsetgrptitle{r34}
\figsetplot{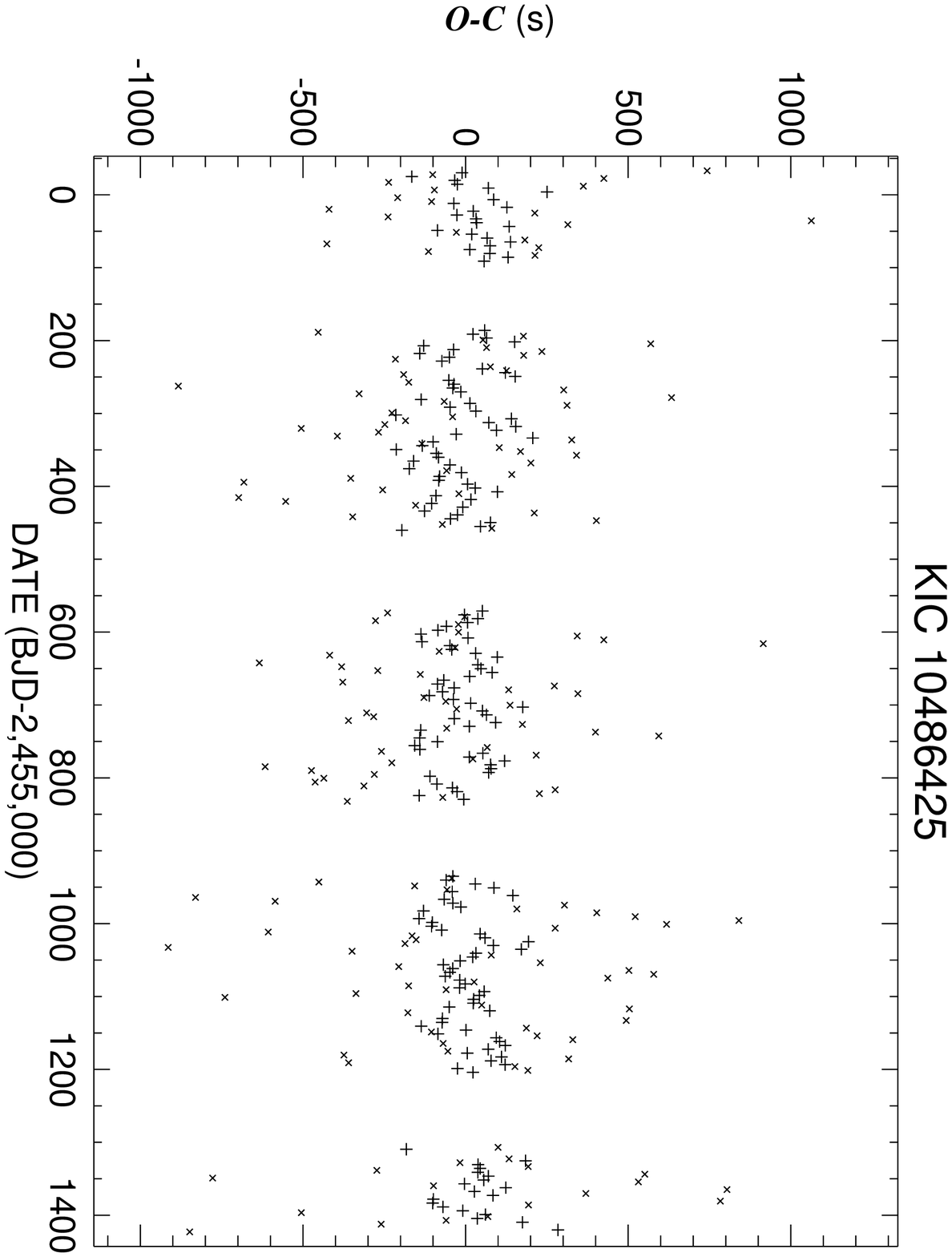}
\figsetgrpnote{The observed minus calculated eclipse times relative to
a linear ephemeris.  The primary and secondary eclipse
times are indicated by $+$ and $\times$ symbols, 
respectively. }
\figsetgrpend

\figsetgrpstart
\figsetgrpnum{1.35}
\figsetgrptitle{r35}
\figsetplot{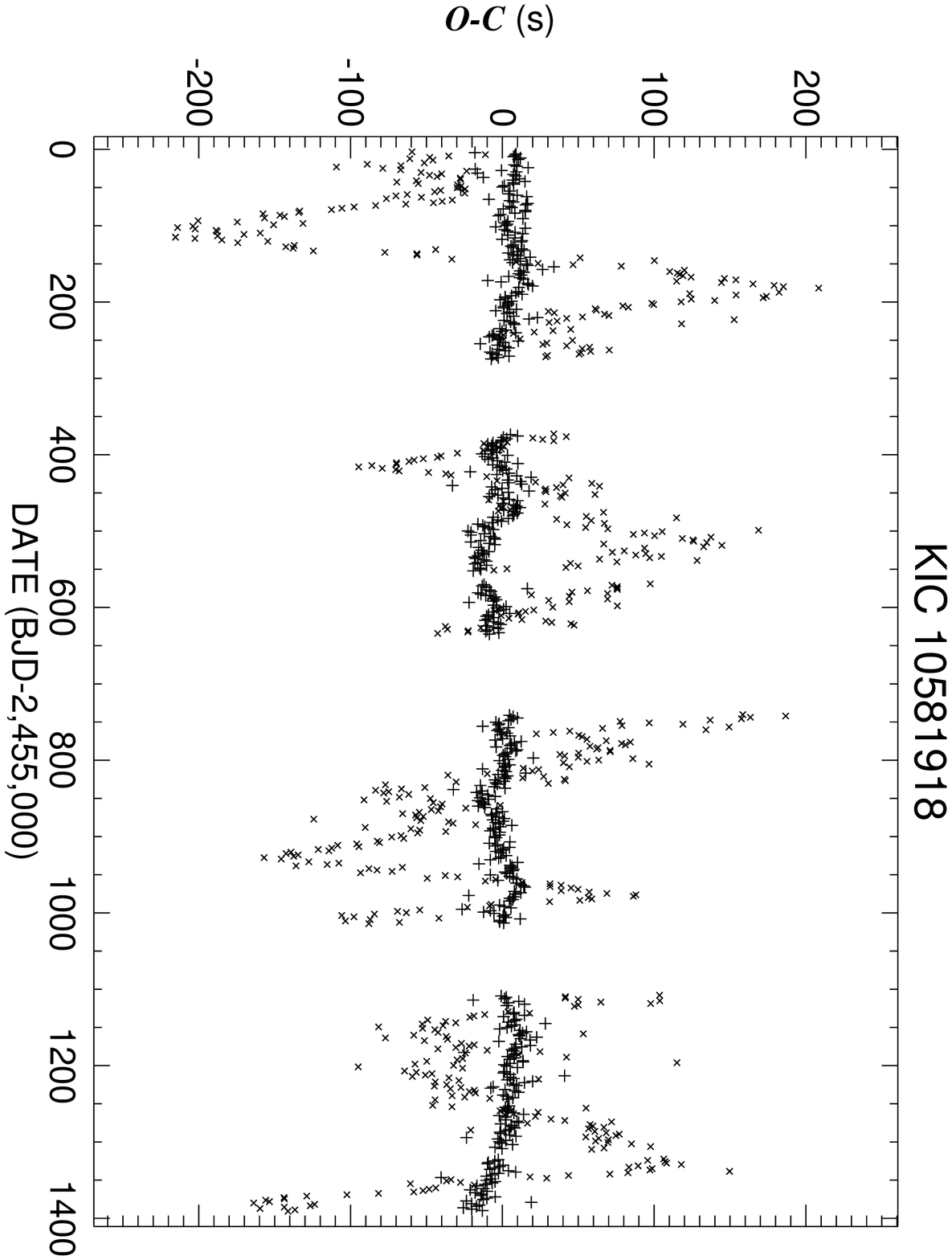}
\figsetgrpnote{The observed minus calculated eclipse times relative to
a linear ephemeris.  The primary and secondary eclipse
times are indicated by $+$ and $\times$ symbols, 
respectively. }
\figsetgrpend

\figsetgrpstart
\figsetgrpnum{1.36}
\figsetgrptitle{r36}
\figsetplot{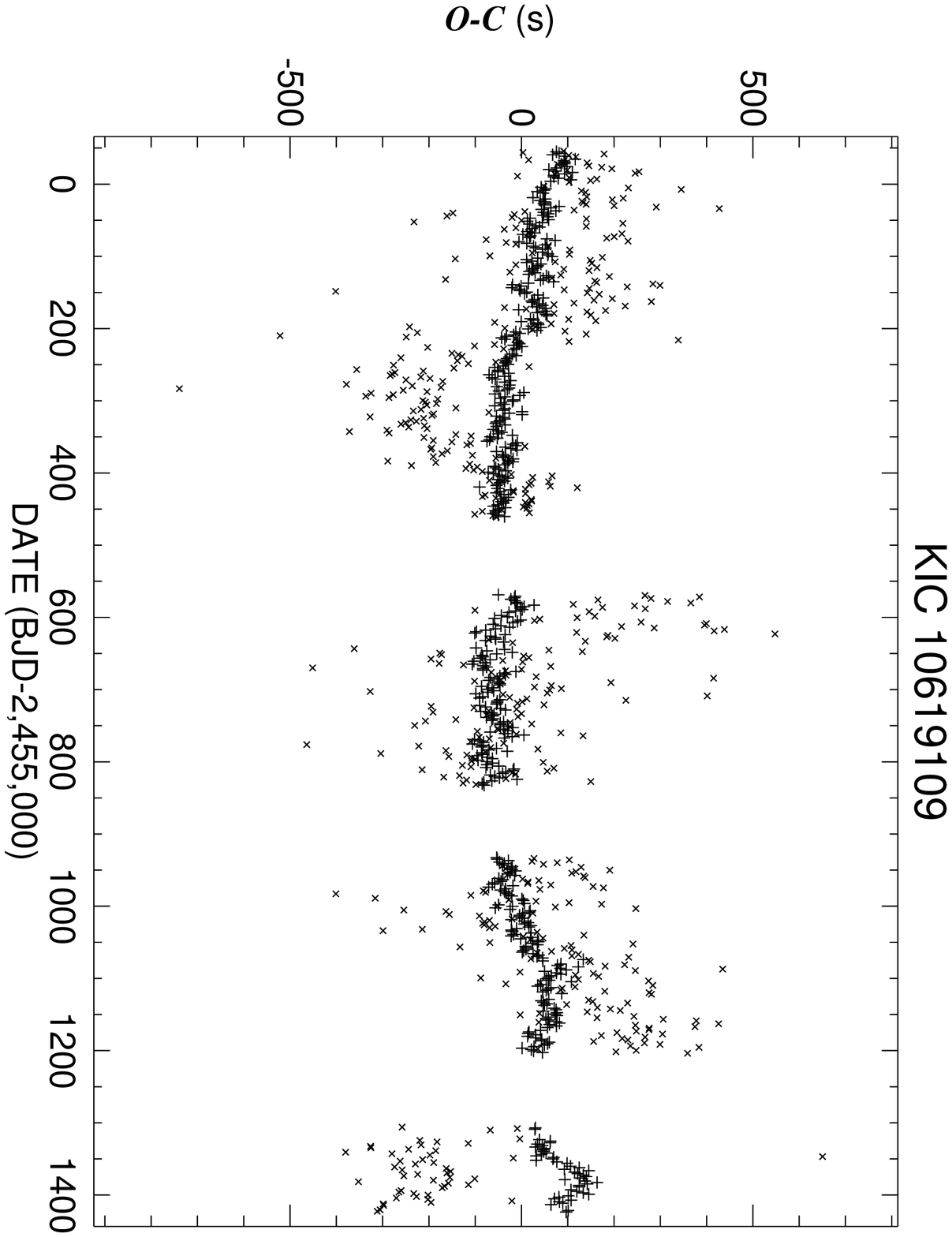}
\figsetgrpnote{The observed minus calculated eclipse times relative to
a linear ephemeris.  The primary and secondary eclipse
times are indicated by $+$ and $\times$ symbols, 
respectively. }
\figsetgrpend

\figsetgrpstart
\figsetgrpnum{1.37}
\figsetgrptitle{r37}
\figsetplot{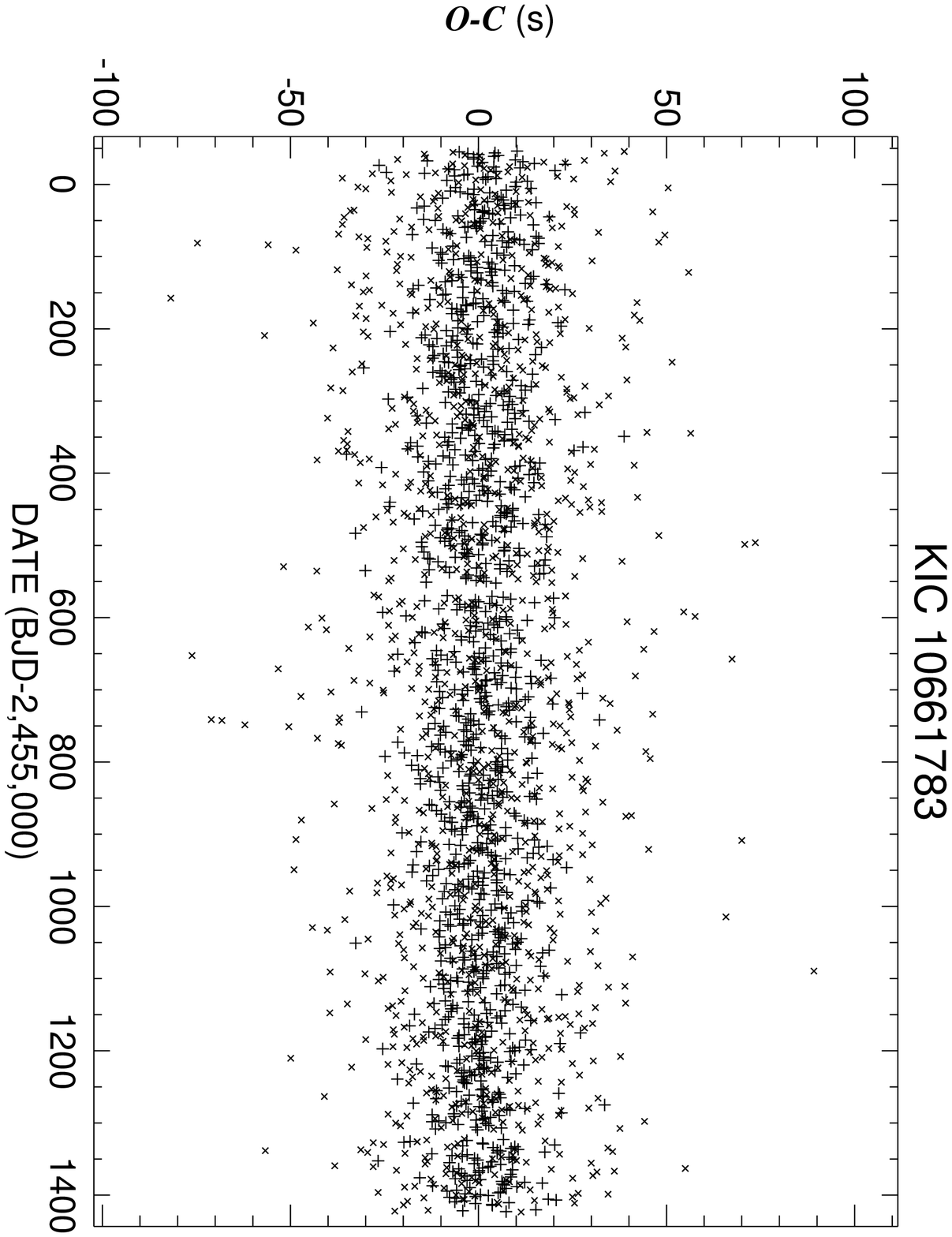}
\figsetgrpnote{The observed minus calculated eclipse times relative to
a linear ephemeris.  The primary and secondary eclipse
times are indicated by $+$ and $\times$ symbols, 
respectively. }
\figsetgrpend

\figsetgrpstart
\figsetgrpnum{1.38}
\figsetgrptitle{r38}
\figsetplot{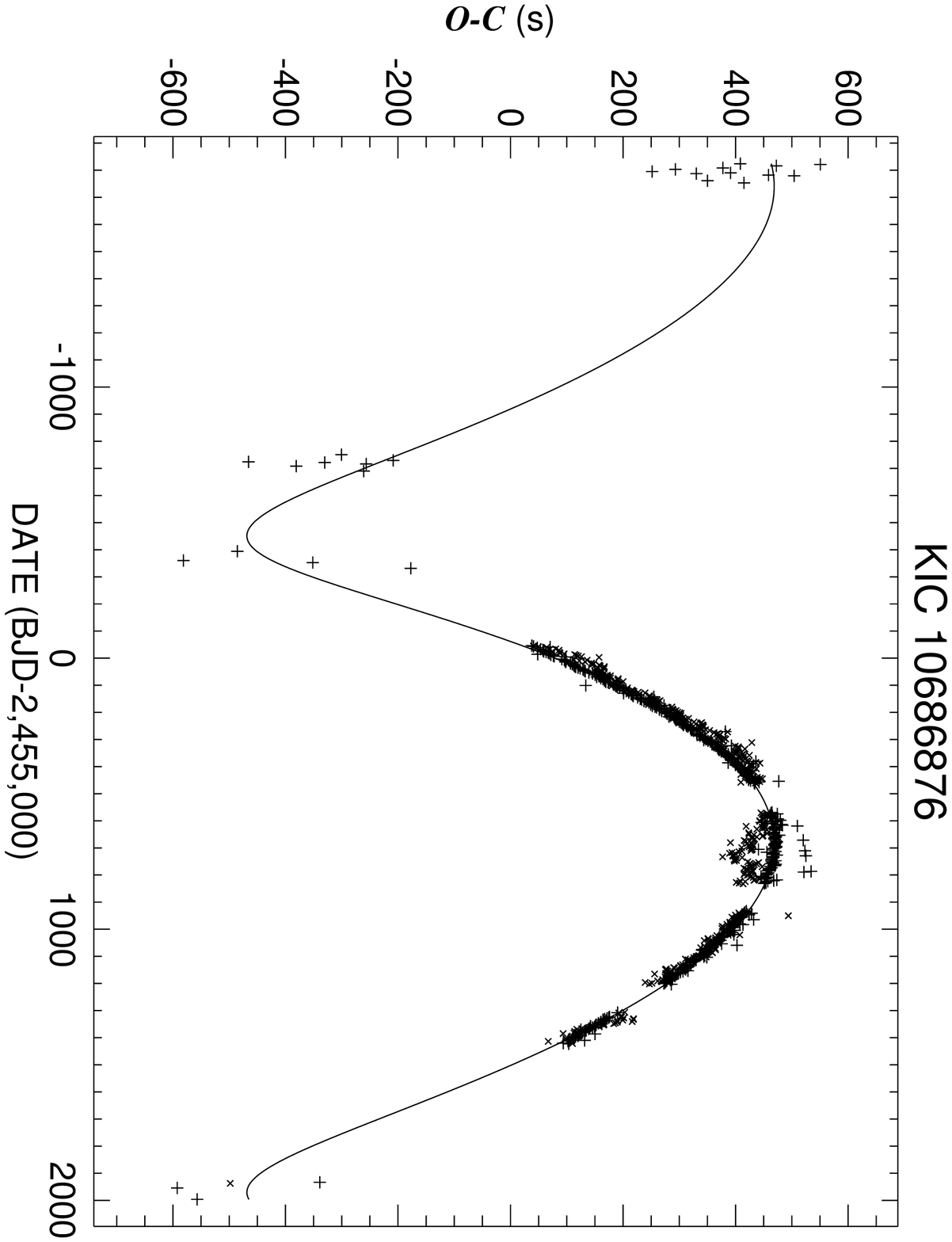}
\figsetgrpnote{The observed minus calculated eclipse times relative to
a linear ephemeris.  The primary and secondary eclipse
times are indicated by $+$ and $\times$ symbols, 
respectively. }
\figsetgrpend

\figsetgrpstart
\figsetgrpnum{1.39}
\figsetgrptitle{r39}
\figsetplot{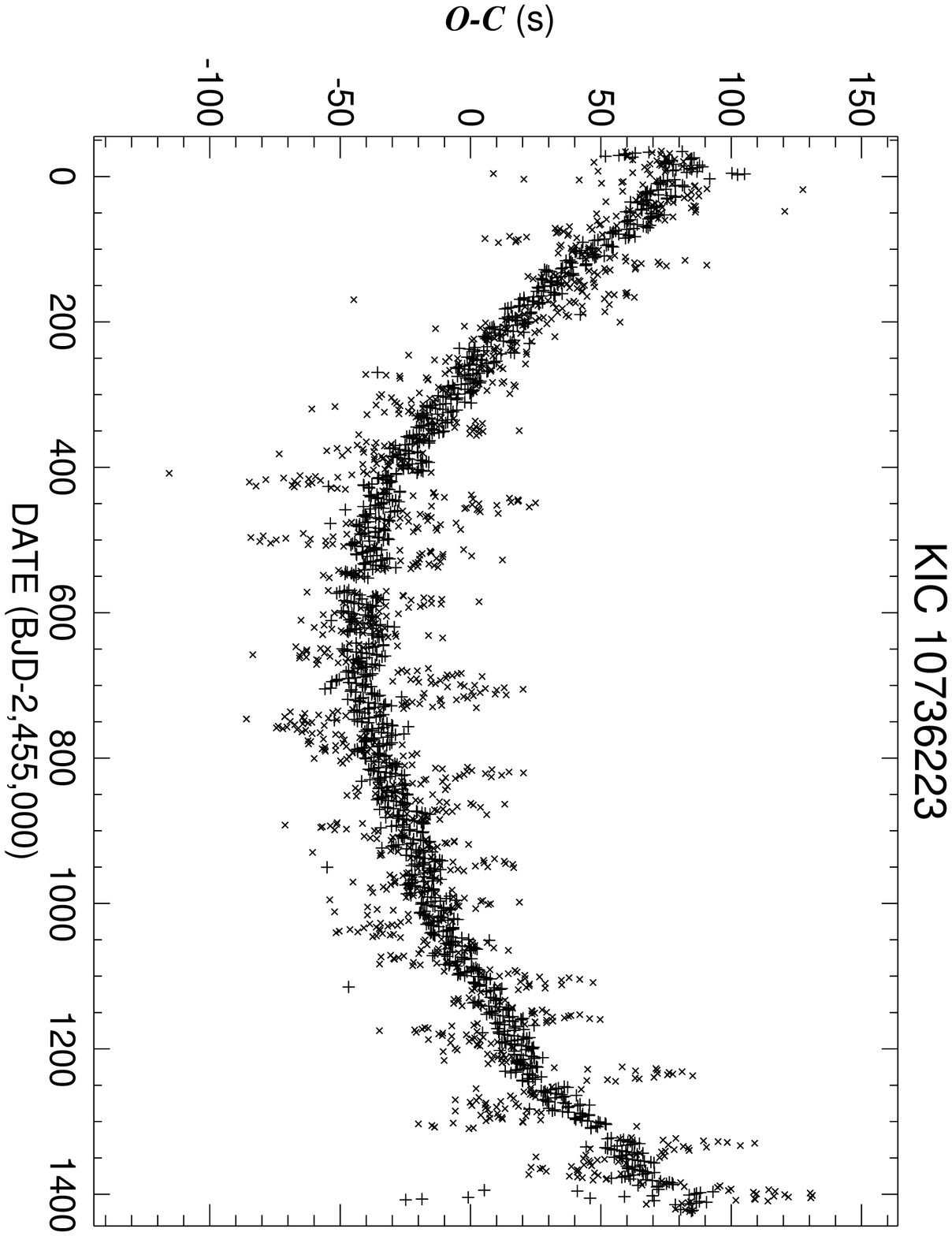}
\figsetgrpnote{The observed minus calculated eclipse times relative to
a linear ephemeris.  The primary and secondary eclipse
times are indicated by $+$ and $\times$ symbols, 
respectively. }
\figsetgrpend

\figsetgrpstart
\figsetgrpnum{1.40}
\figsetgrptitle{r40}
\figsetplot{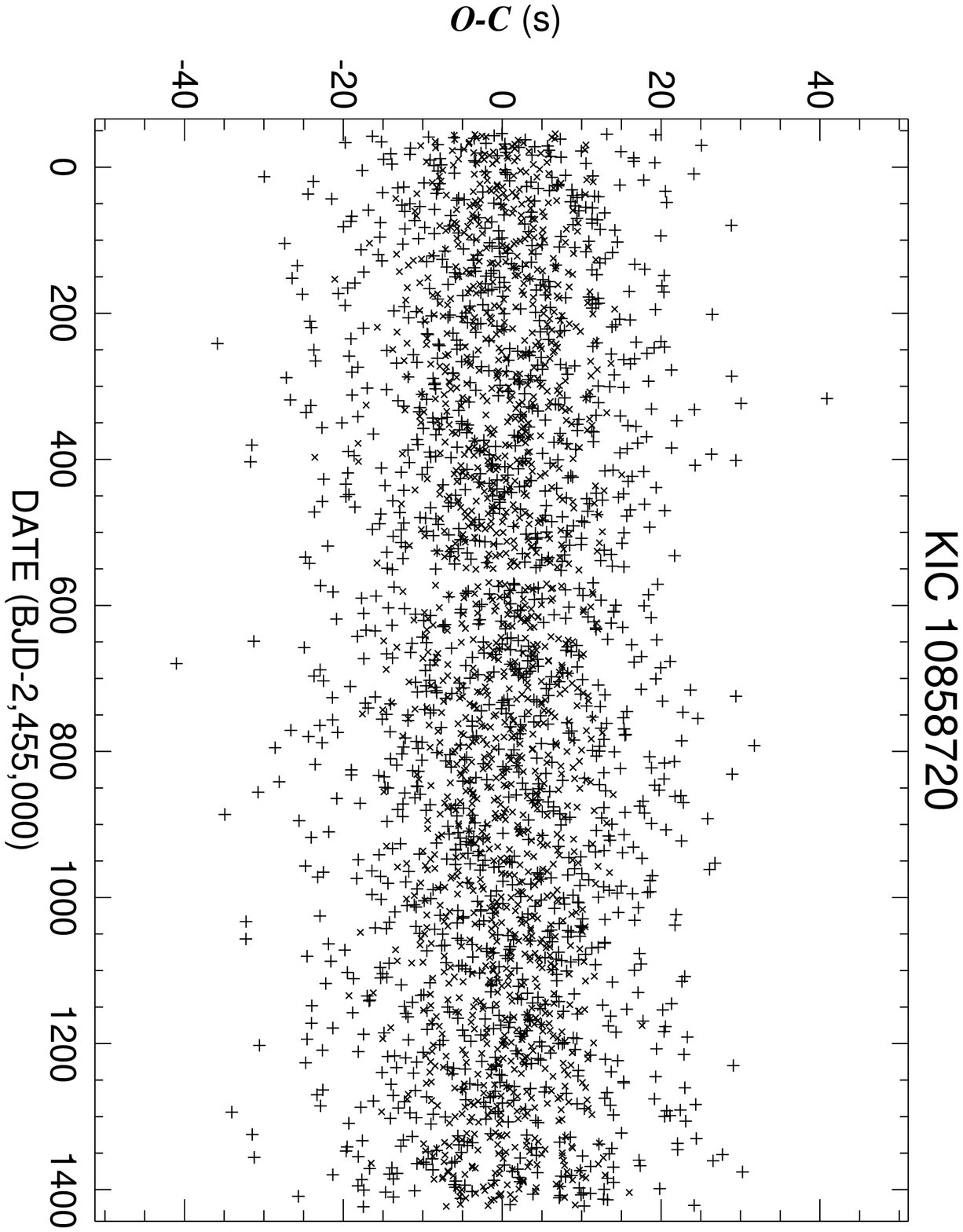}
\figsetgrpnote{The observed minus calculated eclipse times relative to
a linear ephemeris.  The primary and secondary eclipse
times are indicated by $+$ and $\times$ symbols, 
respectively. }
\figsetgrpend

\figsetgrpstart
\figsetgrpnum{1.41}
\figsetgrptitle{r41}
\figsetplot{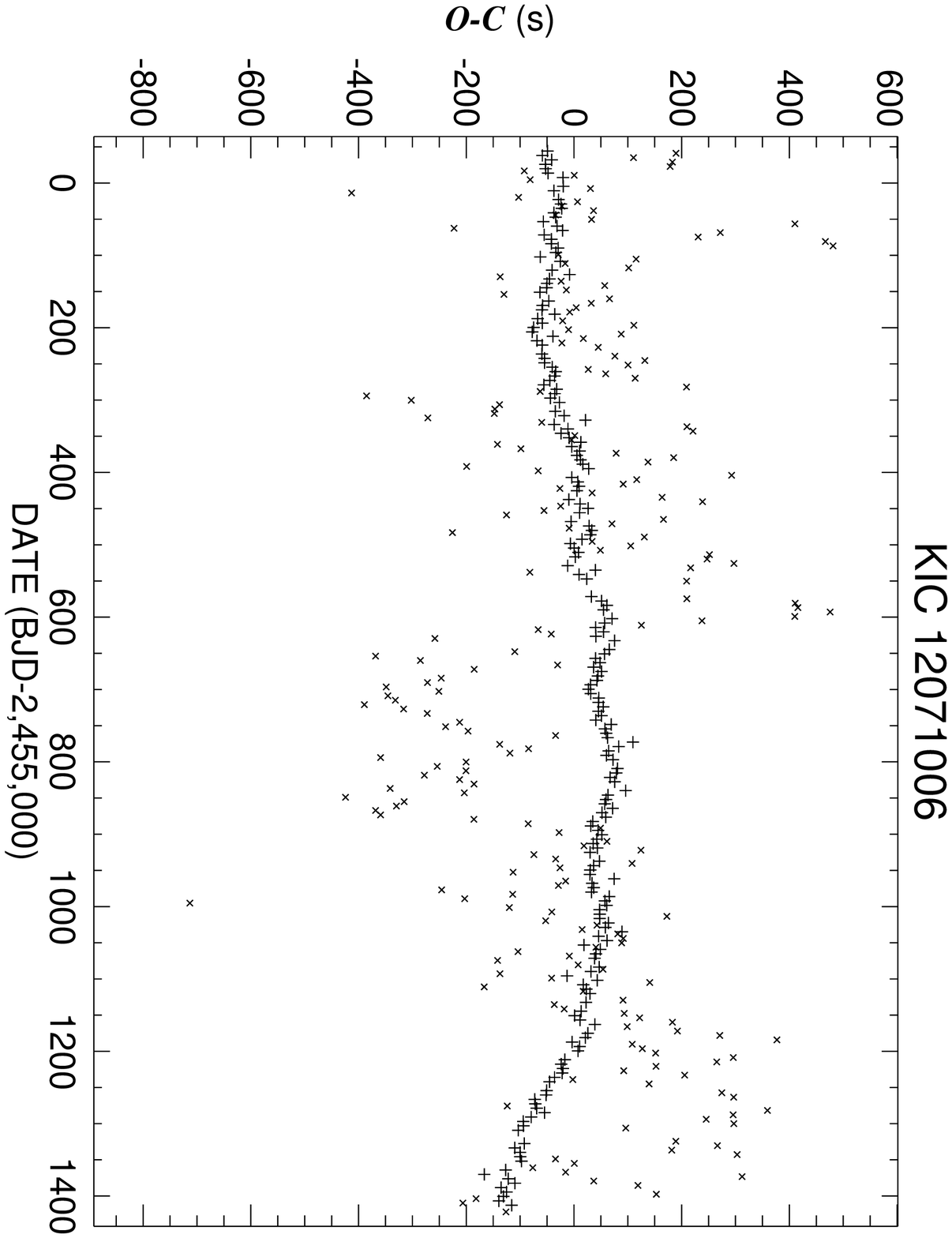}
\figsetgrpnote{The observed minus calculated eclipse times relative to
a linear ephemeris.  The primary and secondary eclipse
times are indicated by $+$ and $\times$ symbols, 
respectively. }
\figsetgrpend

\figsetend



\clearpage
\setcounter{figure}{0}
\renewcommand{\thefigure}{\arabic{figure}.1}
\begin{figure}
\begin{center}
{\includegraphics[angle=90,height=12cm]{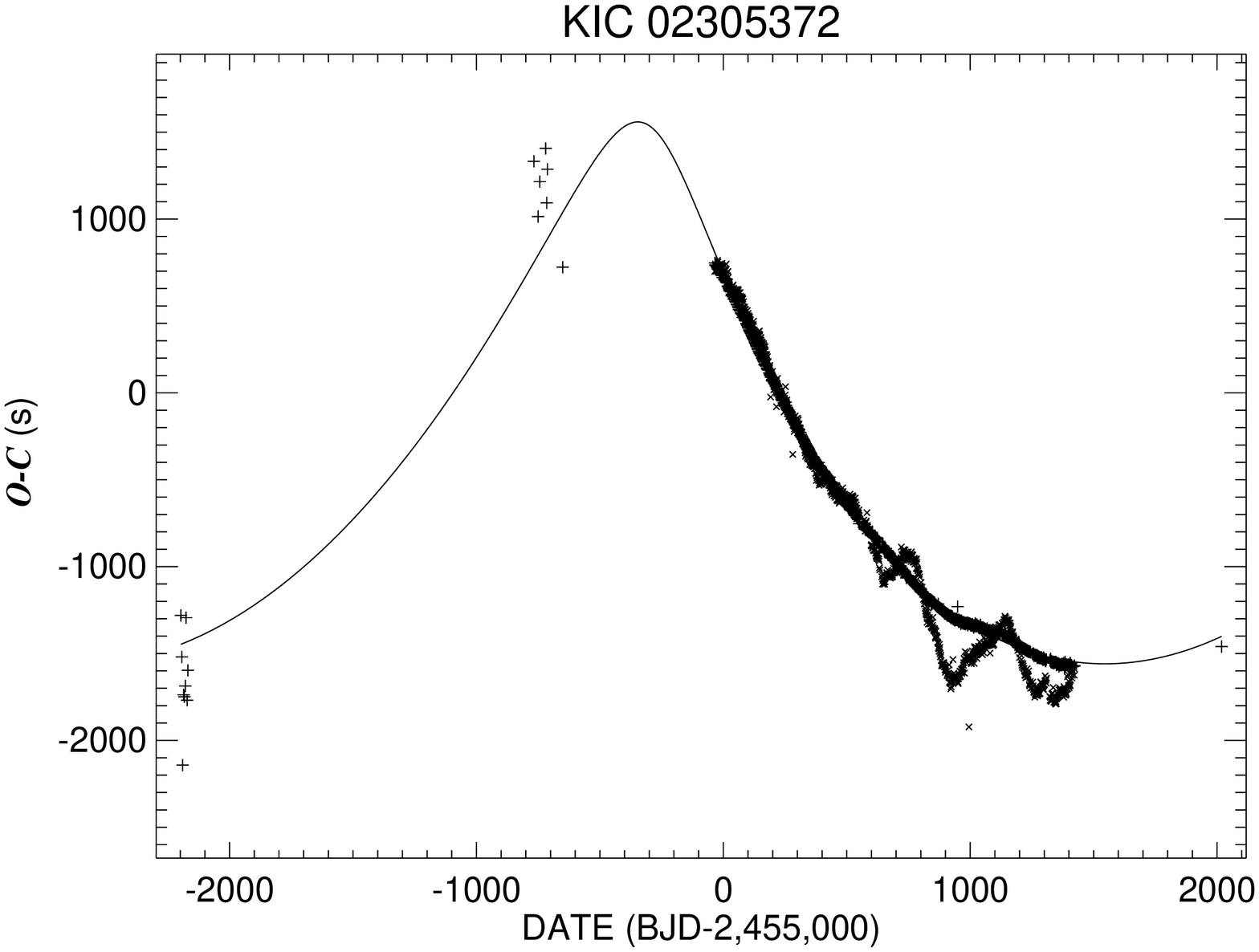}}
\end{center}
\caption{The observed minus calculated eclipse times relative to
a linear ephemeris.  The primary and secondary eclipse
times are indicated by $+$ and $\times$ symbols, 
respectively.}
\end{figure}

\clearpage
\setcounter{figure}{0}
\renewcommand{\thefigure}{\arabic{figure}.2}
\begin{figure}
\begin{center}
{\includegraphics[angle=90,height=12cm]{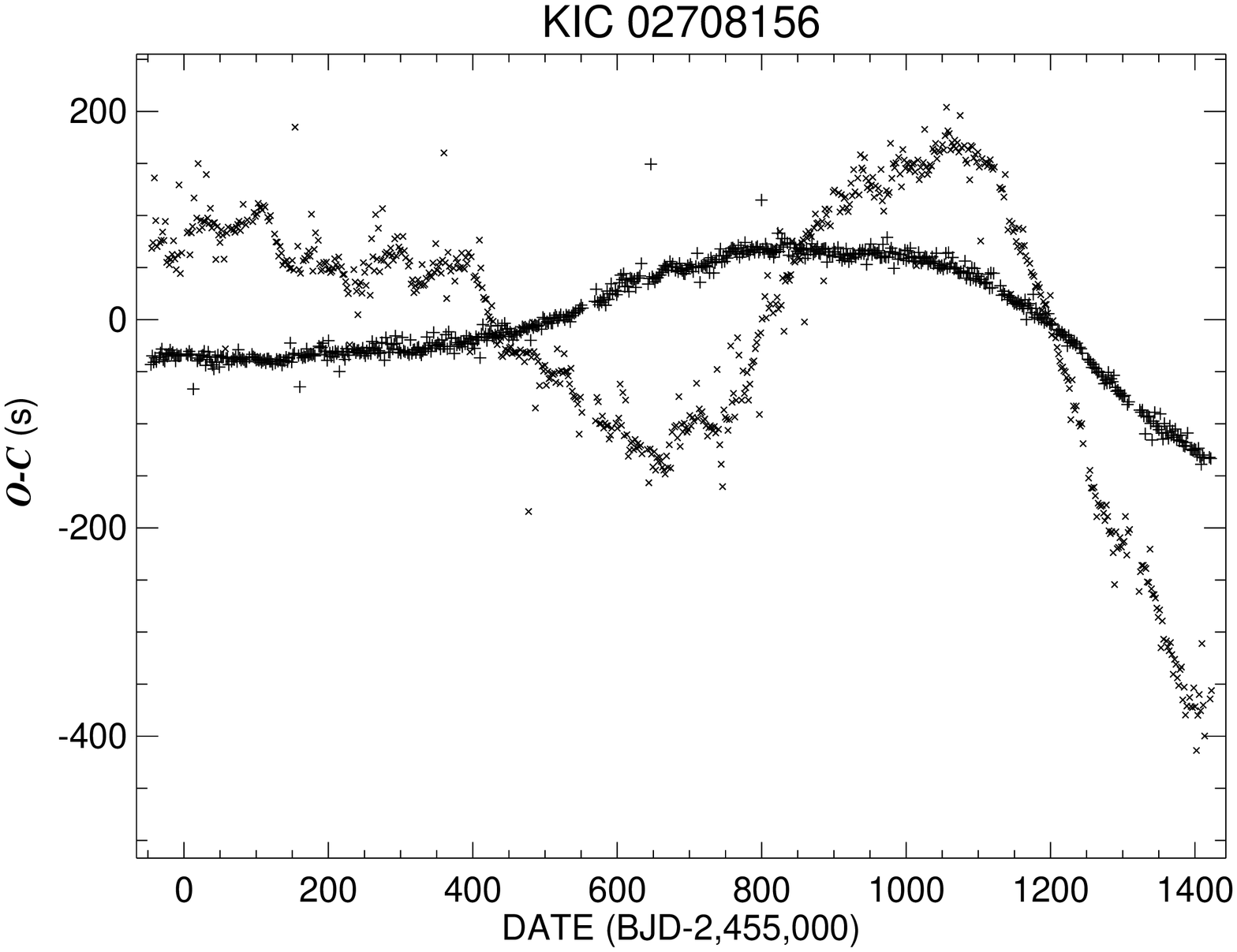}}
\end{center}
\caption{The observed minus calculated eclipse times relative to
a linear ephemeris.  The primary and secondary eclipse
times are indicated by $+$ and $\times$ symbols, 
respectively.}
\end{figure}

\clearpage
\setcounter{figure}{0}
\renewcommand{\thefigure}{\arabic{figure}.3}
\begin{figure}
\begin{center}
{\includegraphics[angle=90,height=12cm]{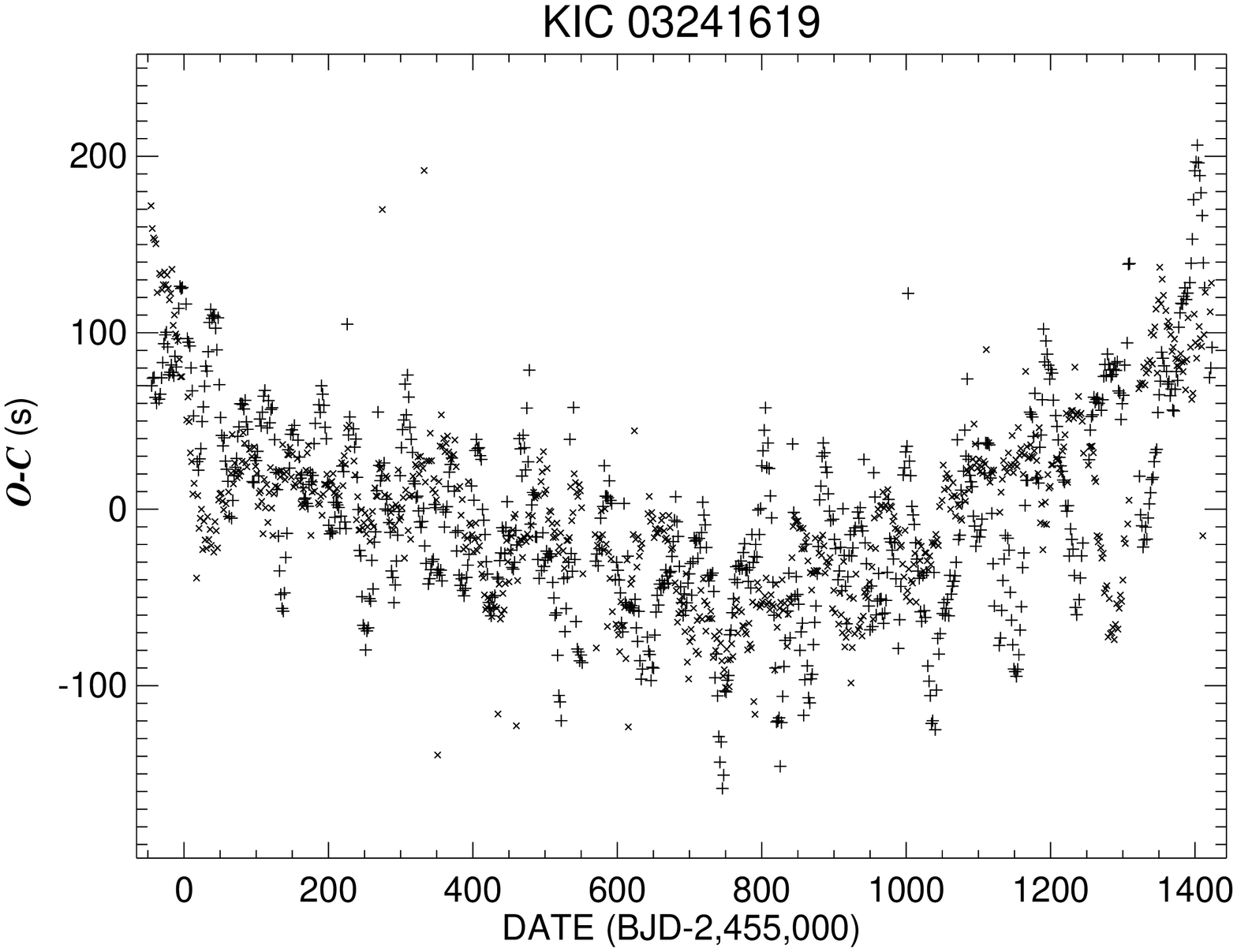}}
\end{center}
\caption{The observed minus calculated eclipse times relative to
a linear ephemeris.  The primary and secondary eclipse
times are indicated by $+$ and $\times$ symbols, 
respectively.}
\end{figure}

\clearpage
\setcounter{figure}{0}
\renewcommand{\thefigure}{\arabic{figure}.4}
\begin{figure}
\begin{center}
{\includegraphics[angle=90,height=12cm]{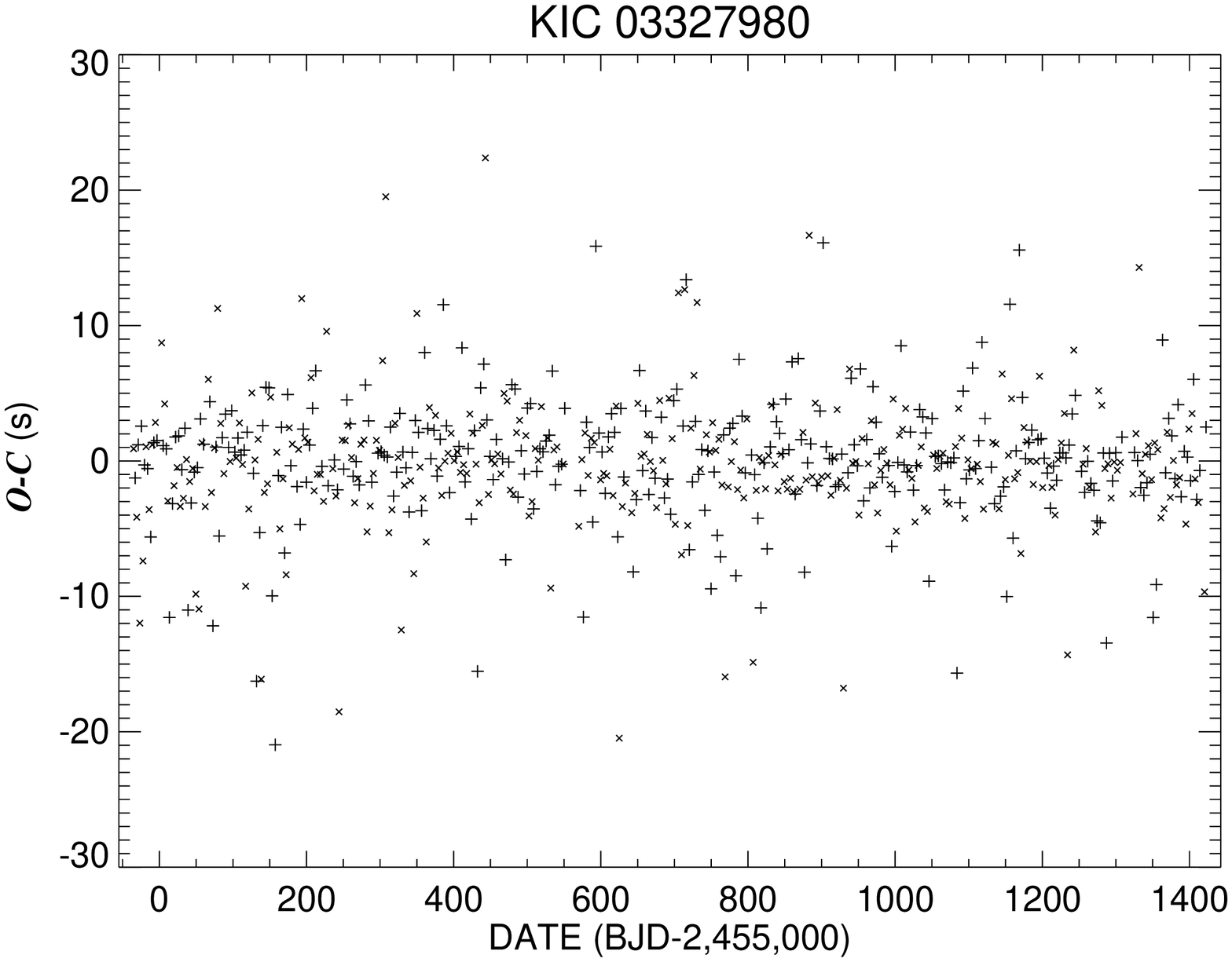}}
\end{center}
\caption{The observed minus calculated eclipse times relative to
a linear ephemeris.  The primary and secondary eclipse
times are indicated by $+$ and $\times$ symbols, 
respectively.}
\end{figure}

\clearpage
\setcounter{figure}{0}
\renewcommand{\thefigure}{\arabic{figure}.5}
\begin{figure}
\begin{center}
{\includegraphics[angle=90,height=12cm]{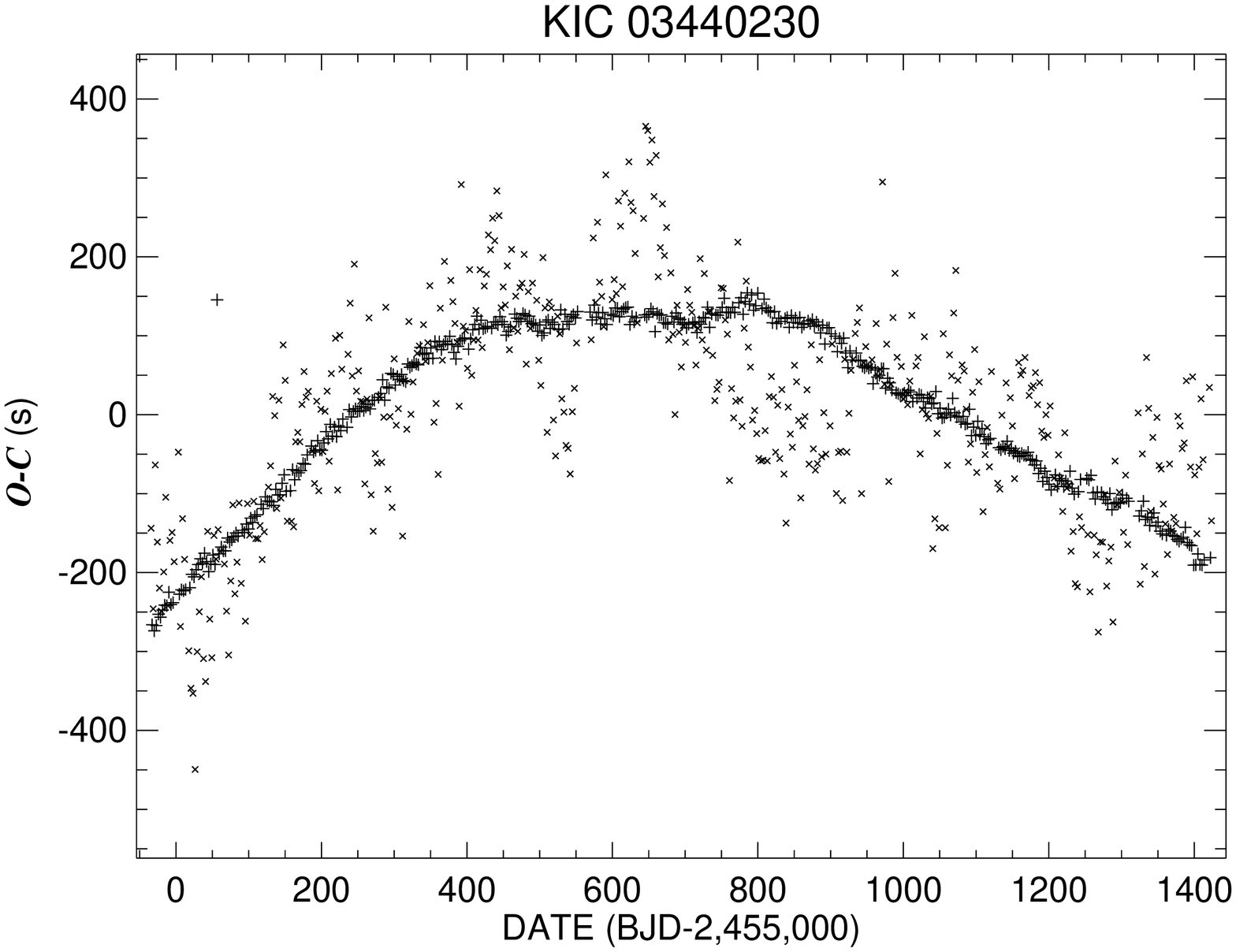}}
\end{center}
\caption{The observed minus calculated eclipse times relative to
a linear ephemeris.  The primary and secondary eclipse
times are indicated by $+$ and $\times$ symbols, 
respectively.}
\end{figure}

\clearpage
\setcounter{figure}{0}
\renewcommand{\thefigure}{\arabic{figure}.6}
\begin{figure}
\begin{center}
{\includegraphics[angle=90,height=12cm]{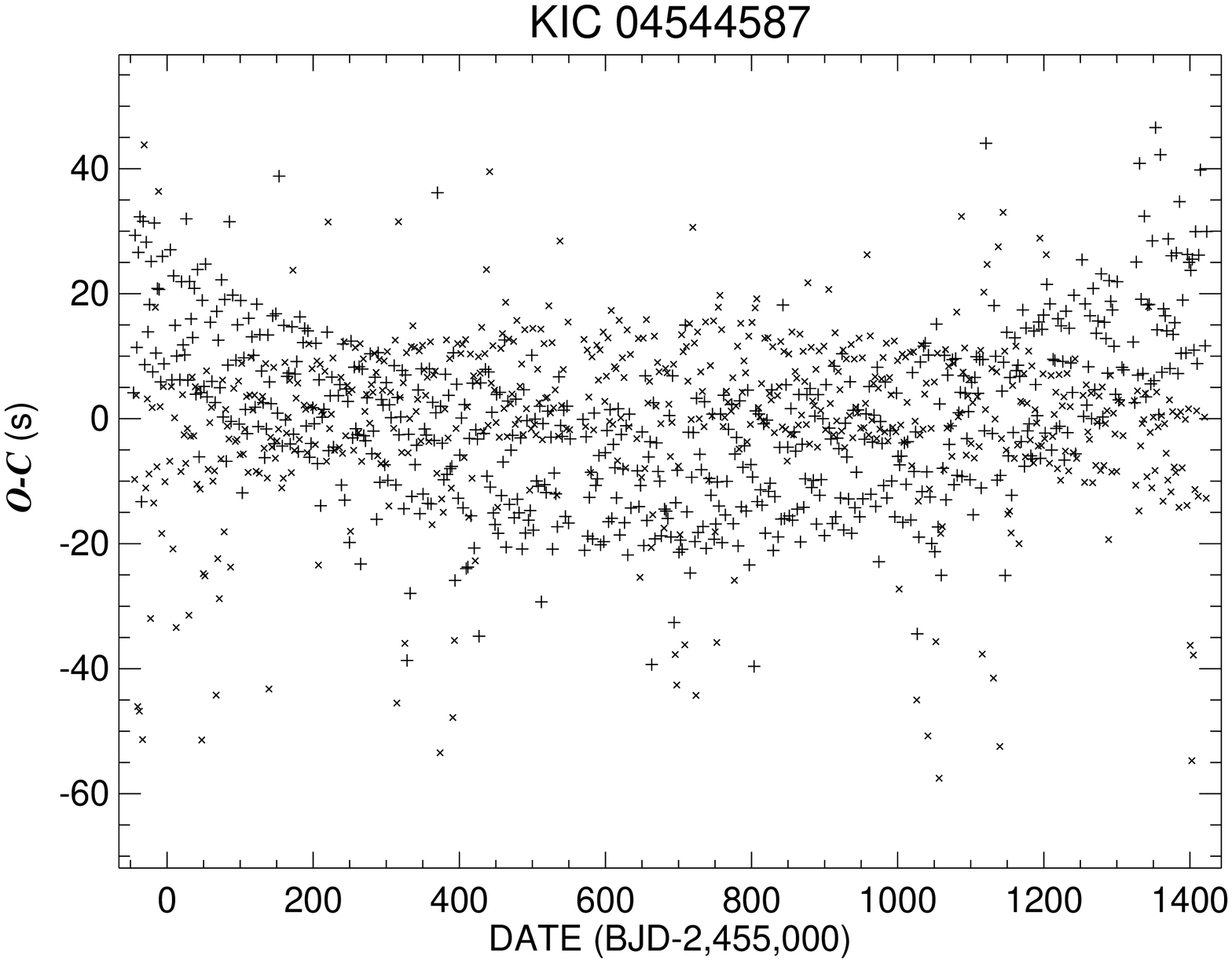}}
\end{center}
\caption{The observed minus calculated eclipse times relative to
a linear ephemeris.  The primary and secondary eclipse
times are indicated by $+$ and $\times$ symbols, 
respectively.}
\end{figure}

\clearpage
\setcounter{figure}{0}
\renewcommand{\thefigure}{\arabic{figure}.7}
\begin{figure}
\begin{center}
{\includegraphics[angle=90,height=12cm]{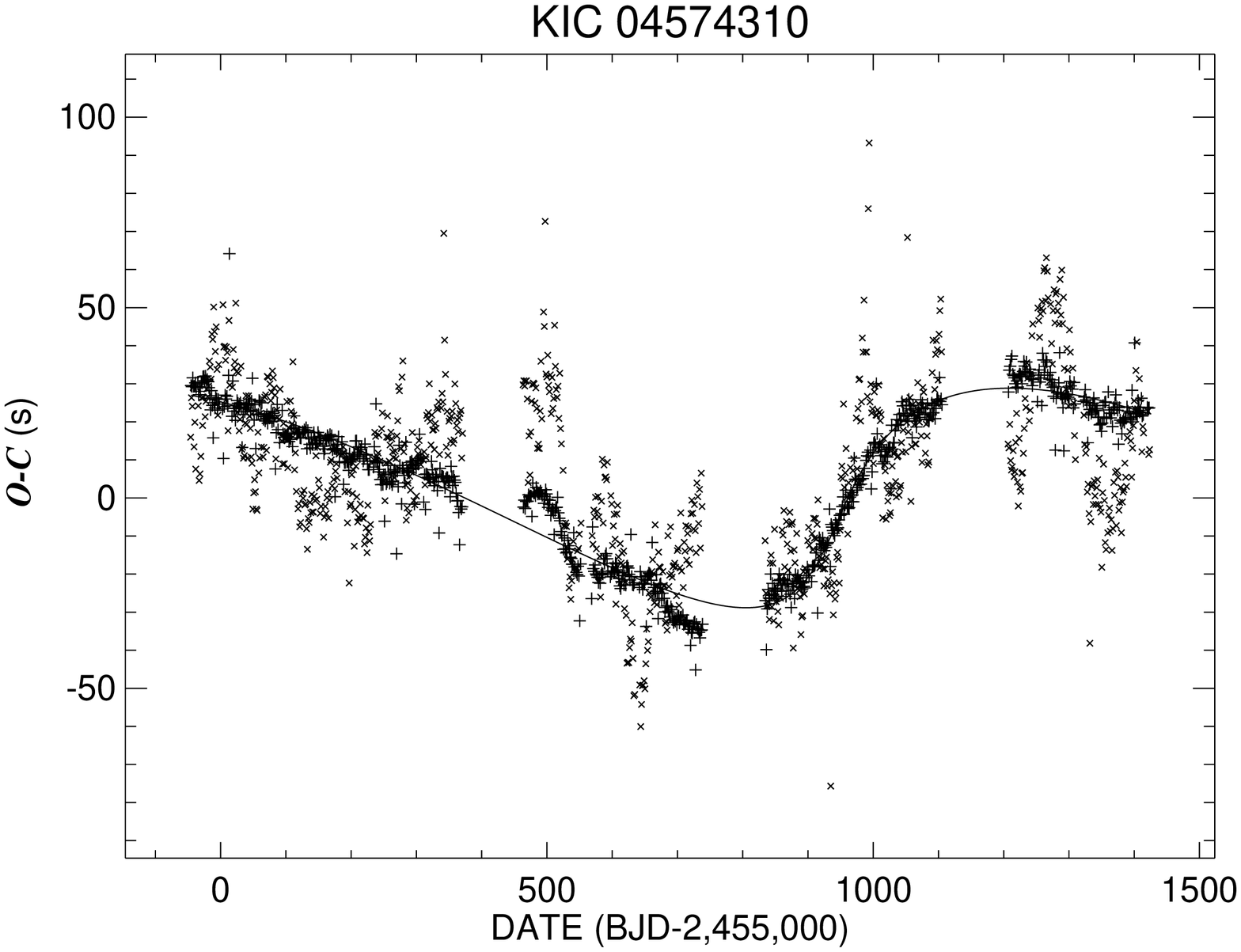}}
\end{center}
\caption{The observed minus calculated eclipse times relative to
a linear ephemeris.  The primary and secondary eclipse
times are indicated by $+$ and $\times$ symbols, 
respectively.}
\end{figure}

\clearpage
\setcounter{figure}{0}
\renewcommand{\thefigure}{\arabic{figure}.8}
\begin{figure}
\begin{center}
{\includegraphics[angle=90,height=12cm]{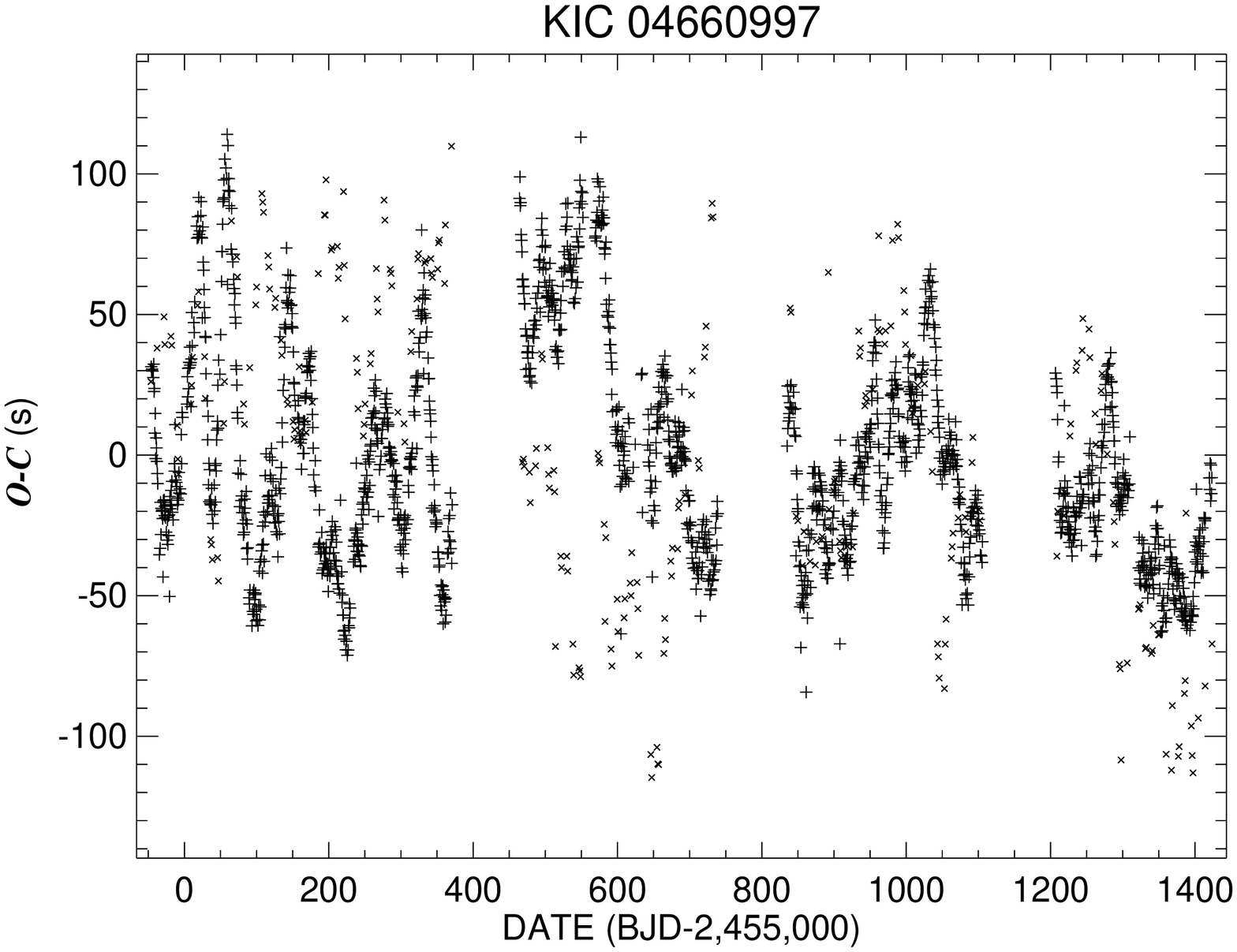}}
\end{center}
\caption{The observed minus calculated eclipse times relative to
a linear ephemeris.  The primary and secondary eclipse
times are indicated by $+$ and $\times$ symbols, 
respectively.}
\end{figure}

\clearpage
\setcounter{figure}{0}
\renewcommand{\thefigure}{\arabic{figure}.9}
\begin{figure}
\begin{center}
{\includegraphics[angle=90,height=12cm]{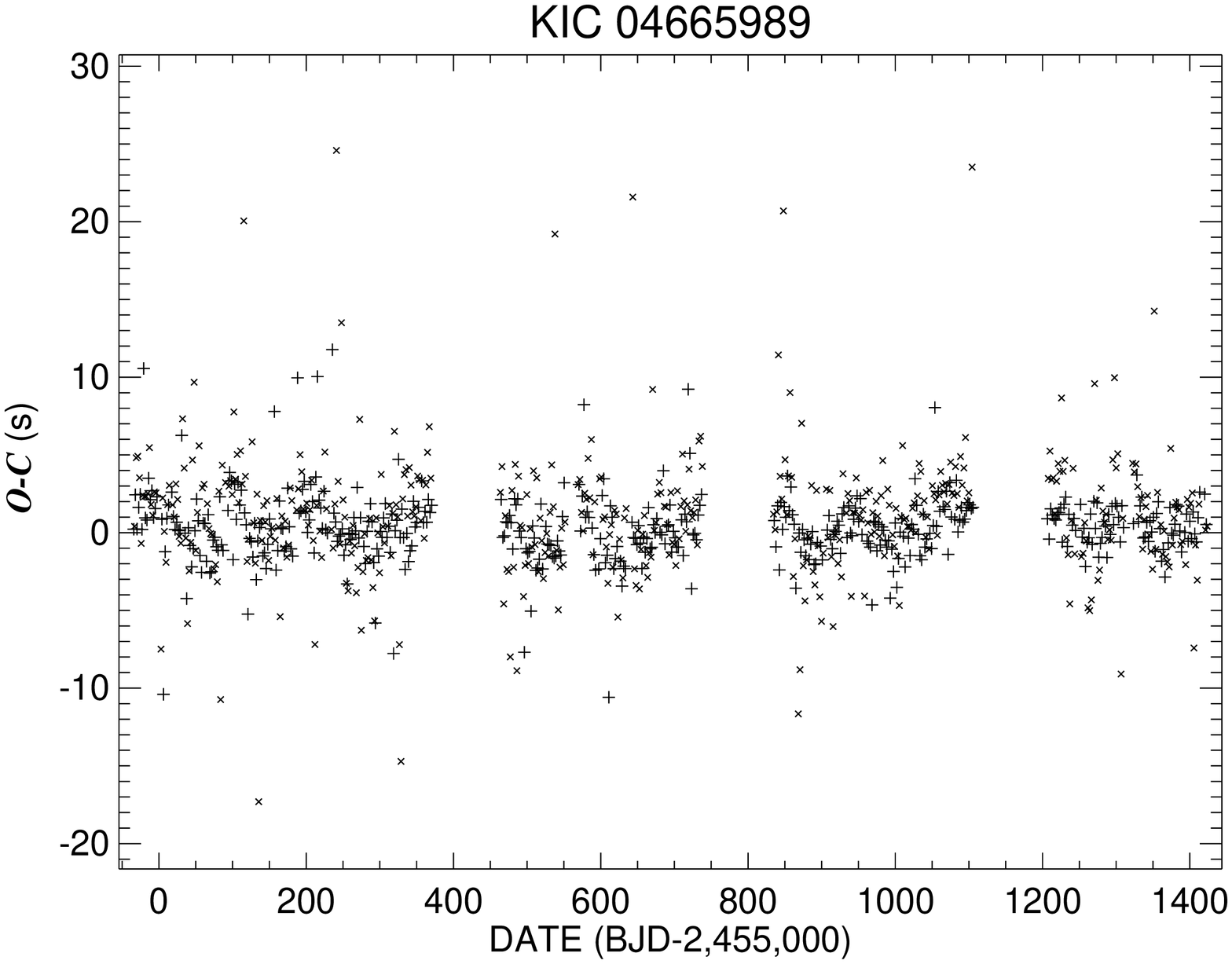}}
\end{center}
\caption{The observed minus calculated eclipse times relative to
a linear ephemeris.  The primary and secondary eclipse
times are indicated by $+$ and $\times$ symbols, 
respectively.}
\end{figure}

\clearpage
\setcounter{figure}{0}
\renewcommand{\thefigure}{\arabic{figure}.10}
\begin{figure}
\begin{center}
{\includegraphics[angle=90,height=12cm]{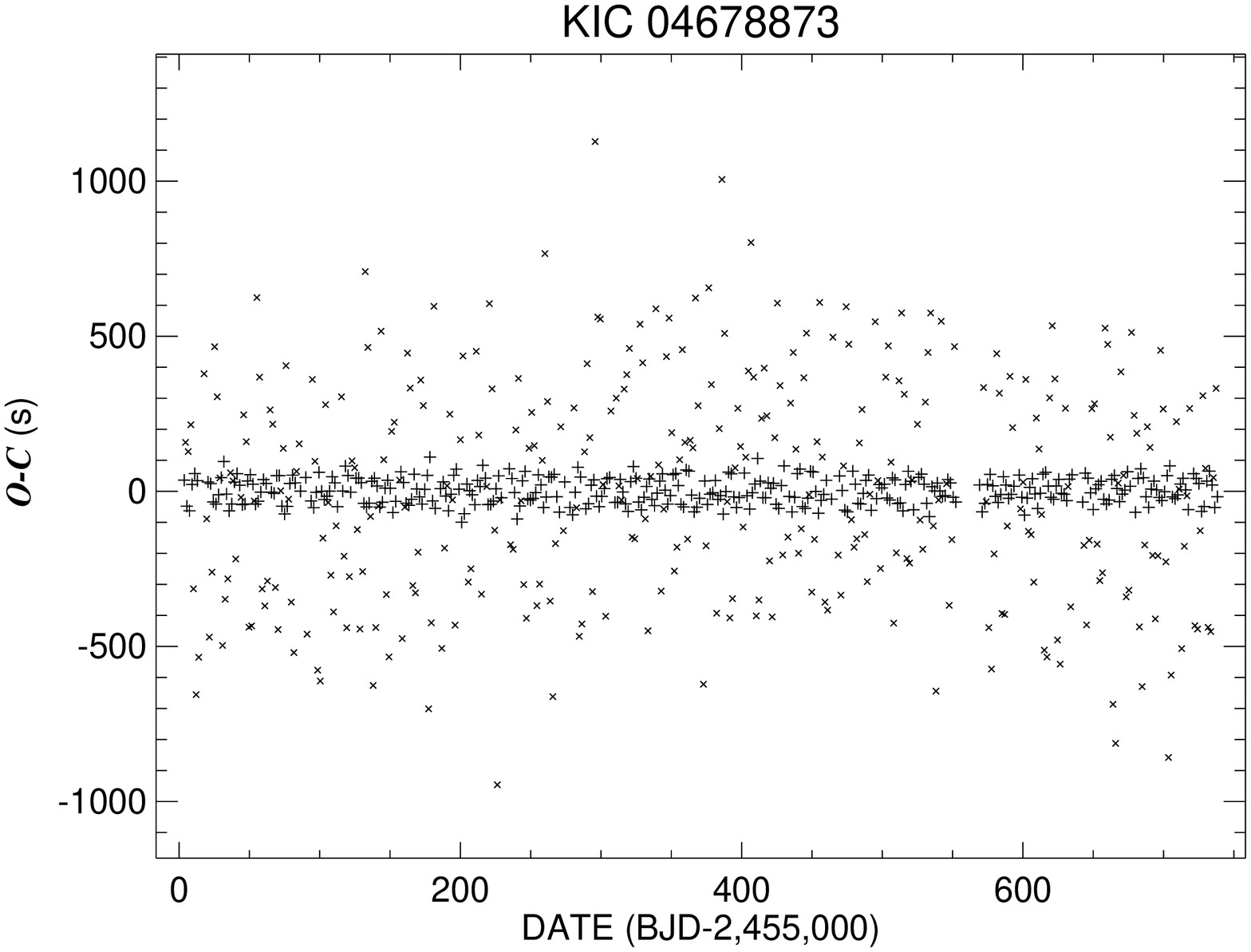}}
\end{center}
\caption{The observed minus calculated eclipse times relative to
a linear ephemeris.  The primary and secondary eclipse
times are indicated by $+$ and $\times$ symbols, 
respectively.}
\end{figure}

\clearpage
\setcounter{figure}{0}
\renewcommand{\thefigure}{\arabic{figure}.11}
\begin{figure}
\begin{center}
{\includegraphics[angle=90,height=12cm]{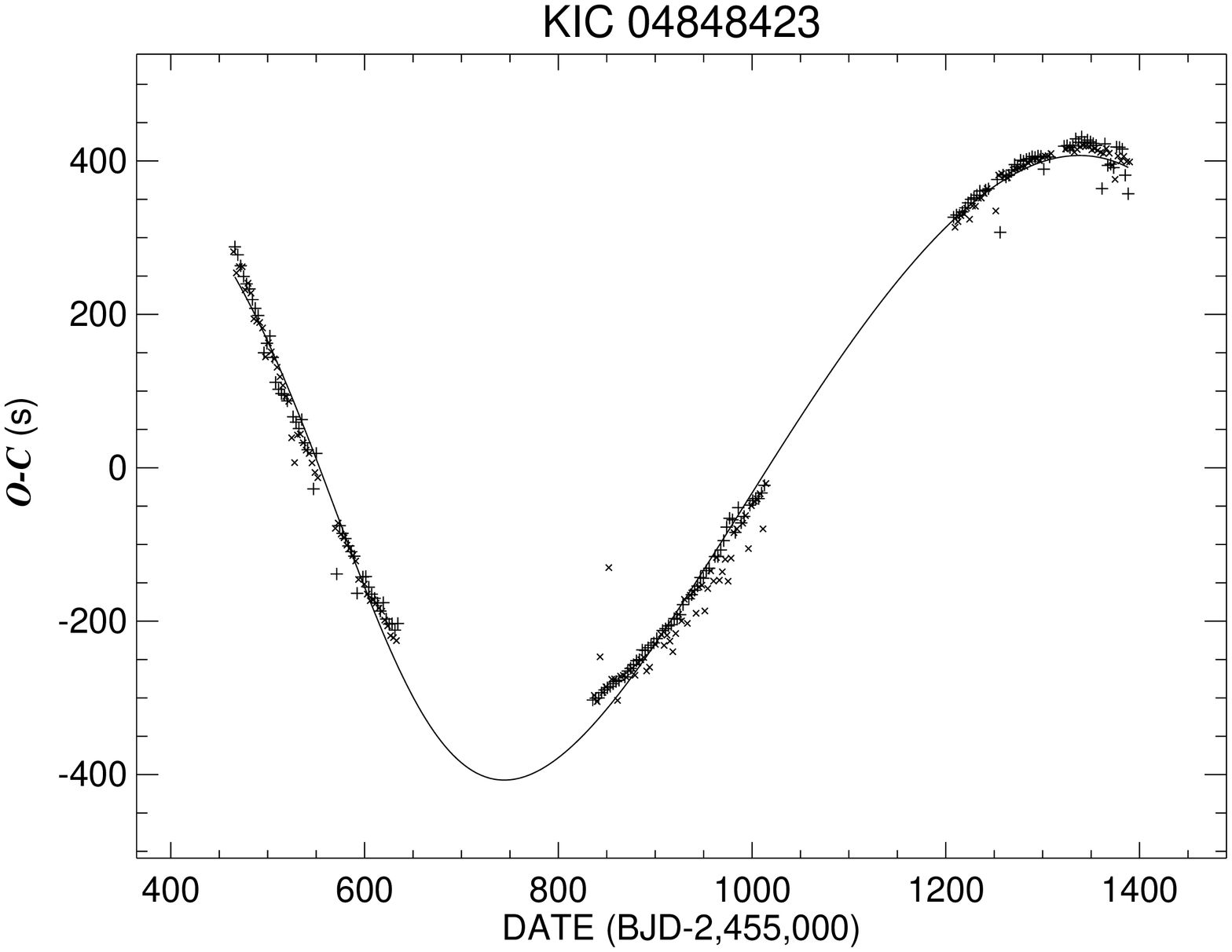}}
\end{center}
\caption{The observed minus calculated eclipse times relative to
a linear ephemeris.  The primary and secondary eclipse
times are indicated by $+$ and $\times$ symbols, 
respectively.}
\end{figure}

\clearpage
\setcounter{figure}{0}
\renewcommand{\thefigure}{\arabic{figure}.12}
\begin{figure}
\begin{center}
{\includegraphics[angle=90,height=12cm]{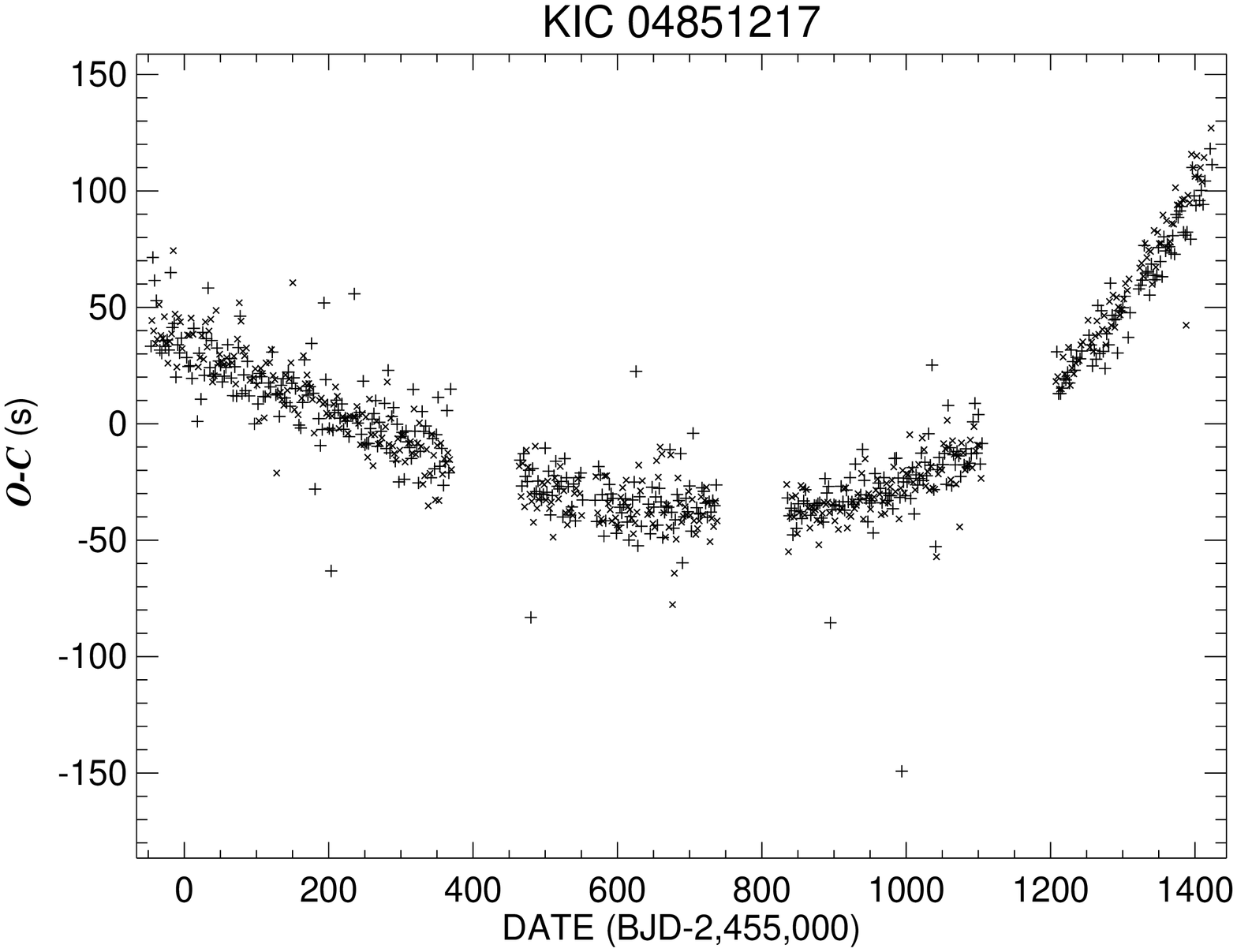}}
\end{center}
\caption{The observed minus calculated eclipse times relative to
a linear ephemeris.  The primary and secondary eclipse
times are indicated by $+$ and $\times$ symbols, 
respectively.}
\end{figure}

\clearpage
\setcounter{figure}{0}
\renewcommand{\thefigure}{\arabic{figure}.13}
\begin{figure}
\begin{center}
{\includegraphics[angle=90,height=12cm]{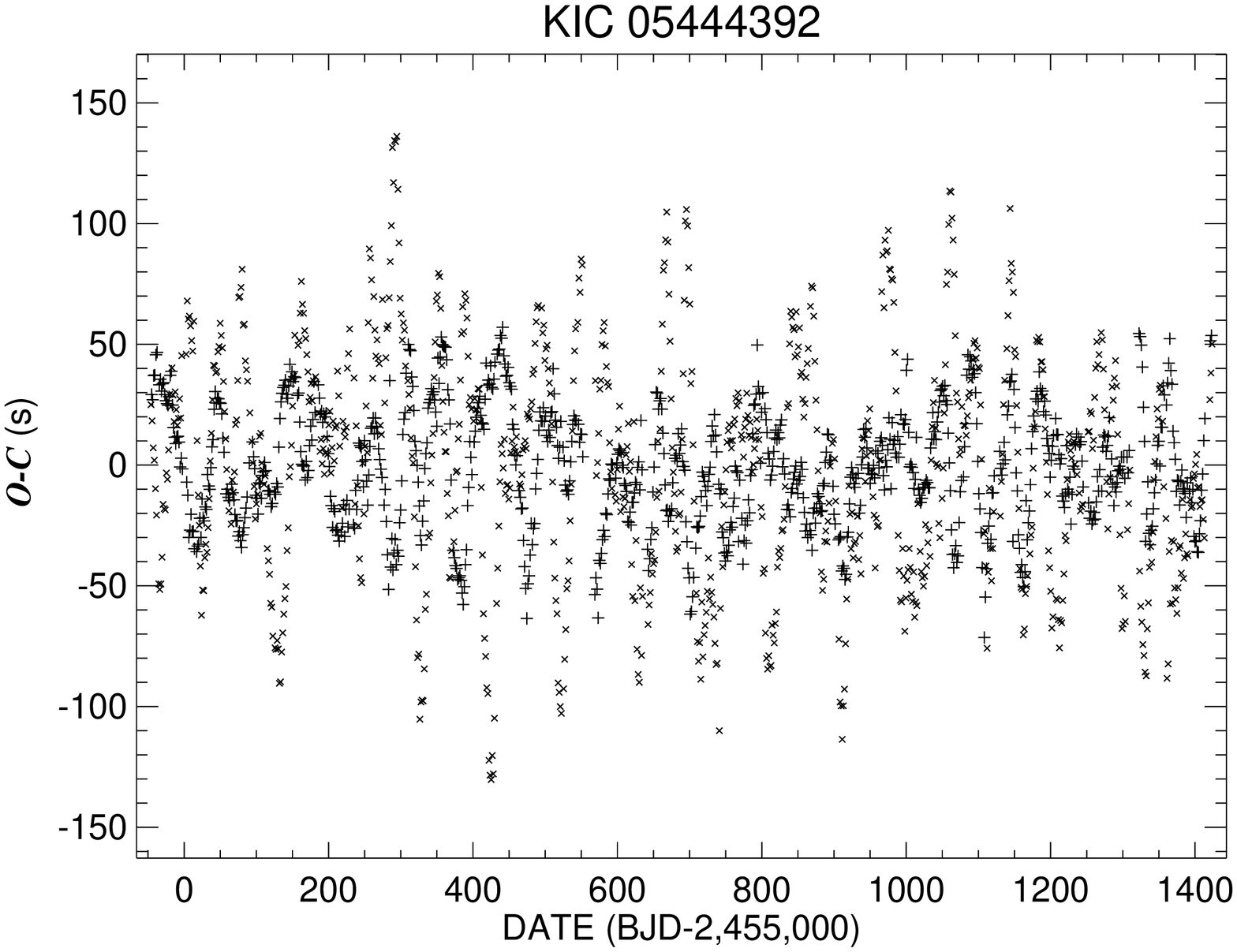}}
\end{center}
\caption{The observed minus calculated eclipse times relative to
a linear ephemeris.  The primary and secondary eclipse
times are indicated by $+$ and $\times$ symbols, 
respectively.}
\end{figure}

\clearpage
\setcounter{figure}{0}
\renewcommand{\thefigure}{\arabic{figure}.14}
\begin{figure}
\begin{center}
{\includegraphics[angle=90,height=12cm]{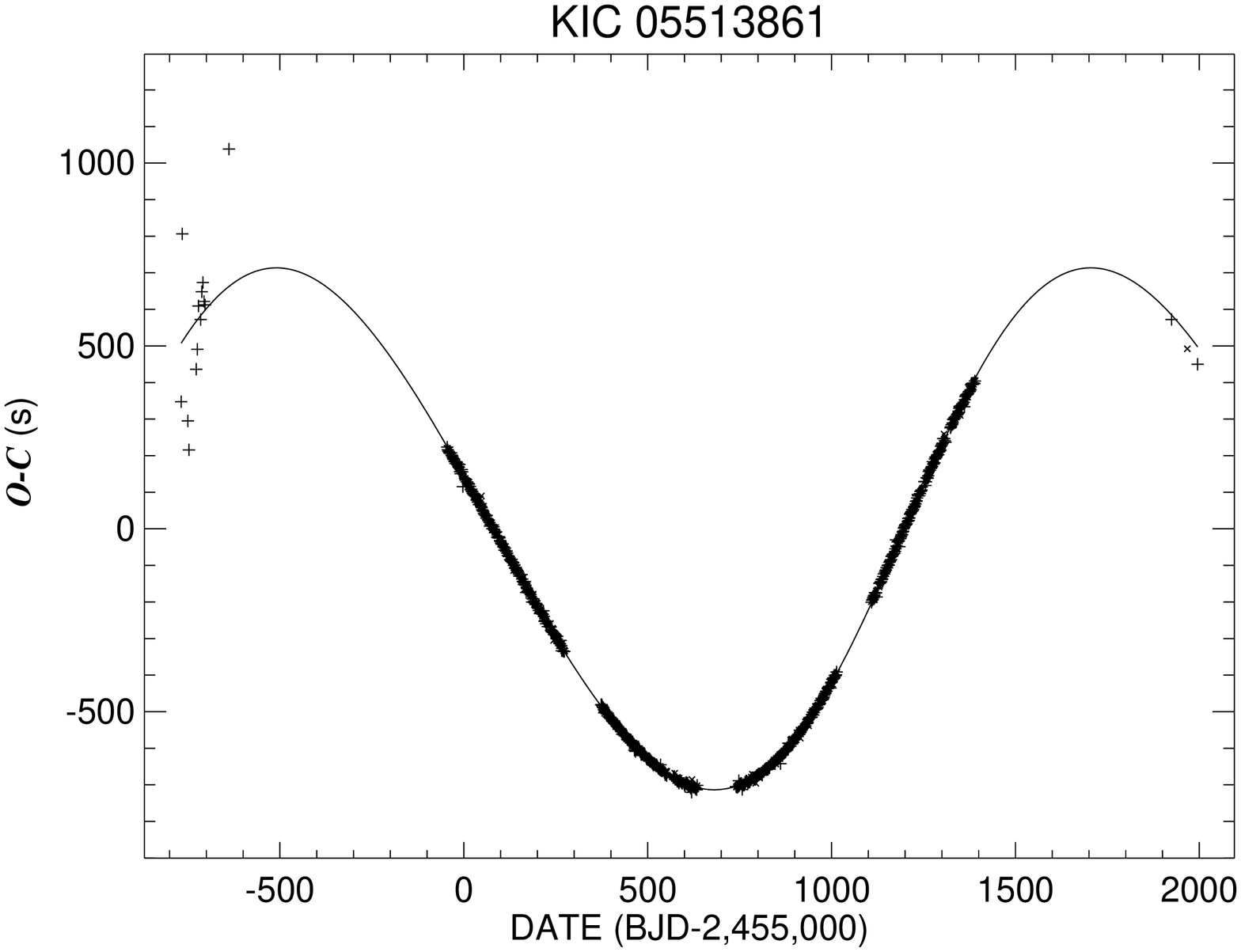}}
\end{center}
\caption{The observed minus calculated eclipse times relative to
a linear ephemeris.  The primary and secondary eclipse
times are indicated by $+$ and $\times$ symbols, 
respectively.}
\end{figure}

\clearpage
\setcounter{figure}{0}
\renewcommand{\thefigure}{\arabic{figure}.15}
\begin{figure}
\begin{center}
{\includegraphics[angle=90,height=12cm]{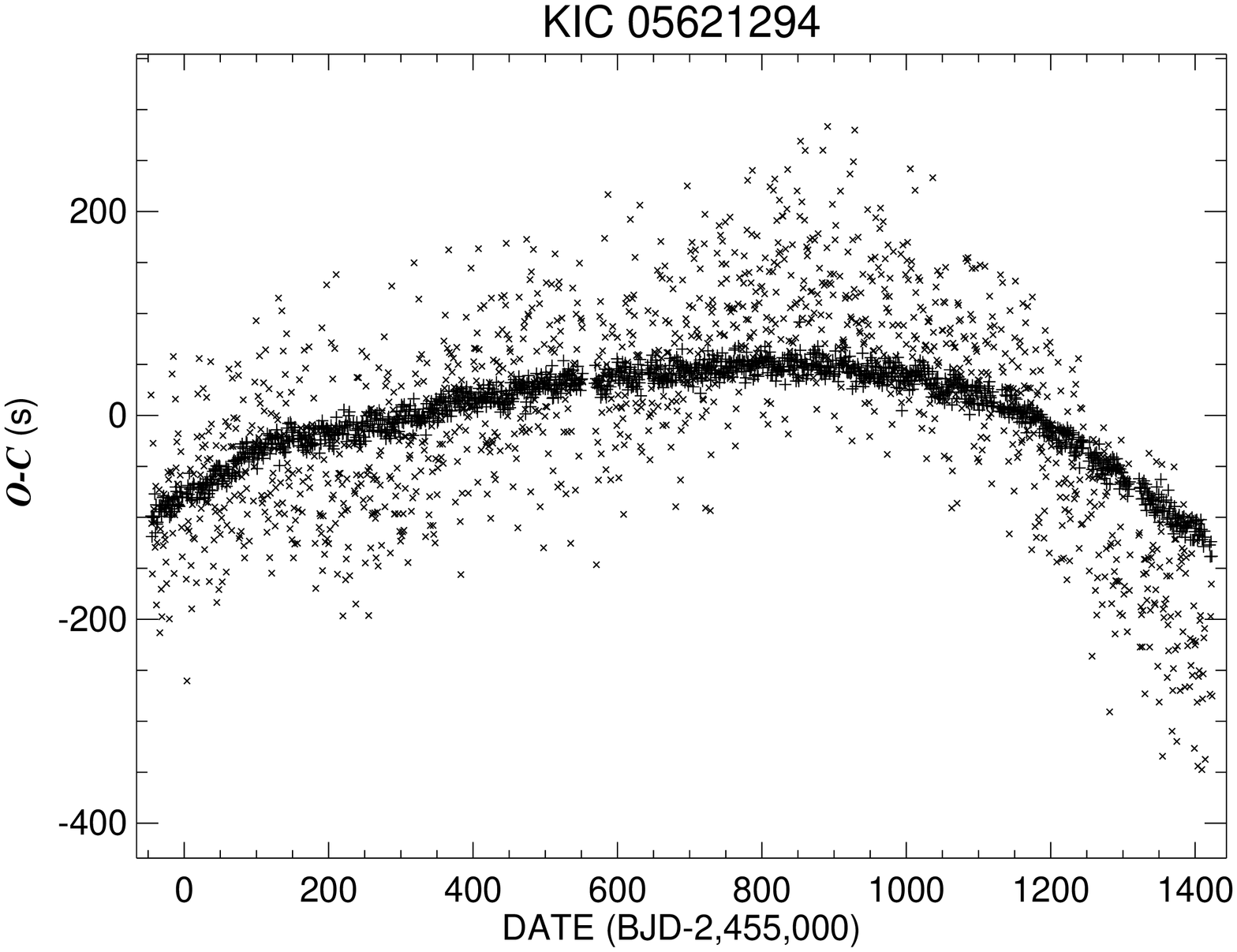}}
\end{center}
\caption{The observed minus calculated eclipse times relative to
a linear ephemeris.  The primary and secondary eclipse
times are indicated by $+$ and $\times$ symbols, 
respectively.}
\end{figure}

\clearpage
\setcounter{figure}{0}
\renewcommand{\thefigure}{\arabic{figure}.16}
\begin{figure}
\begin{center}
{\includegraphics[angle=90,height=12cm]{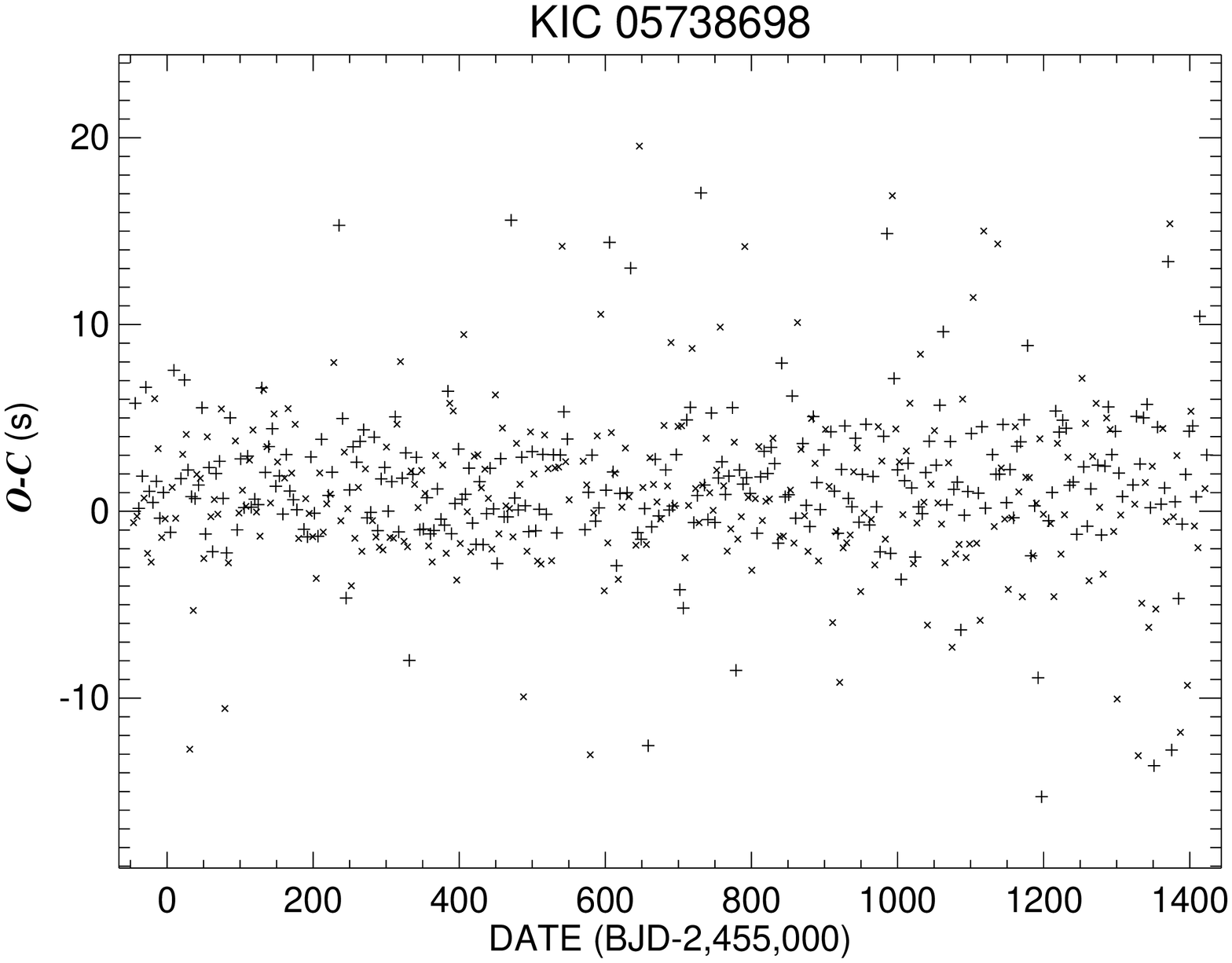}}
\end{center}
\caption{The observed minus calculated eclipse times relative to
a linear ephemeris.  The primary and secondary eclipse
times are indicated by $+$ and $\times$ symbols, 
respectively.}
\end{figure}

\clearpage
\setcounter{figure}{0}
\renewcommand{\thefigure}{\arabic{figure}.17}
\begin{figure}
\begin{center}
{\includegraphics[angle=90,height=12cm]{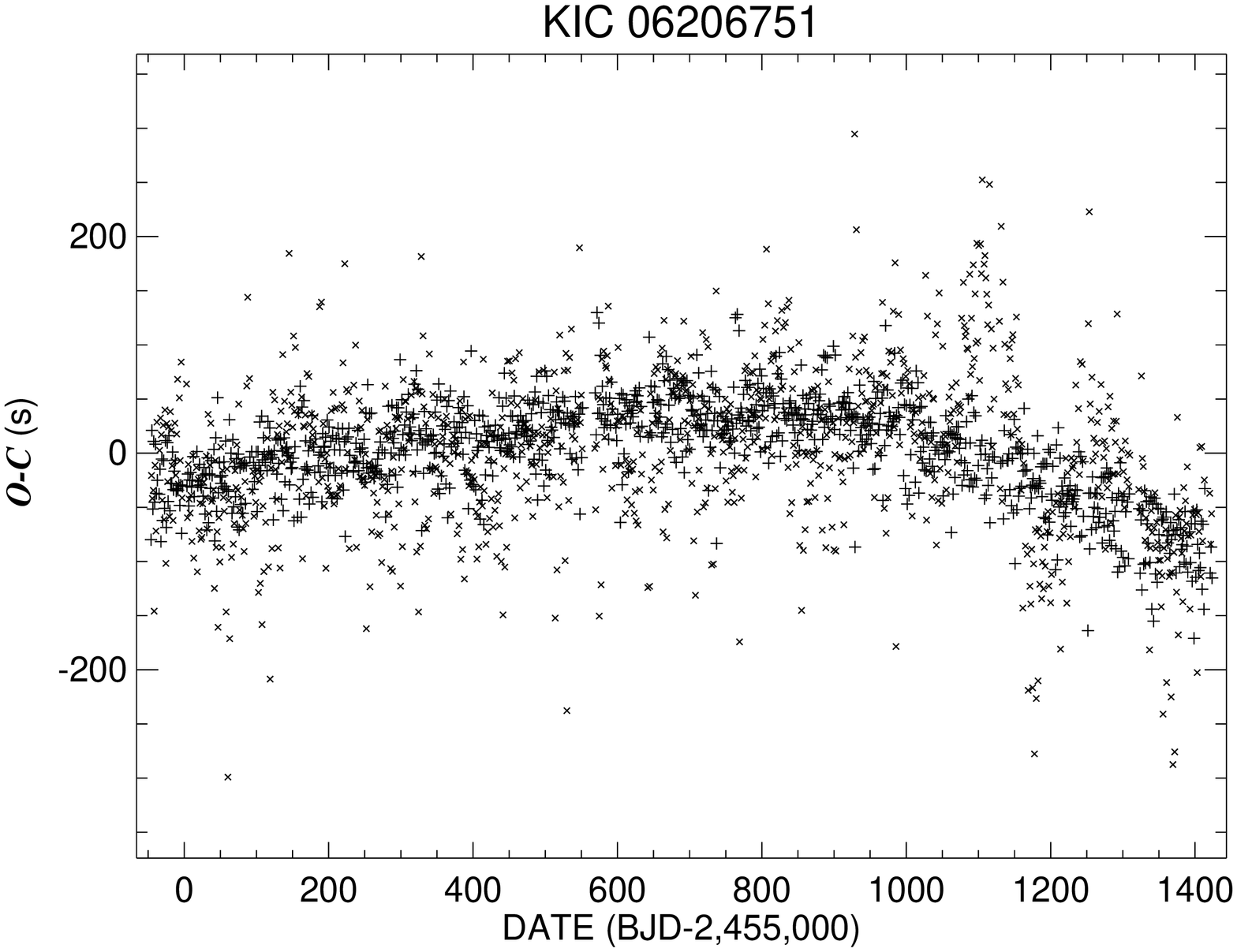}}
\end{center}
\caption{The observed minus calculated eclipse times relative to
a linear ephemeris.  The primary and secondary eclipse
times are indicated by $+$ and $\times$ symbols, 
respectively.}
\end{figure}

\clearpage
\setcounter{figure}{0}
\renewcommand{\thefigure}{\arabic{figure}.18}
\begin{figure}
\begin{center}
{\includegraphics[angle=90,height=12cm]{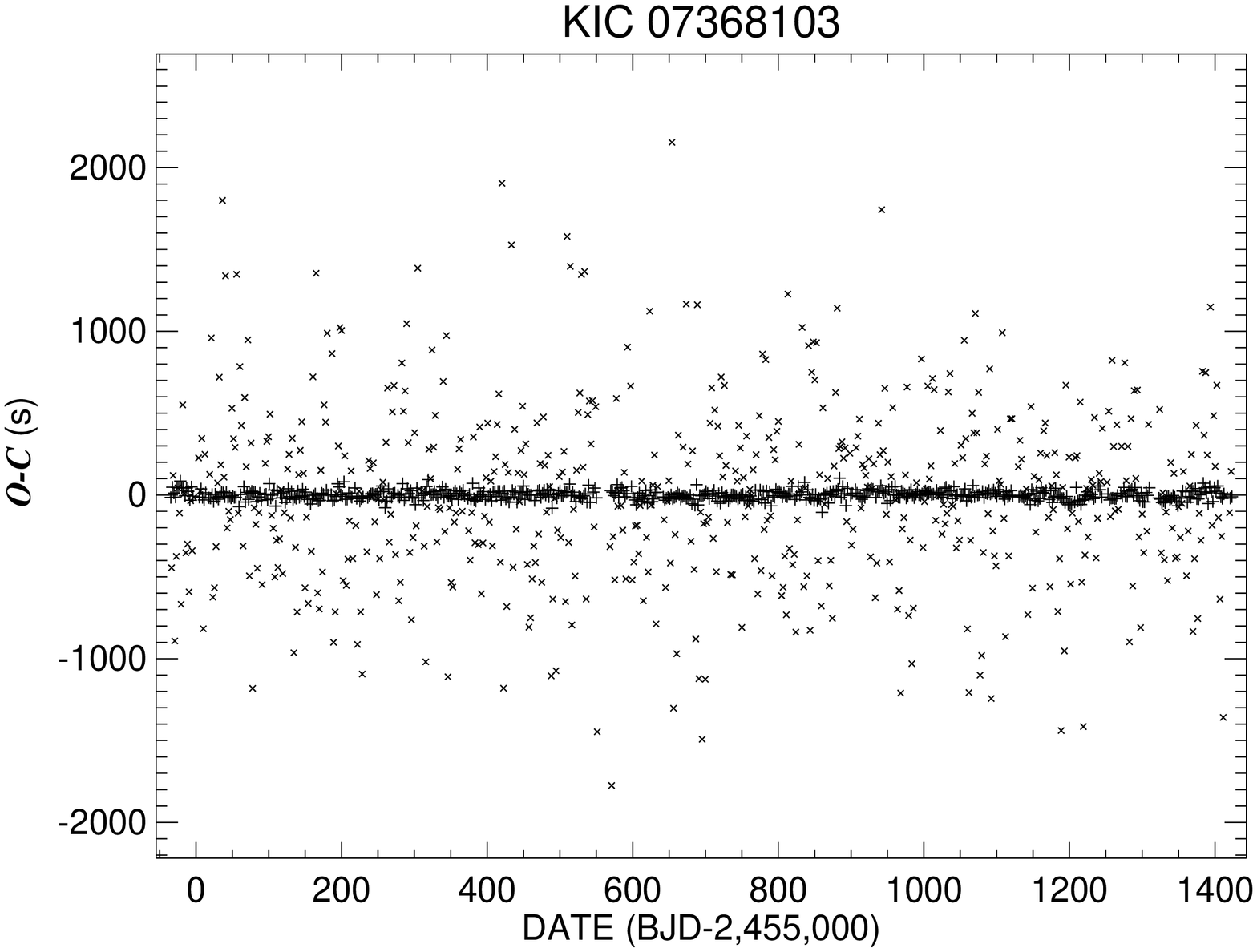}}
\end{center}
\caption{The observed minus calculated eclipse times relative to
a linear ephemeris.  The primary and secondary eclipse
times are indicated by $+$ and $\times$ symbols, 
respectively.}
\end{figure}

\clearpage
\setcounter{figure}{0}
\renewcommand{\thefigure}{\arabic{figure}.19}
\begin{figure}
\begin{center}
{\includegraphics[angle=90,height=12cm]{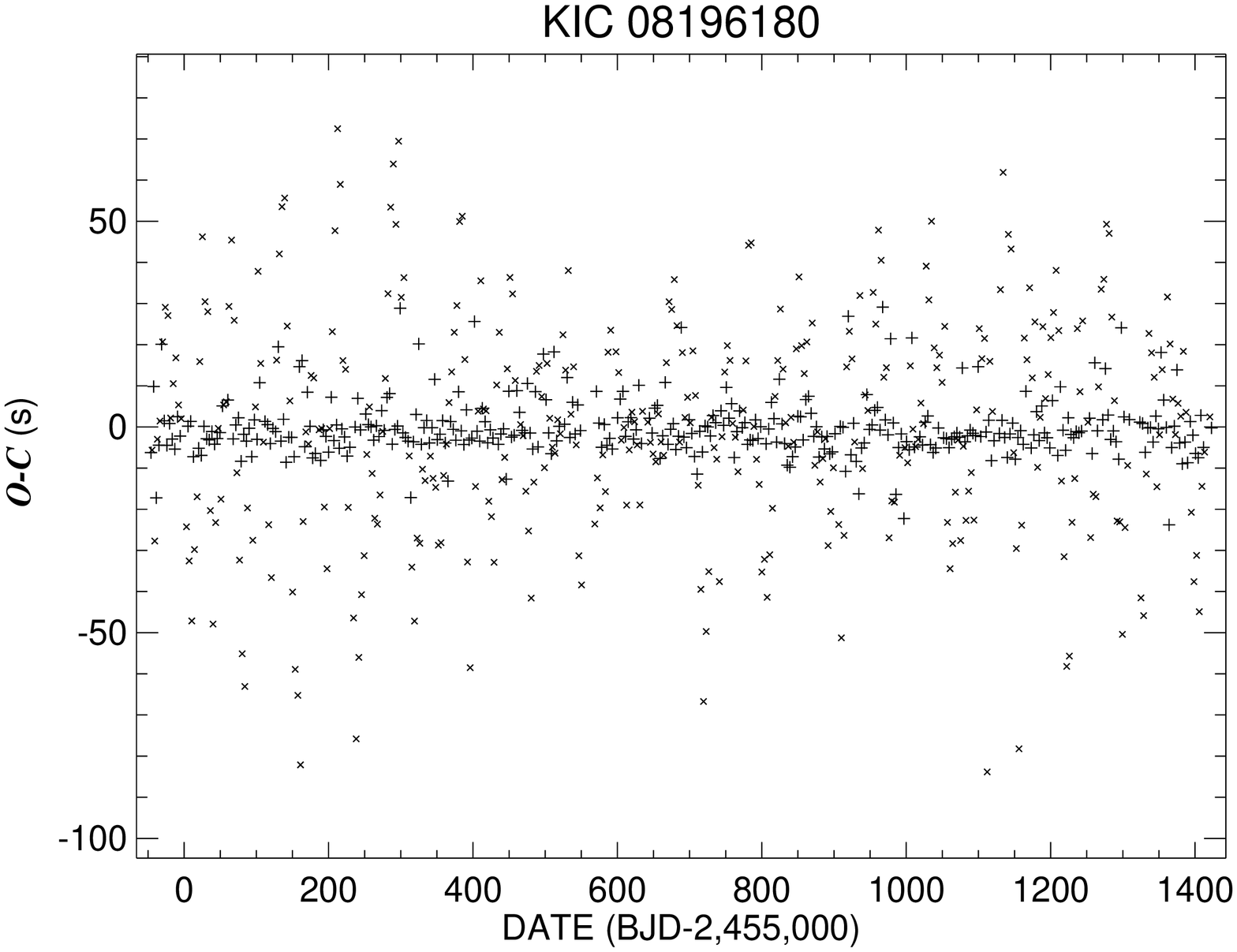}}
\end{center}
\caption{The observed minus calculated eclipse times relative to
a linear ephemeris.  The primary and secondary eclipse
times are indicated by $+$ and $\times$ symbols, 
respectively.}
\end{figure}

\clearpage
\setcounter{figure}{0}
\renewcommand{\thefigure}{\arabic{figure}.20}
\begin{figure}
\begin{center}
{\includegraphics[angle=90,height=12cm]{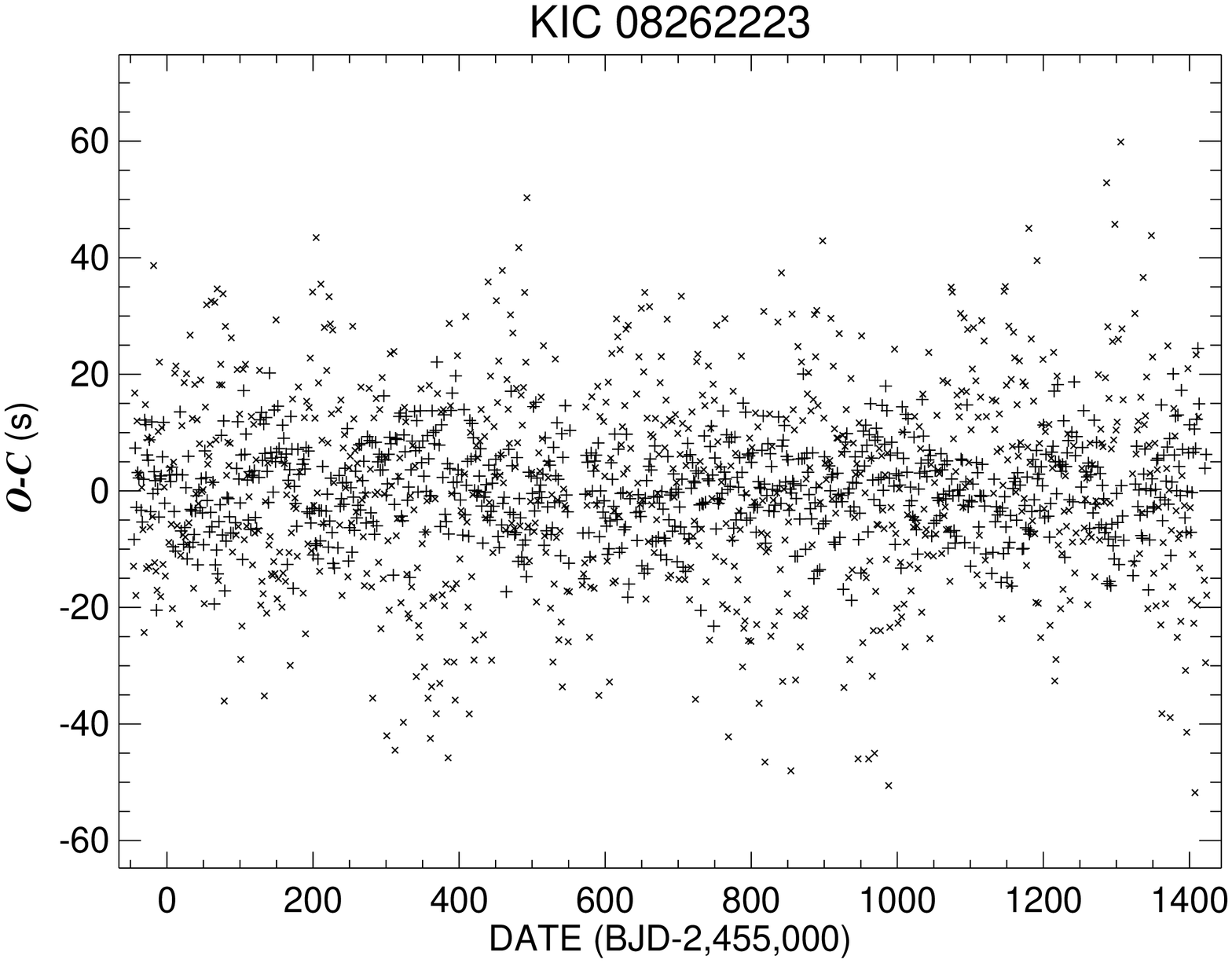}}
\end{center}
\caption{The observed minus calculated eclipse times relative to
a linear ephemeris.  The primary and secondary eclipse
times are indicated by $+$ and $\times$ symbols, 
respectively.}
\end{figure}

\clearpage
\setcounter{figure}{0}
\renewcommand{\thefigure}{\arabic{figure}.21}
\begin{figure}
\begin{center}
{\includegraphics[angle=90,height=12cm]{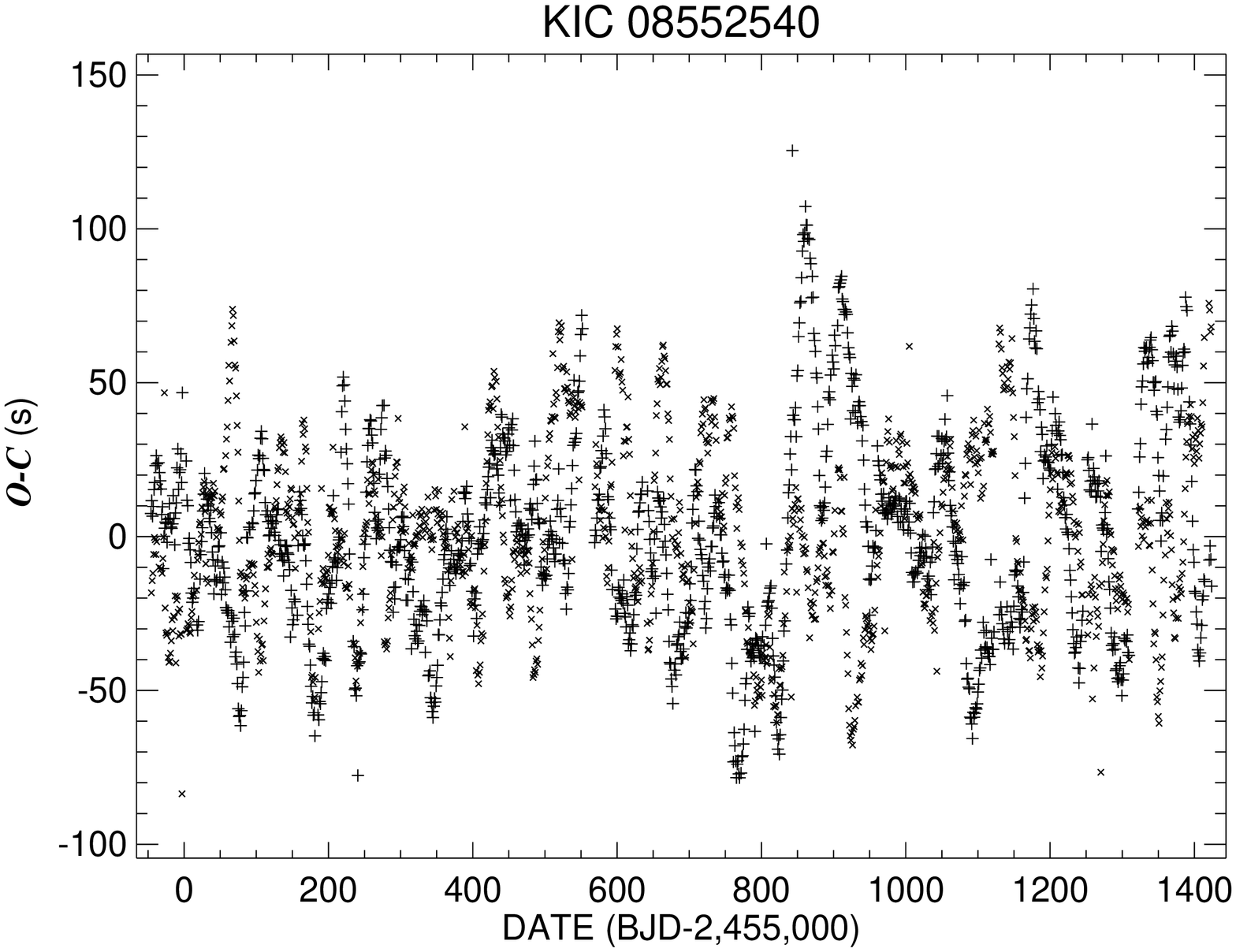}}
\end{center}
\caption{The observed minus calculated eclipse times relative to
a linear ephemeris.  The primary and secondary eclipse
times are indicated by $+$ and $\times$ symbols, 
respectively.}
\end{figure}

\clearpage
\setcounter{figure}{0}
\renewcommand{\thefigure}{\arabic{figure}.22}
\begin{figure}
\begin{center}
{\includegraphics[angle=90,height=12cm]{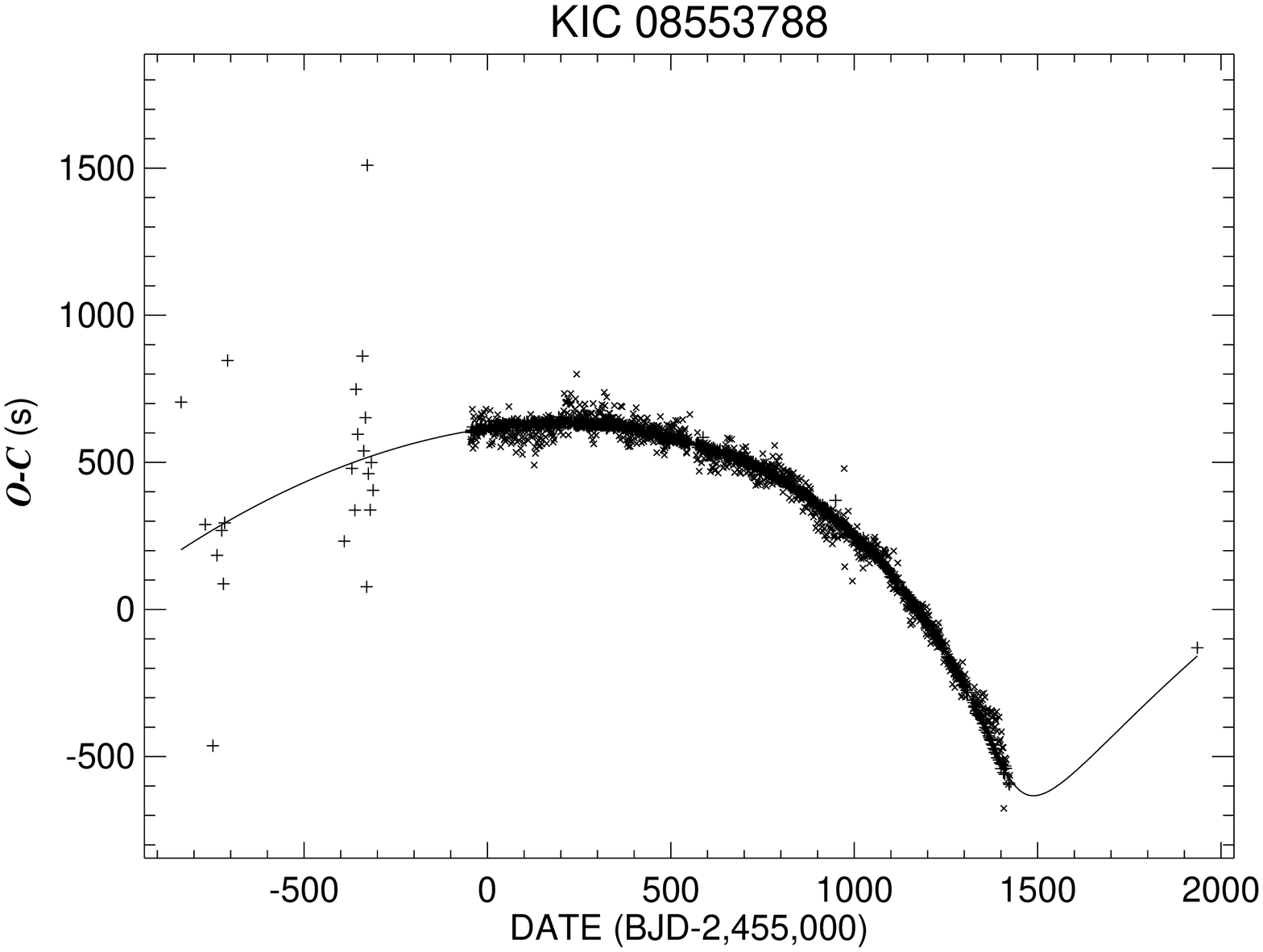}}
\end{center}
\caption{The observed minus calculated eclipse times relative to
a linear ephemeris.  The primary and secondary eclipse
times are indicated by $+$ and $\times$ symbols, 
respectively.}
\end{figure}

\clearpage
\setcounter{figure}{0}
\renewcommand{\thefigure}{\arabic{figure}.23}
\begin{figure}
\begin{center}
{\includegraphics[angle=90,height=12cm]{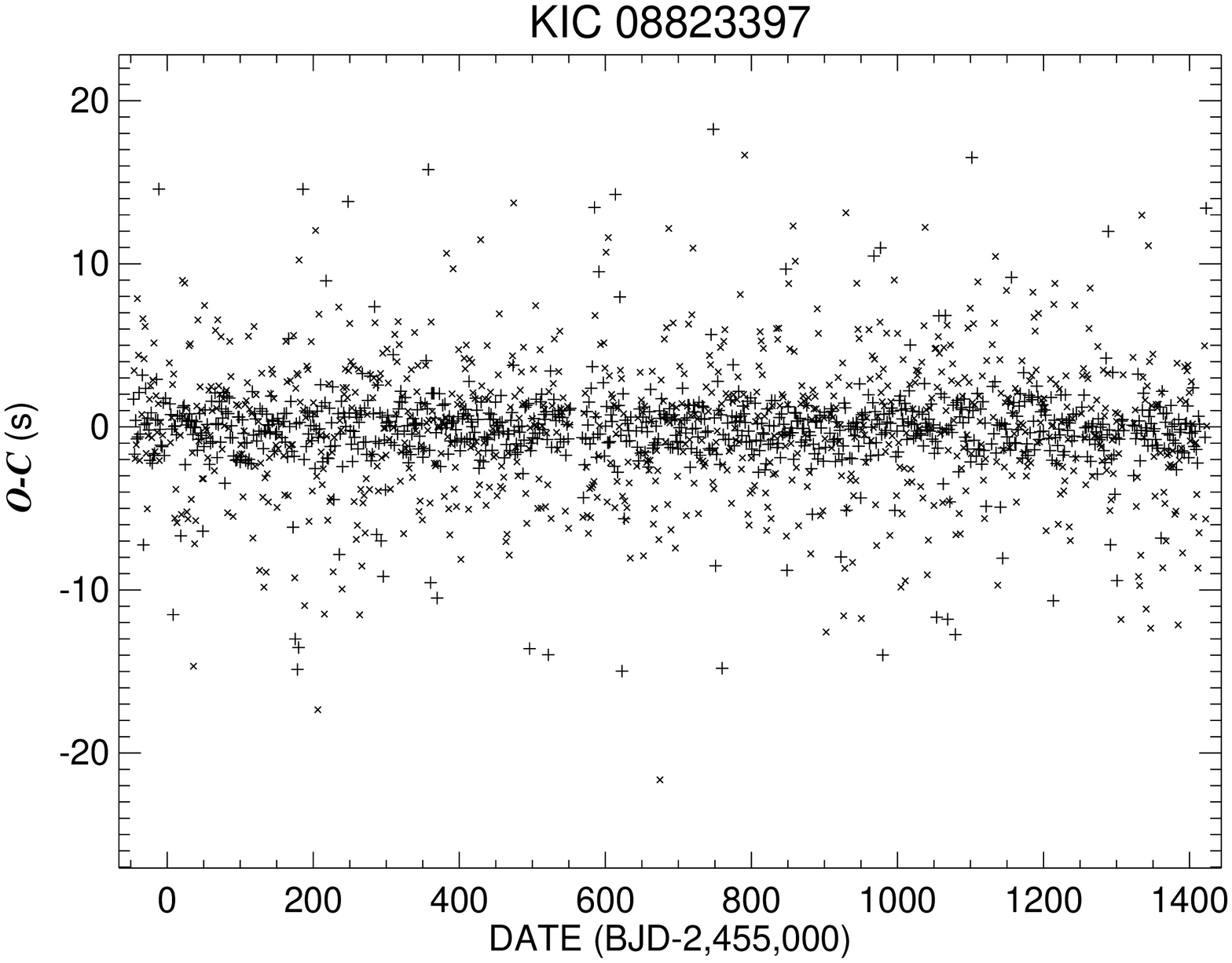}}
\end{center}
\caption{The observed minus calculated eclipse times relative to
a linear ephemeris.  The primary and secondary eclipse
times are indicated by $+$ and $\times$ symbols, 
respectively.}
\end{figure}

\clearpage
\setcounter{figure}{0}
\renewcommand{\thefigure}{\arabic{figure}.24}
\begin{figure}
\begin{center}
{\includegraphics[angle=90,height=12cm]{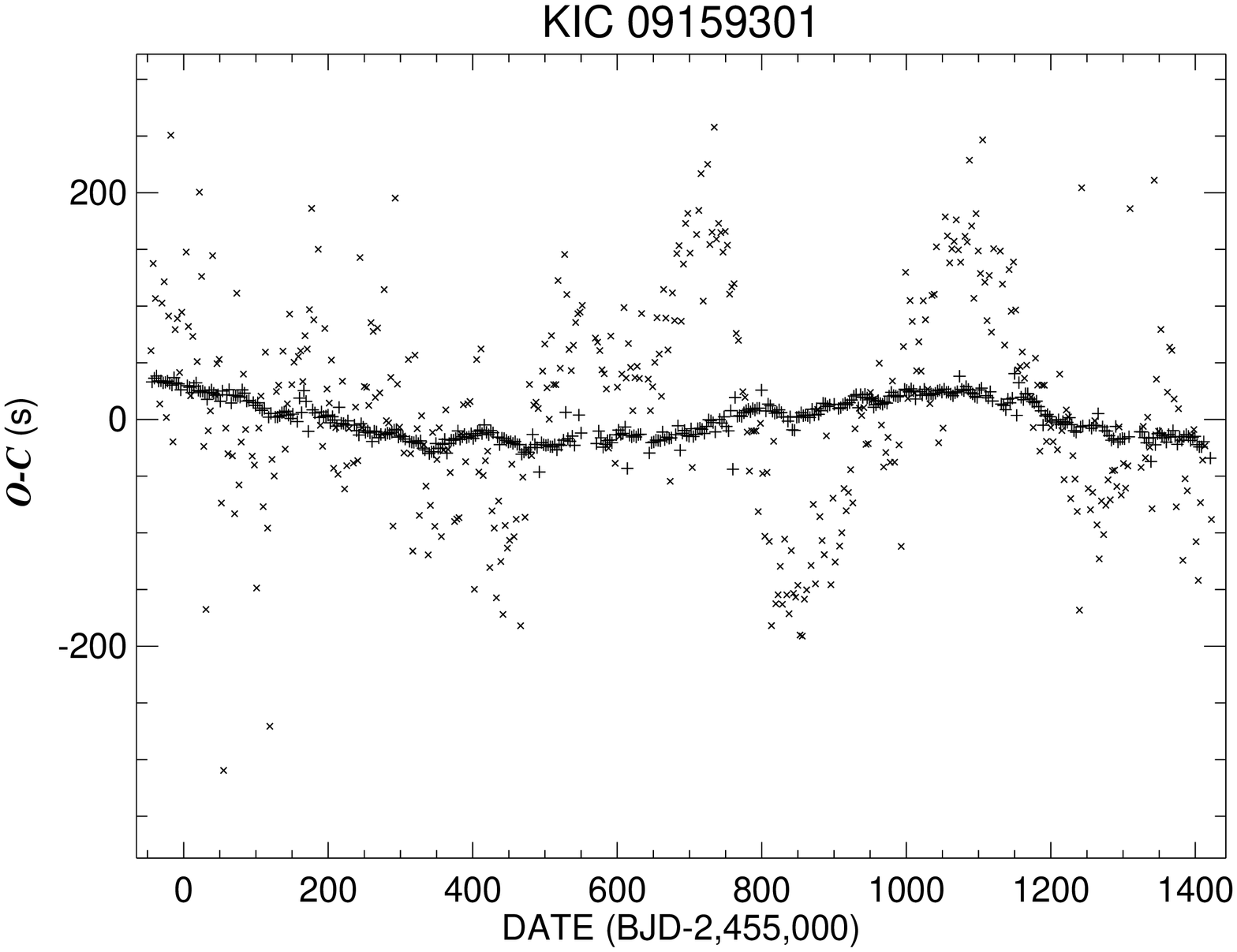}}
\end{center}
\caption{The observed minus calculated eclipse times relative to
a linear ephemeris.  The primary and secondary eclipse
times are indicated by $+$ and $\times$ symbols, 
respectively.}
\end{figure}

\clearpage
\setcounter{figure}{0}
\renewcommand{\thefigure}{\arabic{figure}.25}
\begin{figure}
\begin{center}
{\includegraphics[angle=90,height=12cm]{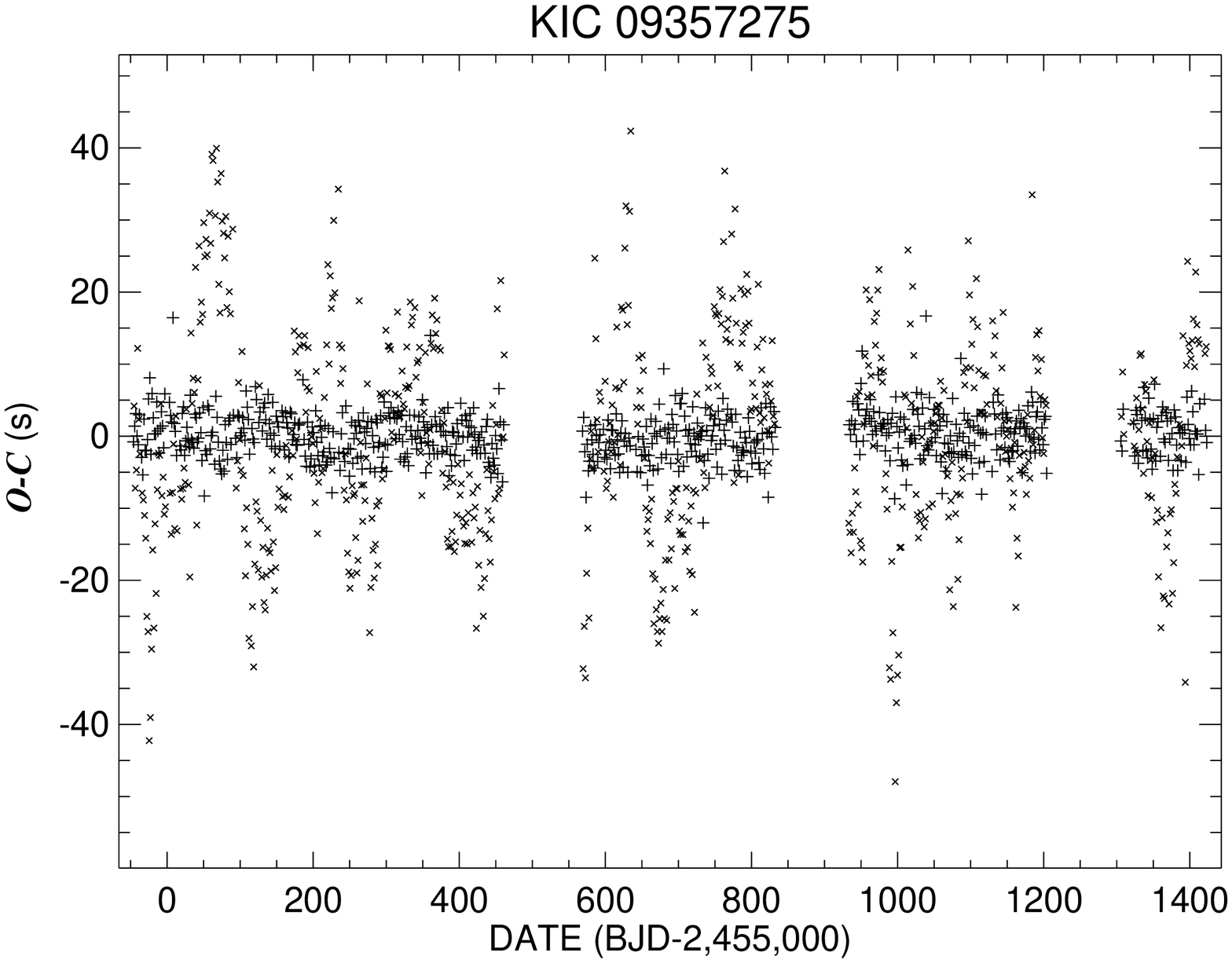}}
\end{center}
\caption{The observed minus calculated eclipse times relative to
a linear ephemeris.  The primary and secondary eclipse
times are indicated by $+$ and $\times$ symbols, 
respectively.}
\end{figure}

\clearpage
\setcounter{figure}{0}
\renewcommand{\thefigure}{\arabic{figure}.26}
\begin{figure}
\begin{center}
{\includegraphics[angle=90,height=12cm]{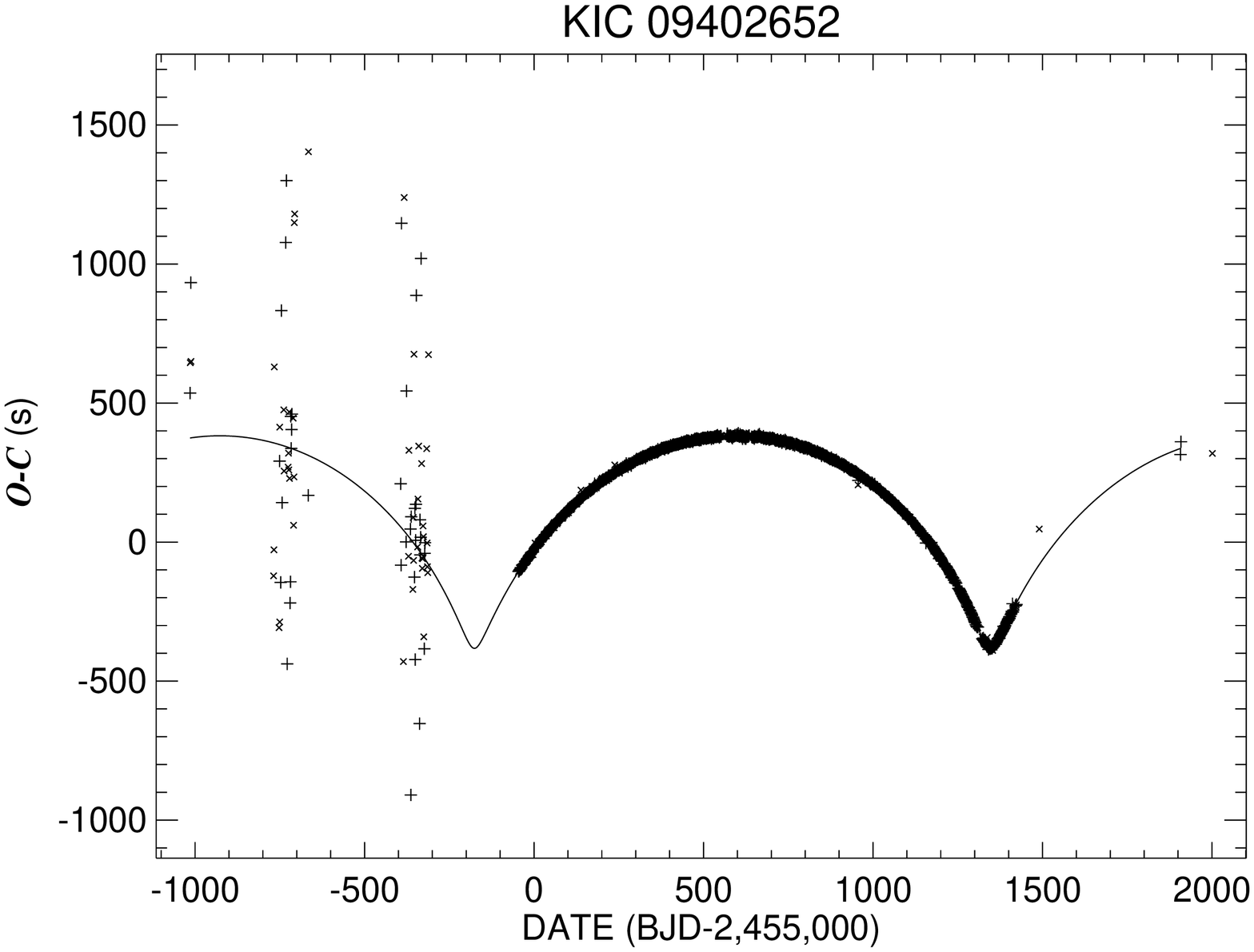}}
\end{center}
\caption{The observed minus calculated eclipse times relative to
a linear ephemeris.  The primary and secondary eclipse
times are indicated by $+$ and $\times$ symbols, 
respectively.}
\end{figure}

\clearpage
\setcounter{figure}{0}
\renewcommand{\thefigure}{\arabic{figure}.27}
\begin{figure}
\begin{center}
{\includegraphics[angle=90,height=12cm]{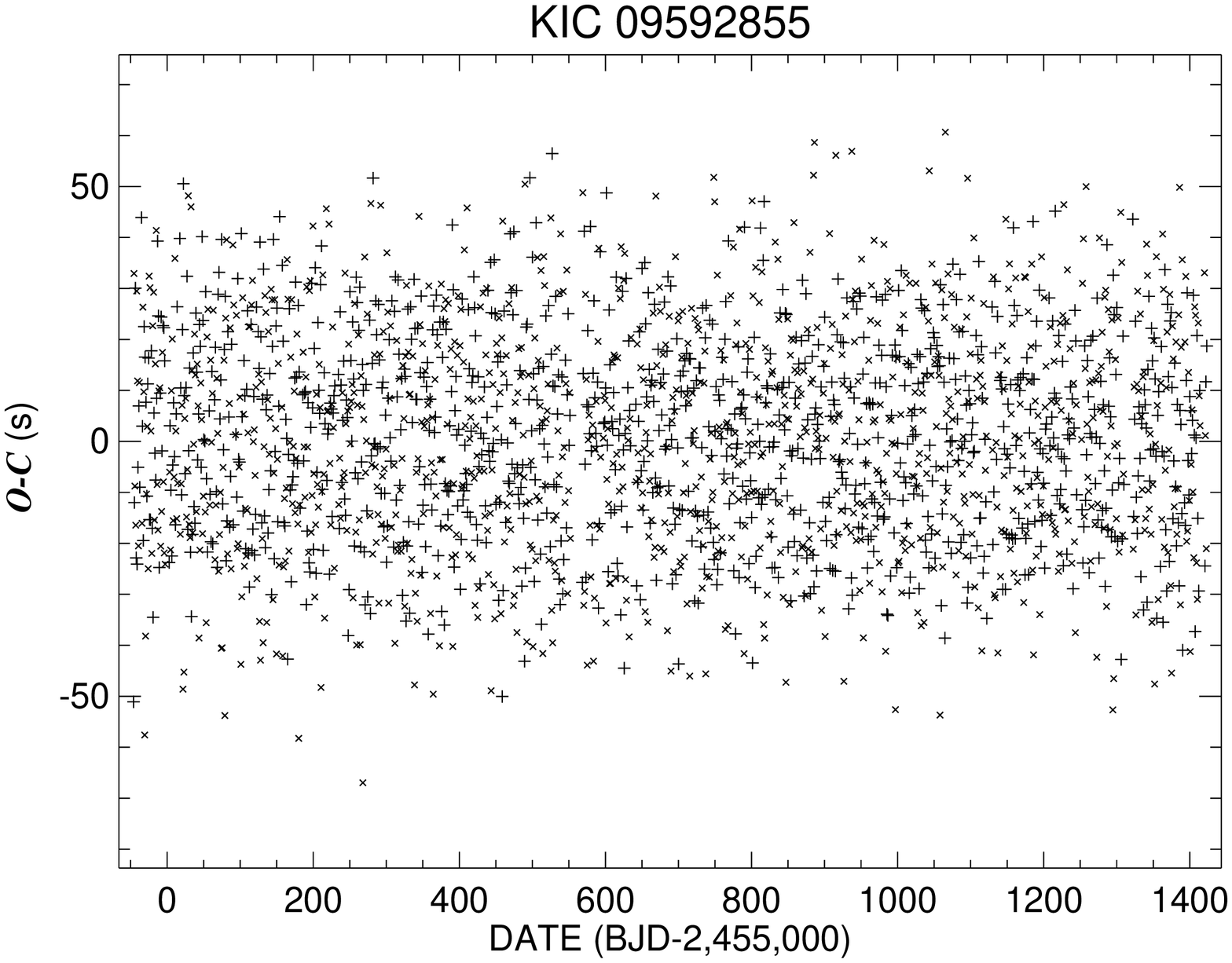}}
\end{center}
\caption{The observed minus calculated eclipse times relative to
a linear ephemeris.  The primary and secondary eclipse
times are indicated by $+$ and $\times$ symbols, 
respectively.}
\end{figure}

\clearpage
\setcounter{figure}{0}
\renewcommand{\thefigure}{\arabic{figure}.28}
\begin{figure}
\begin{center}
{\includegraphics[angle=90,height=12cm]{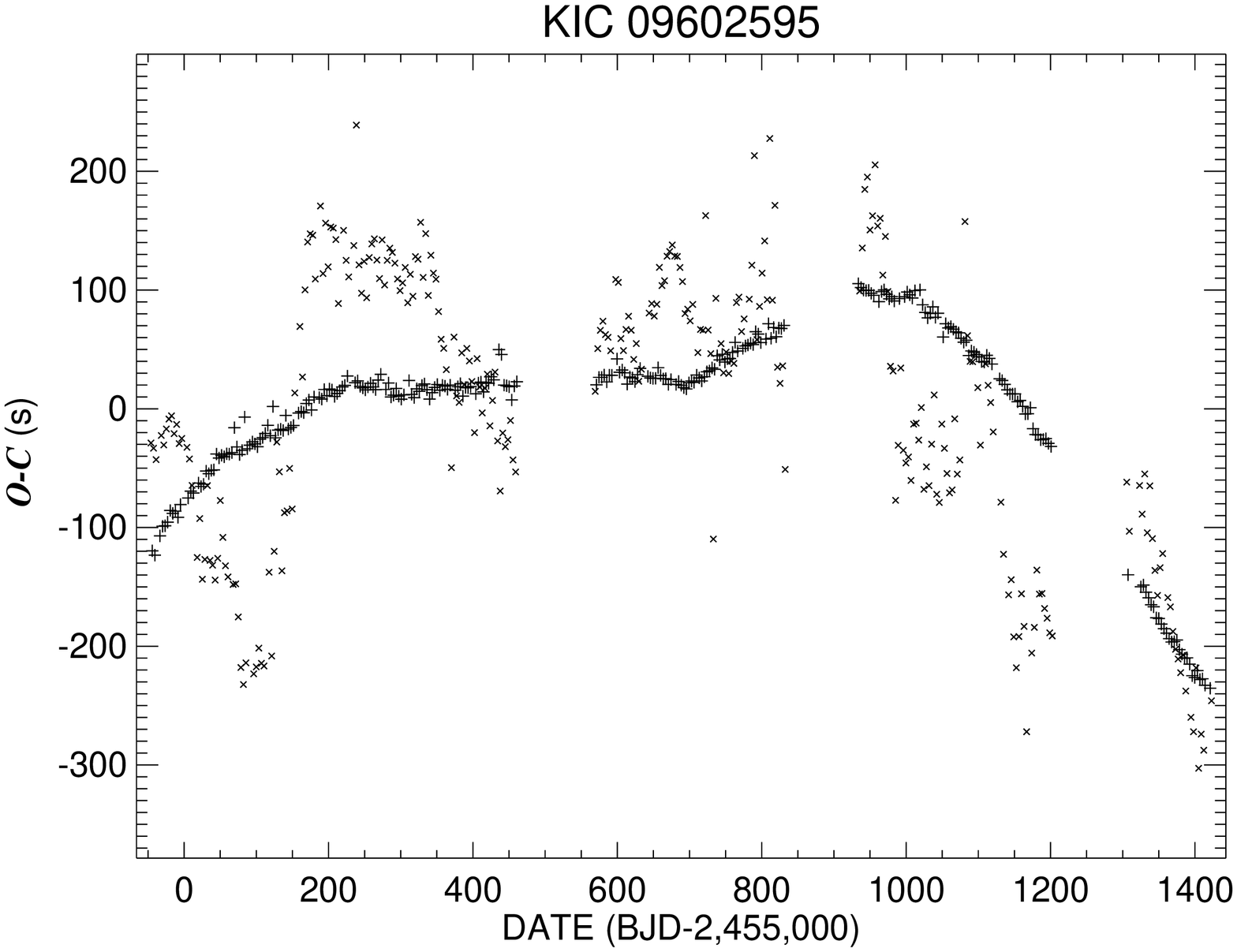}}
\end{center}
\caption{The observed minus calculated eclipse times relative to
a linear ephemeris.  The primary and secondary eclipse
times are indicated by $+$ and $\times$ symbols, 
respectively.}
\end{figure}

\clearpage
\setcounter{figure}{0}
\renewcommand{\thefigure}{\arabic{figure}.29}
\begin{figure}
\begin{center}
{\includegraphics[angle=90,height=12cm]{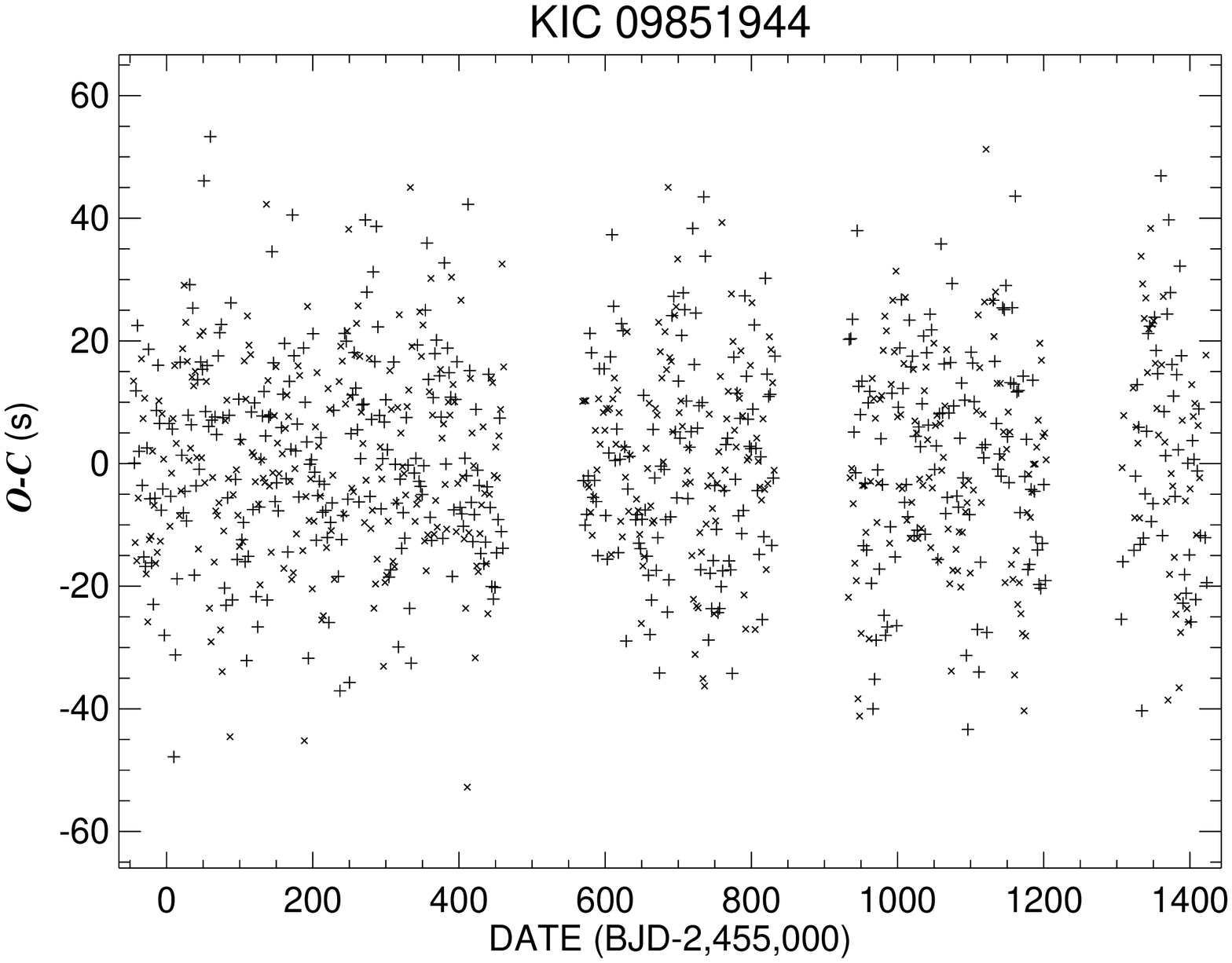}}
\end{center}
\caption{The observed minus calculated eclipse times relative to
a linear ephemeris.  The primary and secondary eclipse
times are indicated by $+$ and $\times$ symbols, 
respectively.}
\end{figure}

\clearpage
\setcounter{figure}{0}
\renewcommand{\thefigure}{\arabic{figure}.30}
\begin{figure}
\begin{center}
{\includegraphics[angle=90,height=12cm]{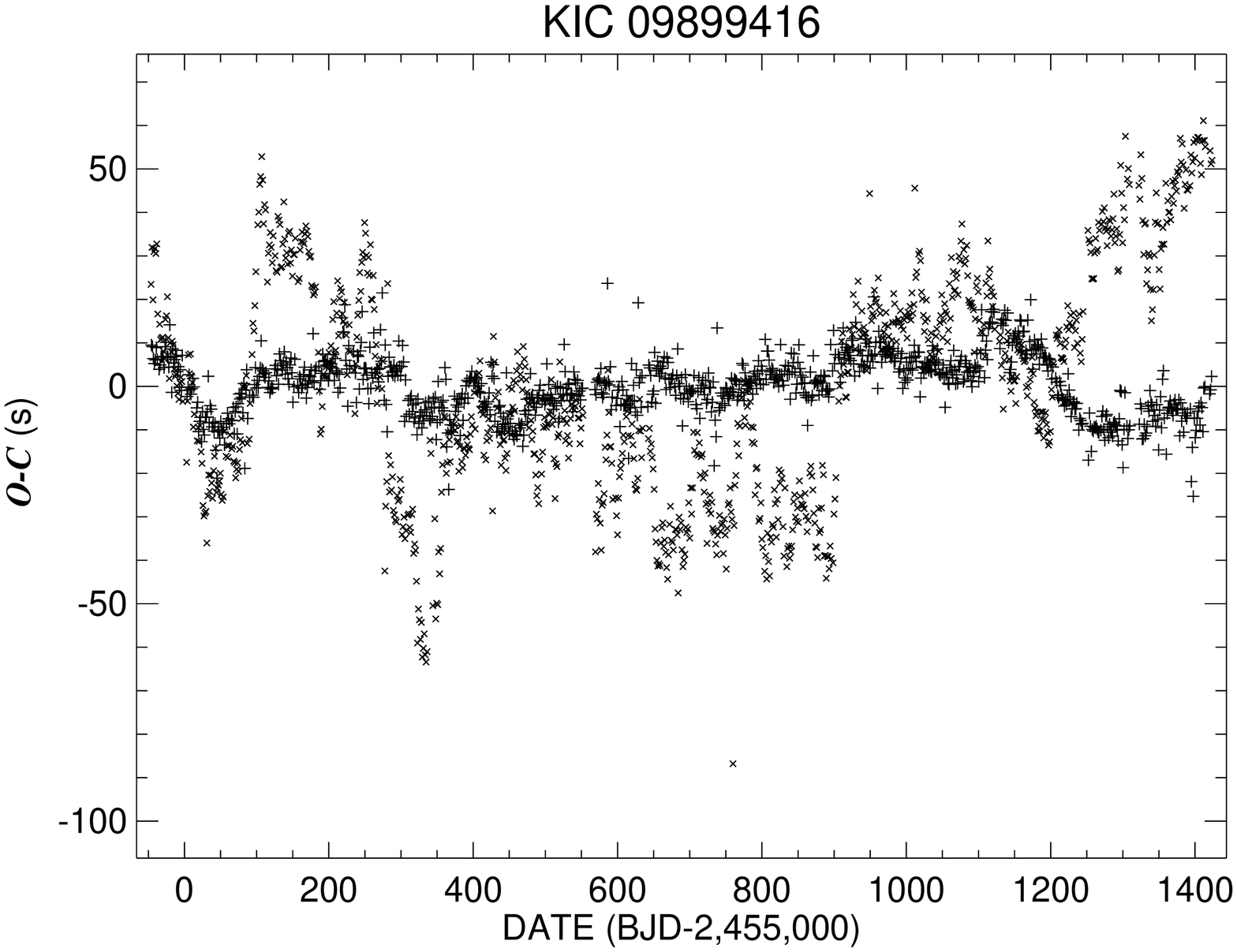}}
\end{center}
\caption{The observed minus calculated eclipse times relative to
a linear ephemeris.  The primary and secondary eclipse
times are indicated by $+$ and $\times$ symbols, 
respectively.}
\end{figure}

\clearpage
\setcounter{figure}{0}
\renewcommand{\thefigure}{\arabic{figure}.31}
\begin{figure}
\begin{center}
{\includegraphics[angle=90,height=12cm]{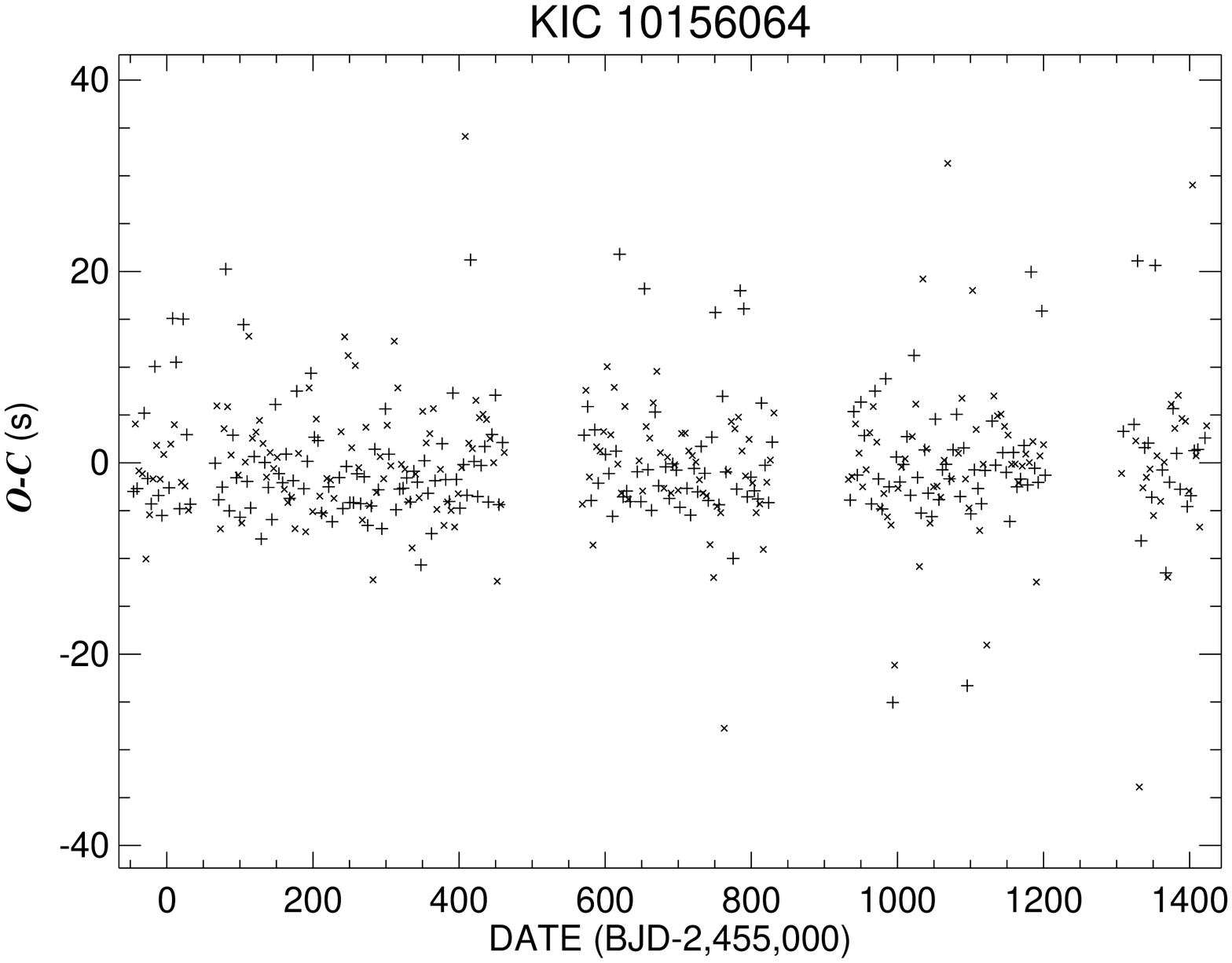}}
\end{center}
\caption{The observed minus calculated eclipse times relative to
a linear ephemeris.  The primary and secondary eclipse
times are indicated by $+$ and $\times$ symbols, 
respectively.}
\end{figure}

\clearpage
\setcounter{figure}{0}
\renewcommand{\thefigure}{\arabic{figure}.32}
\begin{figure}
\begin{center}
{\includegraphics[angle=90,height=12cm]{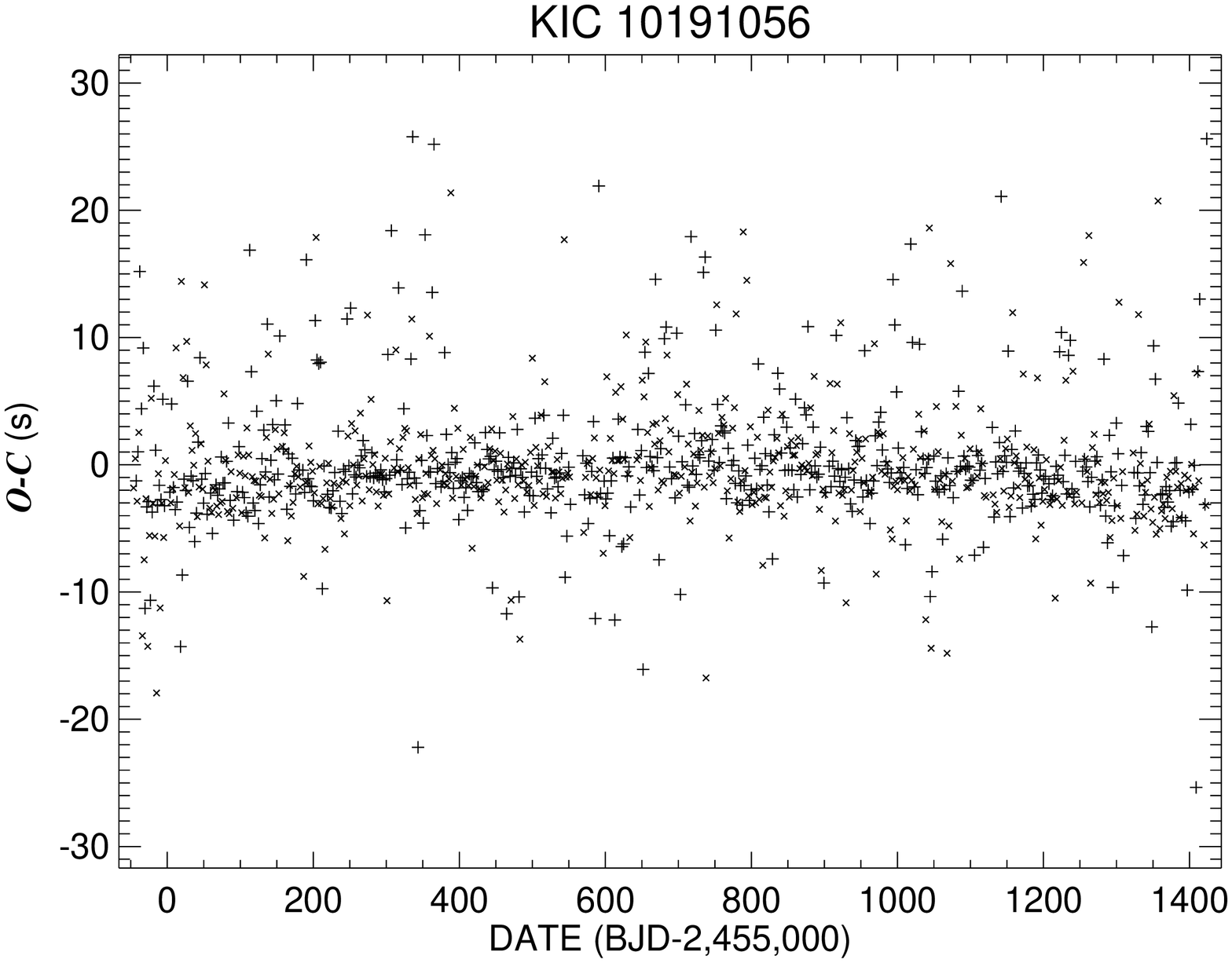}}
\end{center}
\caption{The observed minus calculated eclipse times relative to
a linear ephemeris.  The primary and secondary eclipse
times are indicated by $+$ and $\times$ symbols, 
respectively.}
\end{figure}

\clearpage
\setcounter{figure}{0}
\renewcommand{\thefigure}{\arabic{figure}.33}
\begin{figure}
\begin{center}
{\includegraphics[angle=90,height=12cm]{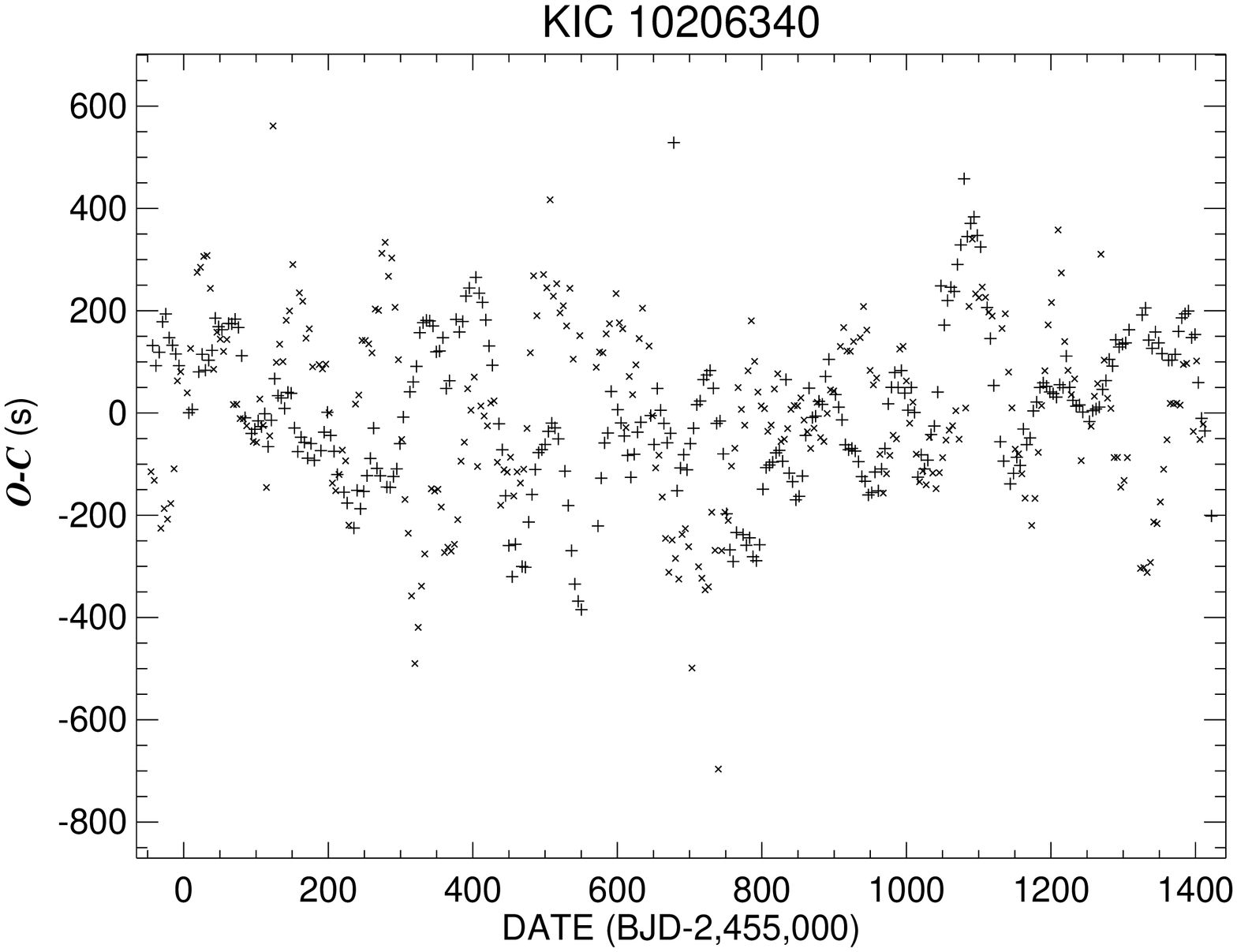}}
\end{center}
\caption{The observed minus calculated eclipse times relative to
a linear ephemeris.  The primary and secondary eclipse
times are indicated by $+$ and $\times$ symbols, 
respectively.}
\end{figure}

\clearpage
\setcounter{figure}{0}
\renewcommand{\thefigure}{\arabic{figure}.34}
\begin{figure}
\begin{center}
{\includegraphics[angle=90,height=12cm]{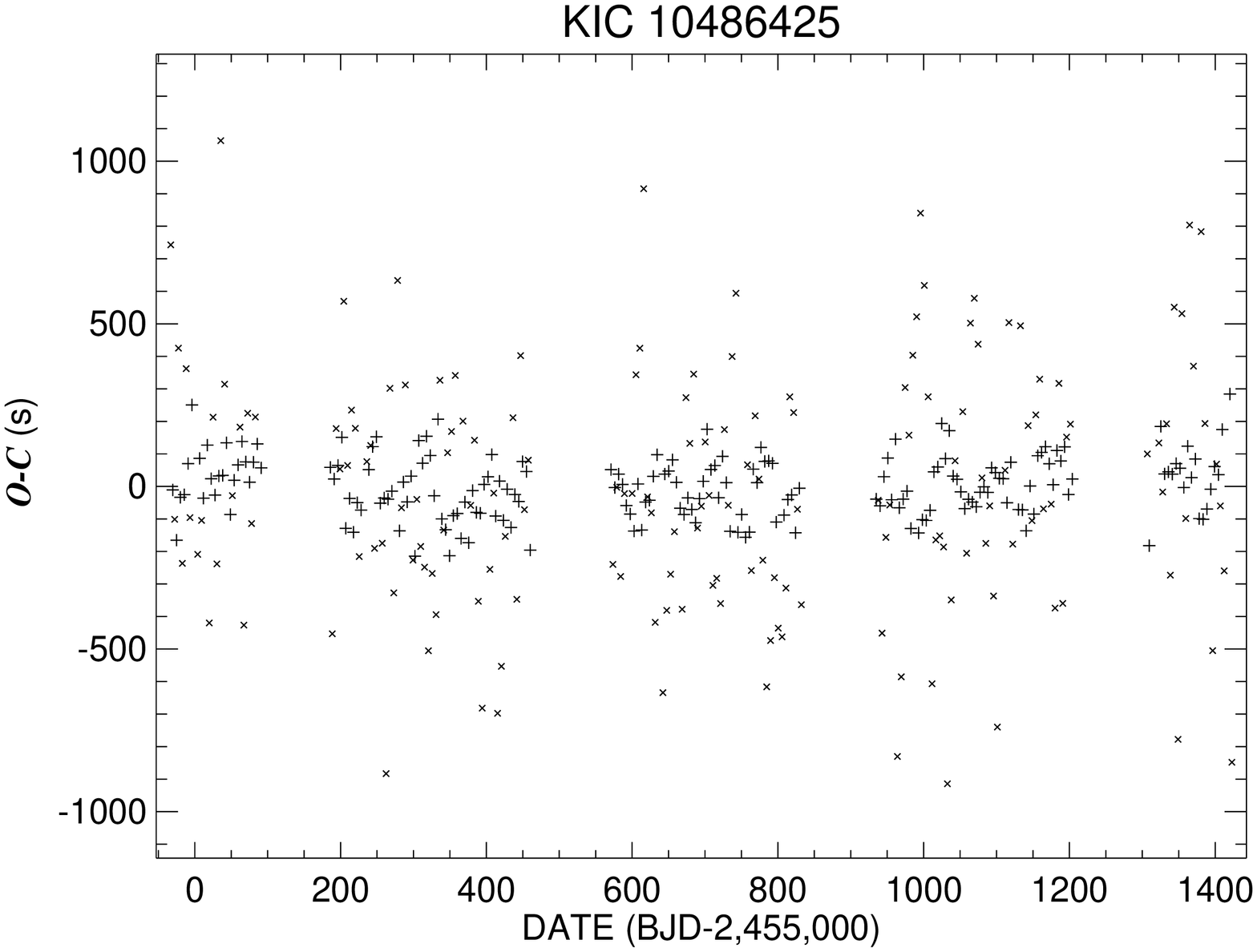}}
\end{center}
\caption{The observed minus calculated eclipse times relative to
a linear ephemeris.  The primary and secondary eclipse
times are indicated by $+$ and $\times$ symbols, 
respectively.}
\end{figure}

\clearpage
\setcounter{figure}{0}
\renewcommand{\thefigure}{\arabic{figure}.35}
\begin{figure}
\begin{center}
{\includegraphics[angle=90,height=12cm]{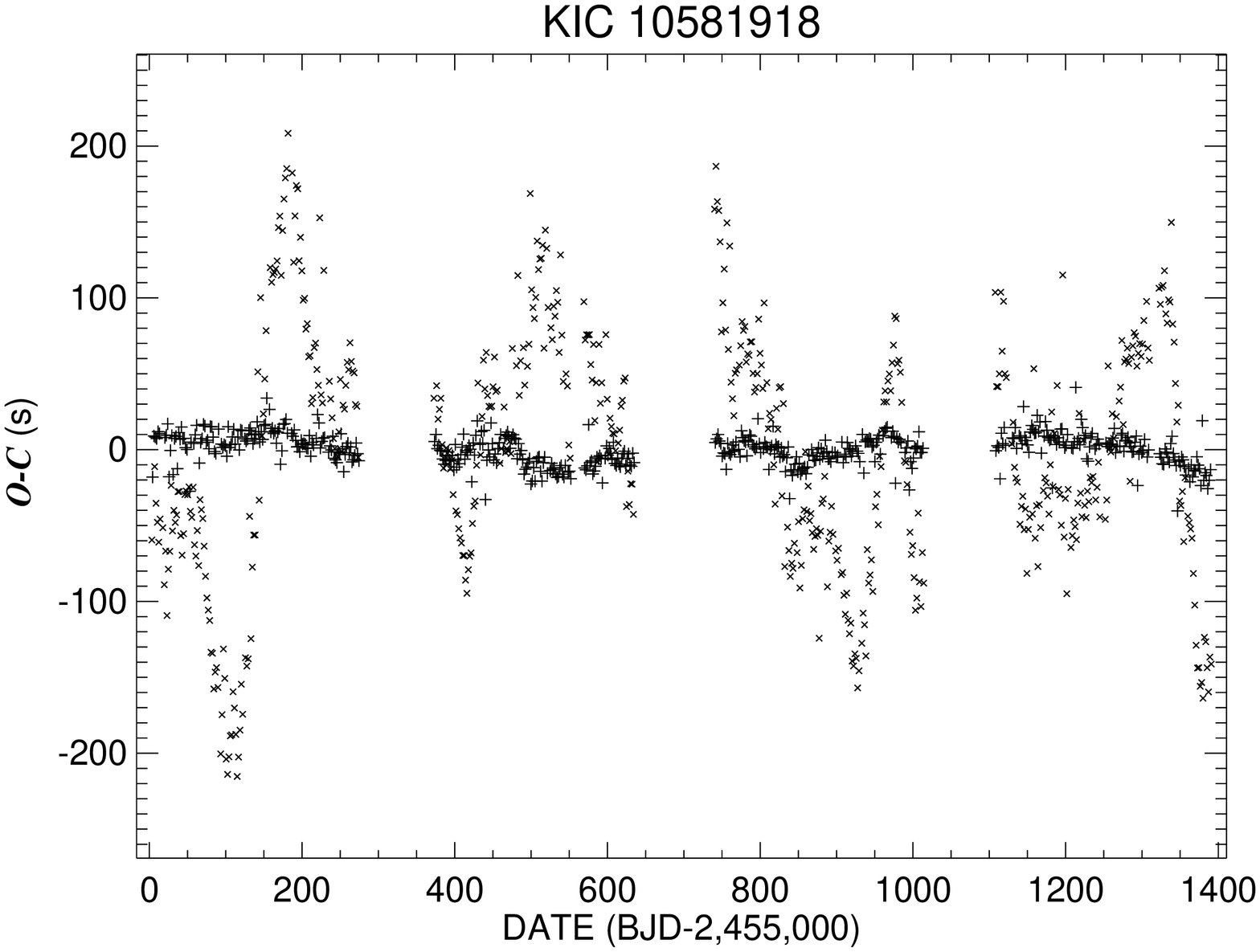}}
\end{center}
\caption{The observed minus calculated eclipse times relative to
a linear ephemeris.  The primary and secondary eclipse
times are indicated by $+$ and $\times$ symbols, 
respectively.}
\end{figure}

\clearpage
\setcounter{figure}{0}
\renewcommand{\thefigure}{\arabic{figure}.36}
\begin{figure}
\begin{center}
{\includegraphics[angle=90,height=12cm]{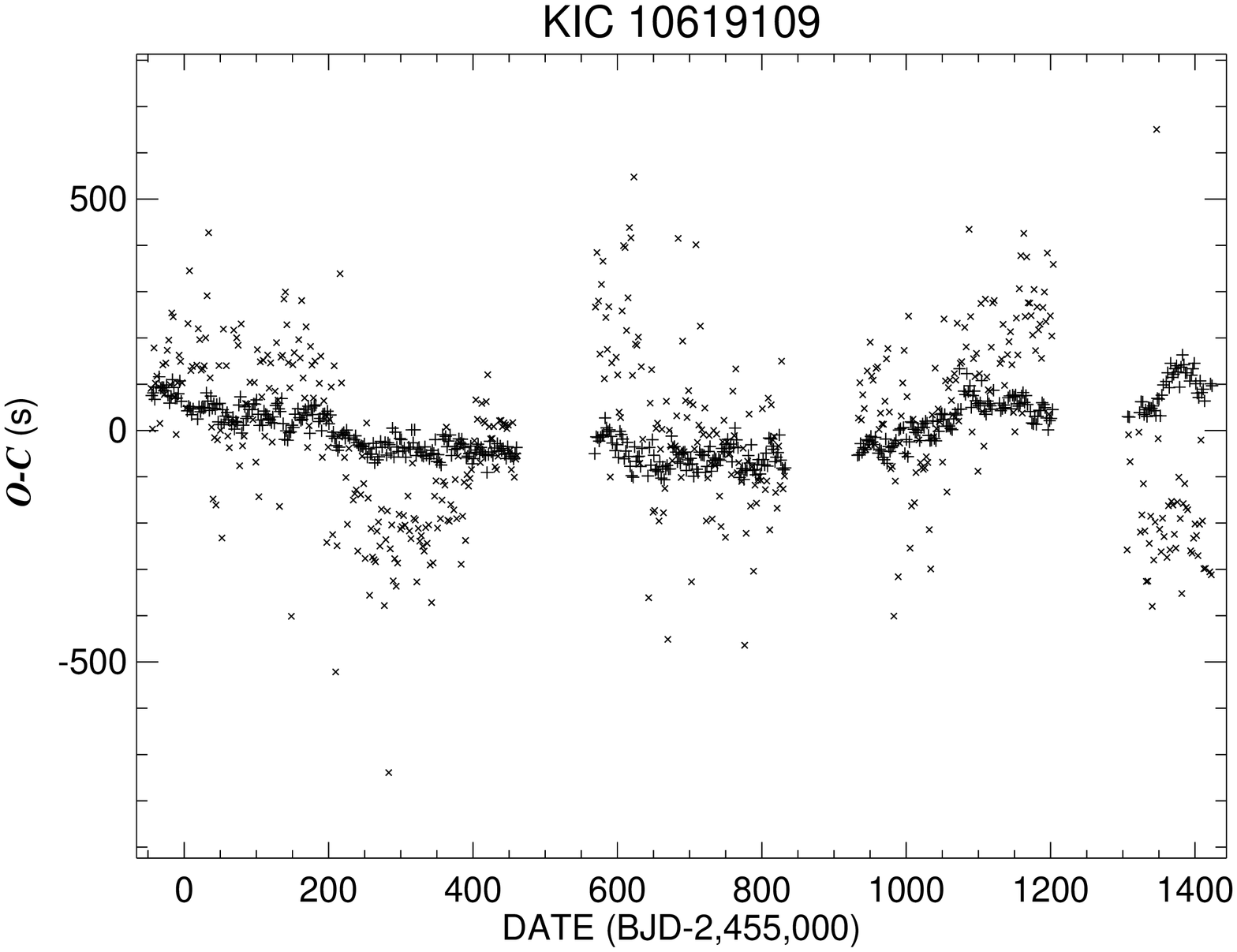}}
\end{center}
\caption{The observed minus calculated eclipse times relative to
a linear ephemeris.  The primary and secondary eclipse
times are indicated by $+$ and $\times$ symbols, 
respectively.}
\end{figure}

\clearpage
\setcounter{figure}{0}
\renewcommand{\thefigure}{\arabic{figure}.37}
\begin{figure}
\begin{center}
{\includegraphics[angle=90,height=12cm]{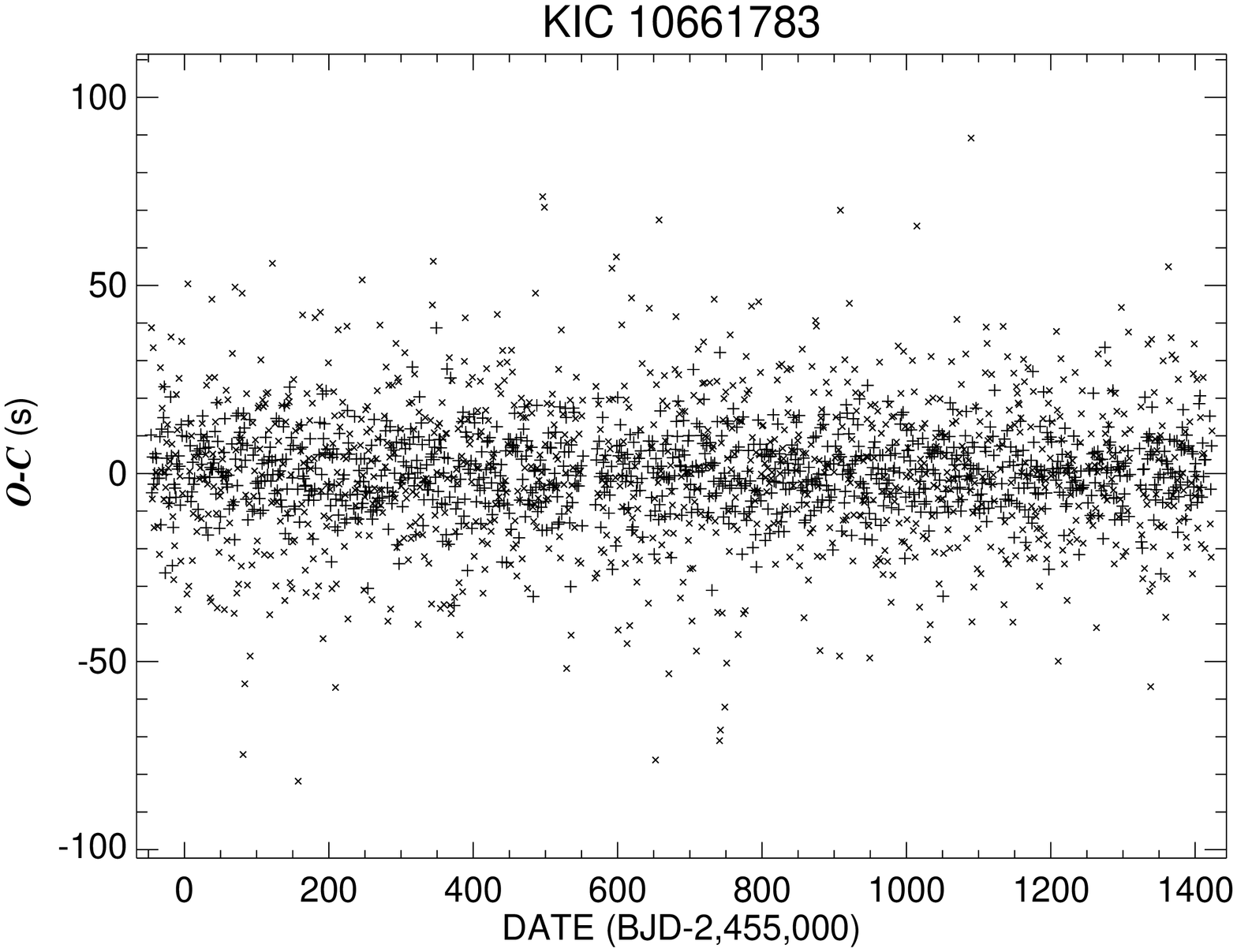}}
\end{center}
\caption{The observed minus calculated eclipse times relative to
a linear ephemeris.  The primary and secondary eclipse
times are indicated by $+$ and $\times$ symbols, 
respectively.}
\end{figure}

\clearpage
\setcounter{figure}{0}
\renewcommand{\thefigure}{\arabic{figure}.38}
\begin{figure}
\begin{center}
{\includegraphics[angle=90,height=12cm]{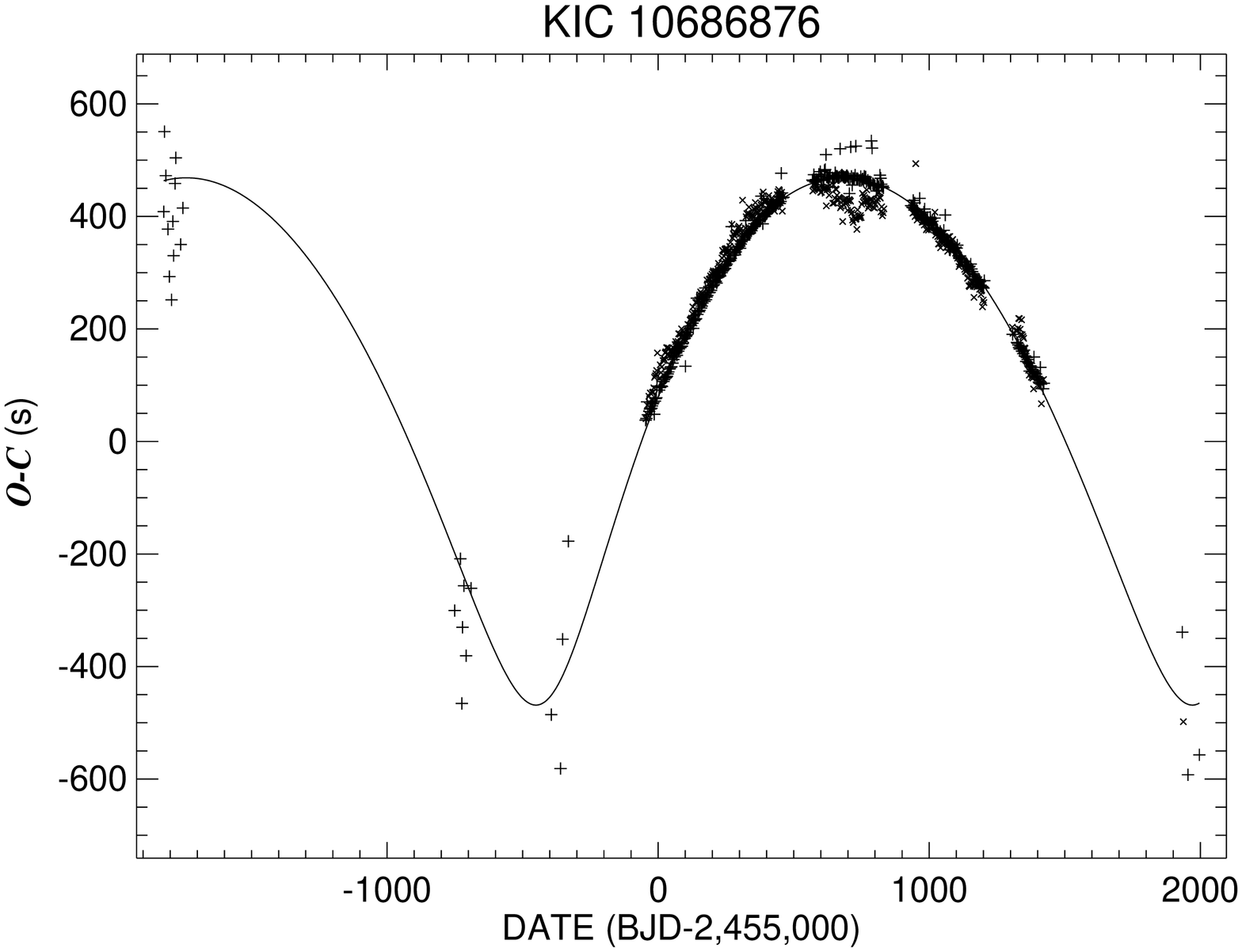}}
\end{center}
\caption{The observed minus calculated eclipse times relative to
a linear ephemeris.  The primary and secondary eclipse
times are indicated by $+$ and $\times$ symbols, 
respectively.}
\end{figure}

\clearpage
\setcounter{figure}{0}
\renewcommand{\thefigure}{\arabic{figure}.39}
\begin{figure}
\begin{center}
{\includegraphics[angle=90,height=12cm]{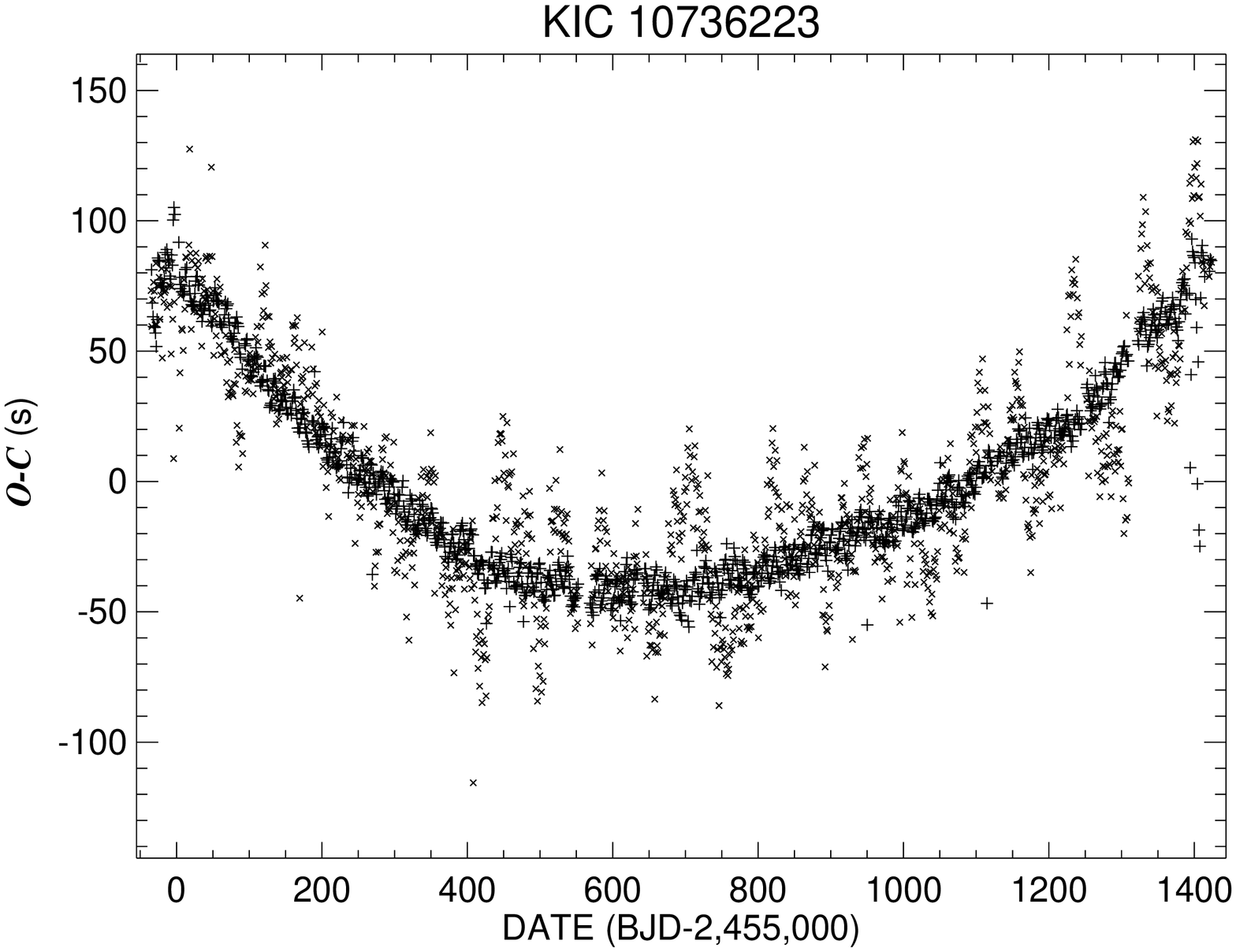}}
\end{center}
\caption{The observed minus calculated eclipse times relative to
a linear ephemeris.  The primary and secondary eclipse
times are indicated by $+$ and $\times$ symbols, 
respectively.}
\end{figure}

\clearpage
\setcounter{figure}{0}
\renewcommand{\thefigure}{\arabic{figure}.40}
\begin{figure}
\begin{center}
{\includegraphics[angle=90,height=12cm]{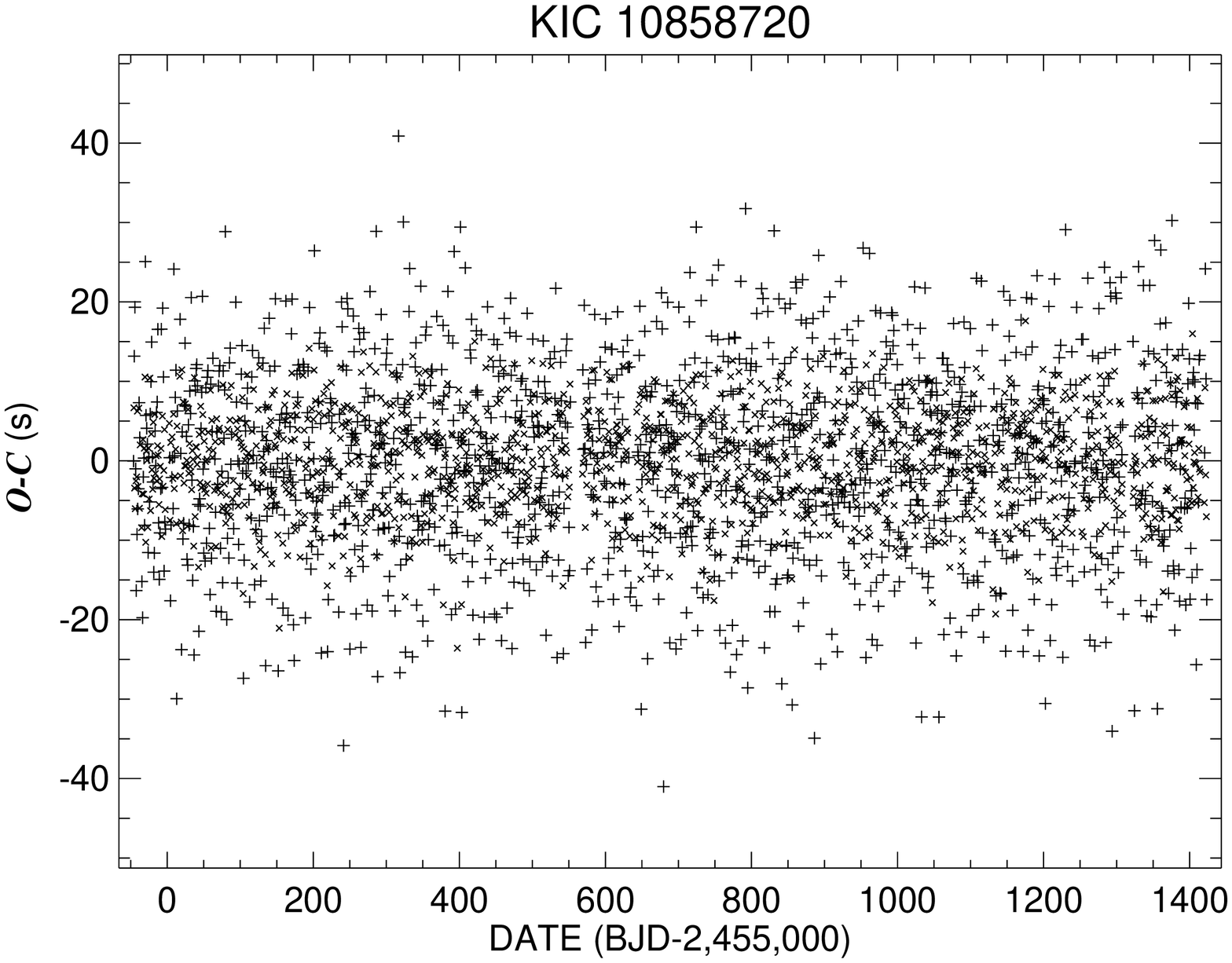}}
\end{center}
\caption{The observed minus calculated eclipse times relative to
a linear ephemeris.  The primary and secondary eclipse
times are indicated by $+$ and $\times$ symbols, 
respectively.}
\end{figure}

\clearpage
\setcounter{figure}{0}
\renewcommand{\thefigure}{\arabic{figure}.41}
\begin{figure}
\begin{center}
{\includegraphics[angle=90,height=12cm]{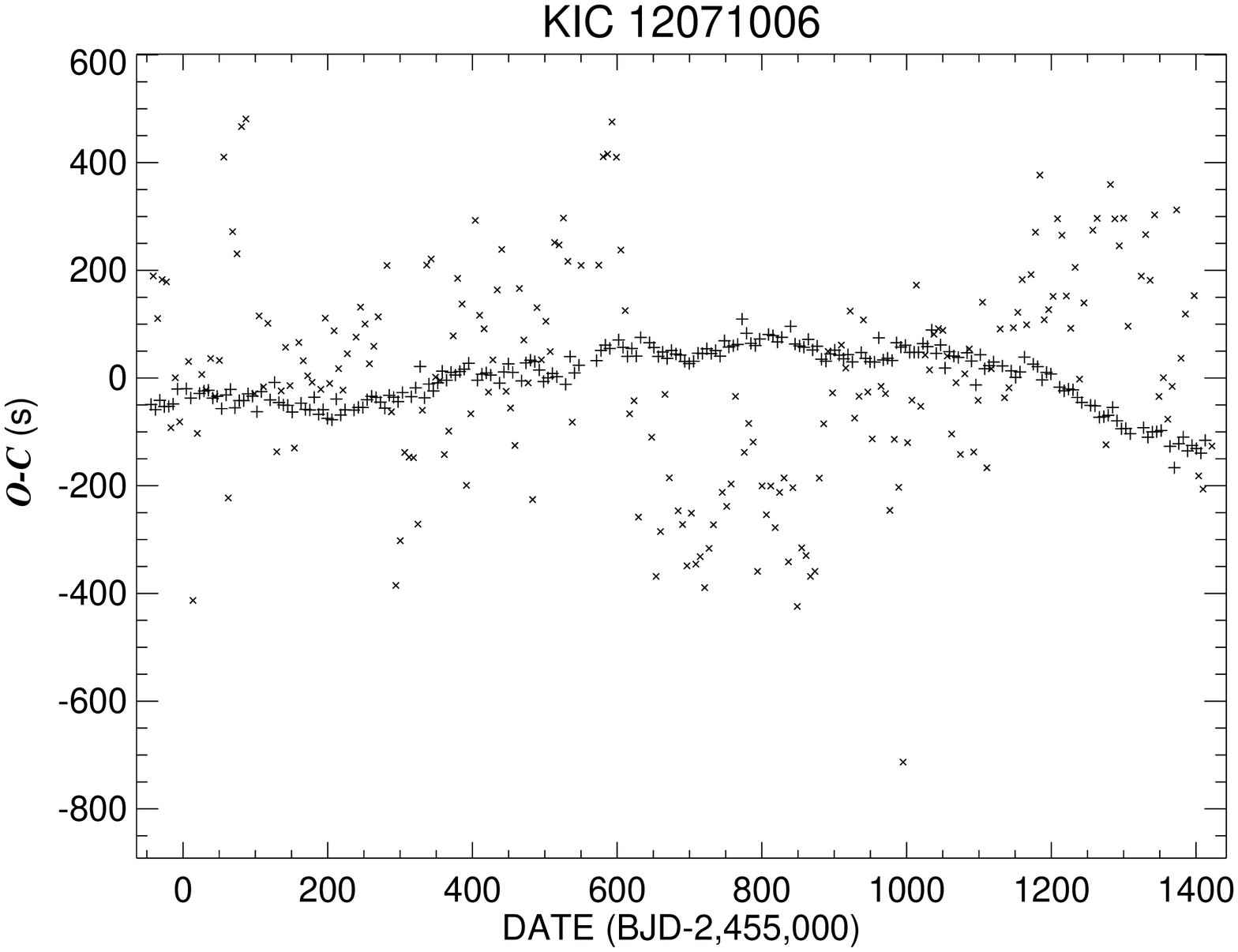}}
\end{center}
\caption{The observed minus calculated eclipse times relative to
a linear ephemeris.  The primary and secondary eclipse
times are indicated by $+$ and $\times$ symbols, 
respectively.}
\end{figure}



\end{document}